# Cosmic Ray Measurements with the KASCADE-Grande Experiment

## Presentations for the
## 31$^{st}$ International Cosmic Ray Conference, Łódź, Poland, July 2009

1. **Results on the cosmic ray energy spectrum measured with KASCADE-Grande**
   *KASCADE-Grande Collaboration, presented by Andreas Haungs*

2. **Cosmic ray energy spectrum based on shower size measurements of KASCADE-Grande**
   *KASCADE-Grande Collaboration, presented by Donghwa Kang*

3. **The Energy Spectrum of Primary Cosmic Rays Reconstructed with the KASCADE-Grande Muon Data.**
   *KASCADE-Grande Collaboration, presented by Juan Carlos Arteaga-Velazquez*

4. **The all particle energy spectrum of KASCADE-Grande in the energy region $10^{16}$ - $10^{18}$ eV by means of the $N_{ch}$ - $N_{\mu}$ technique**
   *KASCADE-Grande Collaboration, presented by Mario Bertaina*

5. **Primary energy reconstruction from the S(500) observable recorded with the KASCADE-Grande detector array**
   *KASCADE-Grande Collaboration, presented by Gabriel Toma*

6. **Performance of the KASCADE-Grande array**
   *KASCADE-Grande Collaboration, presented by Federico Di Pierro*

7. **Muonic Component of Air Showers Measured by the KASCADE-Grande Experiment**
   *KASCADE-Grande Collaboration, presented by Daniel Fuhrmann*

8. **The sensitivity of KASCADE-Grande to the cosmic ray primary composition between $10^{16}$ and $10^{18}$ eV**
   *KASCADE-Grande Collaboration, presented by Elena Cantoni*

9. **A direct measurement of the muon component of air showers by the KASCADE-Grande Experiment**
   *KASCADE-Grande Collaboration, presented by Vitor de Souza*

10. **Study of EAS development with the Muon Tracking Detector in KASCADE-Grande**
    *KASCADE-Grande Collaboration, presented by Janusz Zabierowski*

11. **Lateral distribution of EAS muons measured with the KASCADE-Grande Muon Tracking Detector**
    *KASCADE-Grande Collaboration, presented by Pawel Łuczak*

12. **Muon Production Height and Longitudinal Shower Development in KASCADE-Grande**
    *KASCADE-Grande Collaboration, presented by Paul Doll*

13. **Restoring Azimuthal Symmetry of Lateral Density Distributions of EAS Particles**
    *KASCADE-Grande Collaboration, presented by Octavian Sima*

14. **Quantitative tests of hadronic interaction models with KASCADE-Grande air shower data**
    *KASCADE-Grande Collaboration, presented by Jörg R. Hörandel*

# Results on the cosmic ray energy spectrum measured with KASCADE-Grande

A. Haungs*, W.D. Apel*, J.C. Arteaga†,xi, F. Badea*, K. Bekk*, M. Bertaina‡, J. Blümer*,†,
H. Bozdog* I.M. Brancus§, M. Brüggemann¶, P. Buchholz¶, E. Cantoni‡,||, A. Chiavassa‡,
F. Cossavella†, K. Daumiller*, V. de Souza†,xii, F. Di Pierro‡, P. Doll*, R. Engel*, J. Engler*,
M. Finger*, D. Fuhrmann**, P.L. Ghia||, H.J. Gils*, R. Glasstetter**, C. Grupen¶, D. Heck*,
J.R. Hörandel†,xiii, T. Huege*, P.G. Isar*, K.-H. Kampert**, D. Kang†, D. Kickelbick¶,
H.O. Klages*, P. Łuczak††, H.J. Mathes*, H.J. Mayer*, J. Milke*, B. Mitrica§, C. Morello||,
G. Navarra‡, S. Nehls*, J. Oehlschläger*, S. Ostapchenko*,xiv, S. Over¶, M. Petcu§, T. Pierog*,
H. Rebel*, M. Roth*, H. Schieler*, F. Schröder*, O. Sima‡‡, M. Stümpert†, G. Toma§,
G.C. Trinchero||, H. Ulrich*, A. Weindl*, J. Wochele*, M. Wommer*, J. Zabierowski††

*Institut für Kernphysik, Forschungszentrum Karlsruhe, 76021 Karlsruhe, Germany
†Institut für Experimentelle Kernphysik, Universität Karlsruhe, 76021 Karlsruhe, Germany
‡Dipartimento di Fisica Generale dell'Università, 10125 Torino, Italy
§National Institute of Physics and Nuclear Engineering, 7690 Bucharest, Romania
¶Fachbereich Physik, Universität Siegen, 57068 Siegen, Germany
||Istituto di Fisica dello Spazio Interplanetario, INAF, 10133 Torino, Italy
**Fachbereich Physik, Universität Wuppertal, 42097 Wuppertal, Germany
††Soltan Institute for Nuclear Studies, 90950 Lodz, Poland
‡‡Department of Physics, University of Bucharest, 76900 Bucharest, Romania
xi now at: Universidad Michoacana, Morelia, Mexico
xii now at: Universidade de São Paulo, Instituto de Física de São Carlos, Brasil
xiii now at: Dept. of Astrophysics, Radboud University Nijmegen, The Netherlands
xiv now at: University of Trondheim, Norway

*Abstract*. **KASCADE-Grande is an extensive air shower experiment at Forschungszentrum Karlsruhe, Germany. The present contribution attempts to provide a synopsis of the actual results of the reconstruction of the all-particle energy spectrum in the range of $10^{16}$ eV to $10^{18}$ eV based on four different methods with partly different sources of systematic uncertainties. Since the calibration of the observables in terms of the primary energy depends on Monte-Carlo simulations, we compare the results of the various methods applied to the same sample of measured data.**

## I. KASCADE-Grande

Main parts of the experiment are the Grande array spread over an area of $700 \times 700 \, \text{m}^2$, the original KASCADE array covering $200 \times 200 \, \text{m}^2$ with unshielded and shielded detectors, and additional muon tracking devices. This multi-detector system allows us to investigate the energy spectrum, composition, and anisotropies of cosmic rays in the energy range up to 1 EeV. The estimation of energy and mass of the primary particles is based on the combined investigation of the charged particle, the electron, and the muon components measured by the detector arrays of Grande and KASCADE.

The multi-detector experiment KASCADE [1] (located at $49.1°$n, $8.4°$e, 110 m a.s.l.) was extended to KASCADE-Grande in 2003 by installing a large array of 37 stations consisting of $10 \, \text{m}^2$ scintillation detectors each (fig. 1). KASCADE-Grande [2] provides an area of $0.5 \, \text{km}^2$ and operates jointly with the existing KASCADE detectors. The joint measurements with the KASCADE muon tracking devices are ensured by an additional cluster (Piccolo) close to the center of KASCADE-Grande for fast trigger purposes. While the Grande detectors are sensitive to charged particles, the KASCADE array detectors measure the electromagnetic component and the muonic component separately. The muon detectors enable to reconstruct the total number of muons on an event-by-event basis also for Grande triggered events.

## II. Reconstruction

Basic shower observables like the core position, angle-of-incidence, and total number of charged particles are provided by the measurements of the Grande stations. A core position resolution of $\approx 5 \, \text{m}$, a direction resolution of $\approx 0.7°$, and a resolution of the total particle number in the showers of $\approx 15\%$ is reached [3]. The total number of muons ($N_\mu$ resolution $\approx 25\%$) is calculated using the core position determined by the Grande array and the muon densities measured by the KASCADE muon array detectors [4]. Full efficiency for triggering and reconstruction of air-showers is reached at primary energy of $\approx 2 \cdot 10^{16}$ eV, slightly varying on the cuts needed for the reconstruction of the different observables.

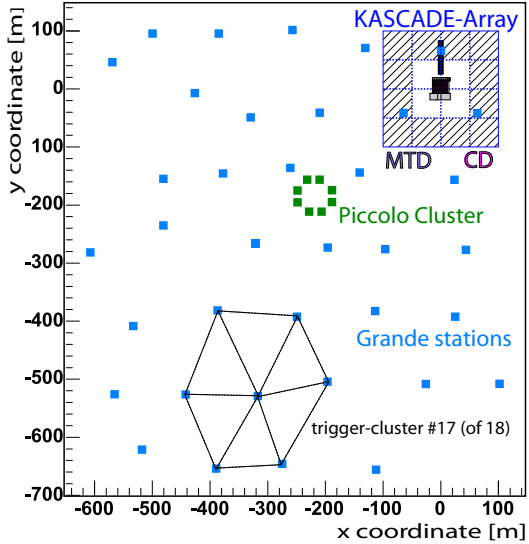

Fig. 1: Layout of the KASCADE-Grande experiment: The original KASCADE, the distribution of the 37 stations of the Grande array, and the small Piccolo cluster for fast trigger purposes are shown. The outer 12 clusters of the KASCADE array consist of $\mu$- and $e/\gamma$-detectors, the inner 4 clusters of $e/\gamma$-detectors, only.

Applying different methods to the same data sample has advantages in various aspects: One would expect the same result for the energy spectrum by all methods when the measurements are accurate enough, when the reconstructions work without failures, and when the Monte-Carlo simulations describe correctly the shower development. But, the fact that the various observables have substantial differences in their composition sensitivity hampers a straightforward analysis. However, investigating results of different methods can be used to

- cross-check the measurements by different sub-detectors;
- cross-check the reconstruction procedures;
- cross-check the influence of systematic uncertainties;
- test the sensitivity of the observables to the elemental composition;
- test the validity of hadronic interaction models underlying the simulations.

### III. ANALYSIS

The estimation of the all-particle energy spectrum is presently based on four different methods using different observables of KASCADE-Grande:

- $N_{ch}$-method: The reconstructed charge particle shower size per individual event is corrected for attenuation by the constant intensity cut method and calibrated by Monte-Carlo simulations under the assumption of a dependence $E_0 \propto N_{ch}^{\alpha_{ch}}$ and a particular primary composition [5].
- $N_\mu$-method: The reconstructed muon shower size per individual event is corrected for attenuation

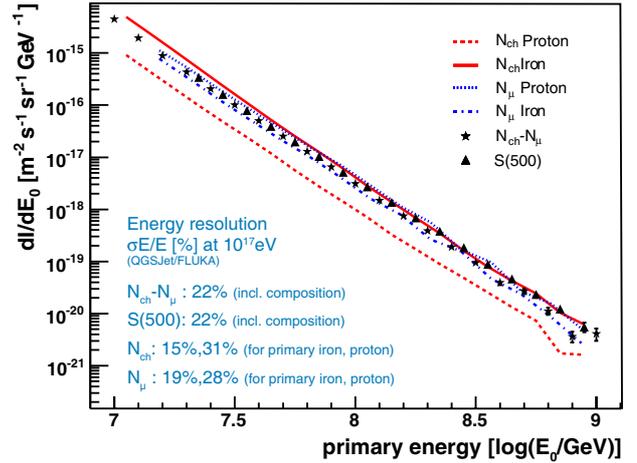

Fig. 2: Reconstructed all-particle energy spectrum by four different methods applied to KASCADE-Grande data. Given are also the energy resolution for the methods.

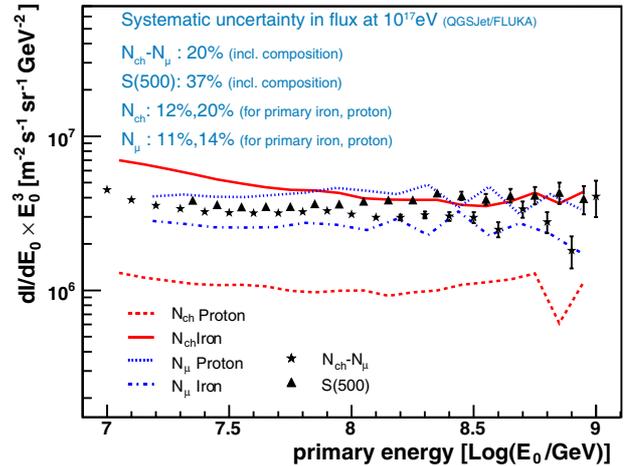

Fig. 3: Same as figure 2, but the flux multiplied by $E_0^3$. Values for the uncertainty in the flux determination are given for the different methods.

and calibrated by Monte-Carlo simulations under the assumption of a dependence $E_0 \propto N_\mu^{\alpha_\mu}$ and a particular primary composition [6].
- $N_{ch}-N_\mu$-method: This method combines the information provided by the two observables. By help of Monte-Carlo simulations a formula is obtained to calculate the primary energy per individual shower on basis of $N_{ch}$ and $N_\mu$. The formula takes into account the mass sensitivity in order to minimize the composition dependence. The attenuation is corrected for by deriving the formula for different zenith angle intervals independently and combining the energy spectrum afterwards [7].
- $S(500)$-method: The reconstructed particle density at the specific distance to the shower axis of 500 m per individual event is corrected for attenuation and calibrated by Monte-Carlo simulations under the assumption of a dependence $E_0 \propto S(500)^{\alpha_{S(500)}}$.

The distance of 500 m is chosen to have a minimum influence of the primary particle type, therefore a smaller dependence on primary composition is expected [8].

In figures 2 and 3 the resulting spectra are compiled. Due to the different procedures, the results for the first two methods are shown under proton and iron assumption, respectively, only, whereas for the other two methods the resulting all-particle energy spectrum is displayed. Figure 3 shows the same results but with the flux multiplied by a factor of $E^{3.0}$.

### A. Systematic uncertainties and attenuation

The application of the different methods allows us to compare and cross-check the influence of various sources of systematic uncertainties. The $N_{ch}$-method uses the basic measurements of the Grande array only, resulting in a high accuracy of $N_{ch}$ with better than 15% over the whole range of shower size, without any energy dependent bias. But, using only one observable, there is a large dependence on the primary elemental composition, reflected by the distance between the spectra obtained for proton and iron assumption at the calibration. The $N_\mu$-method on the other hand is based on the muon shower size, which can be estimated less accurate (25% with a little bias dependent on the distance of the shower core to the muon detectors which is corrected for), but with a much less composition dependence. The $N_{ch}$-$N_\mu$-method, due to the combination of the reconstruction uncertainty of two variables shows basically a larger uncertainty in the reconstruction, but this is compensated by taking into account the correlation of these observables at individual events. Furthermore, by this procedure the composition dependence is strikingly decreased. The $S(500)$ value by construction yields a larger uncertainty of the variable reconstruction, but has also a minor composition dependence.

For all methods, the energy resolution is estimated using full Monte-Carlo simulations and comparing the reconstructed with the simulated primary energy (for instance figure 2 gives the numbers for an energy of $E_0 = 10^{17}$ eV). Values of systematic uncertainties in the flux determination for the different methods are shown in fig. 3 (again for $E_0 = 10^{17}$ eV). These uncertainties are to a large amount due to the reconstruction of the observables, but there are additional sources of systematics which belong to all methods: e.g., concerning the Monte-Carlo statistics, the assumed Monte-Carlo spectral slope, or the fits of the calibration procedures. The different attenuation (and its handling to correct for) of the various observables ($\Lambda(N_{ch}) \approx 495 \pm 20$ g/cm$^2$; $\Lambda(N_\mu) \approx 1100 \pm 100$ g/cm$^2$; $\Lambda(S(500)) \approx 347 \pm 22$ g/cm$^2$ at $E_0 = 10^{17}$ eV) however, lead again to slightly different contribution to the total systematic uncertainty. The total uncertainties (energy resolution and systematics) for the various methods are discussed in refs. [5], [6], [7], [8] and can be displayed as a band surrounding the reconstructed energy spectrum (e.g., see fig. 4).

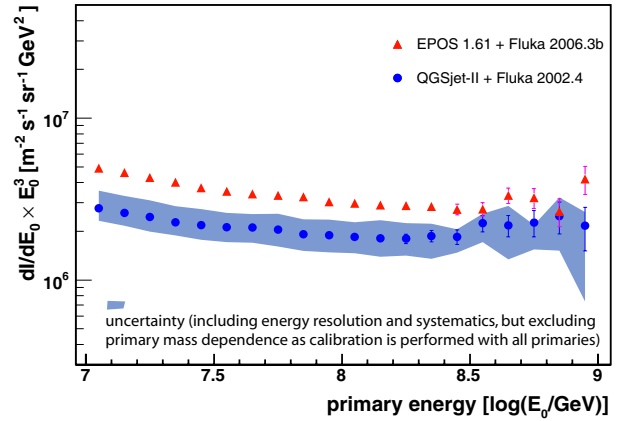

Fig. 4: Reconstructed all-particle energy spectrum with the $N_{ch}$-method and the calibration function obtained by assuming mixed composition, but based on two different hadronic interaction models.

### B. Discussion

Taking into account the systematic uncertainties, there is a fair agreement between the all-particle energy spectra of the different applications (fig. 3).

Of particular interest is the fact that by using $N_{ch}$, the iron assumption predicts a higher flux than the proton assumption, whereas using $N_\mu$ the opposite is the case. That means that the 'true' spectrum has to be a solution in between the range spanned by both methods. If one has only the possibility of applying one method, than there is a large variance in possible solutions (everything in the range spanned by proton and iron line, not even parallel to these lines). However, more detailed investigations have shown, that a structure in the spectrum or a sudden change in composition would be retained in the resulting spectrum, even if the calibration is performed with an individual primary, only. Interestingly, over the whole energy range there is only little room for a solution satisfying both ranges, spanned by $N_{ch}$ and $N_\mu$, and this solution has to be of relative heavy composition - in the framework of the QGSJet-II hadronic interaction model. The narrower range for a solution provided by the $N_\mu$-method compared to $N_{ch}$ confirms the finding of KASCADE that at sea-level the number of mostly low-energy muons $N_\mu$ is a very good and composition insensitive energy estimator.

The results of the composition independent $N_{ch}$-$N_\mu$-, and $S(500)$-methods lie inside the area spanned by the composition dependent methods, which is a very promising result. The $S(500)$-method results in a slightly higher flux than the $N_{ch}$-$N_\mu$-method, but the two spectra are consistent taking into account the systematic uncertainties.

All the discussed results show a smooth all-particle energy spectrum without any hint to a distinct structure over the whole energy range from 10 PeV to 1 EeV. Another conclusion is that, taking into account the systematic uncertainties for all methods, the underlying hadronic interaction model (QGSJet-II/FLUKA) is intrinsically consistent, i.e. the correlation between the different observables, respectively the particle components can describe the global features of our measurements.

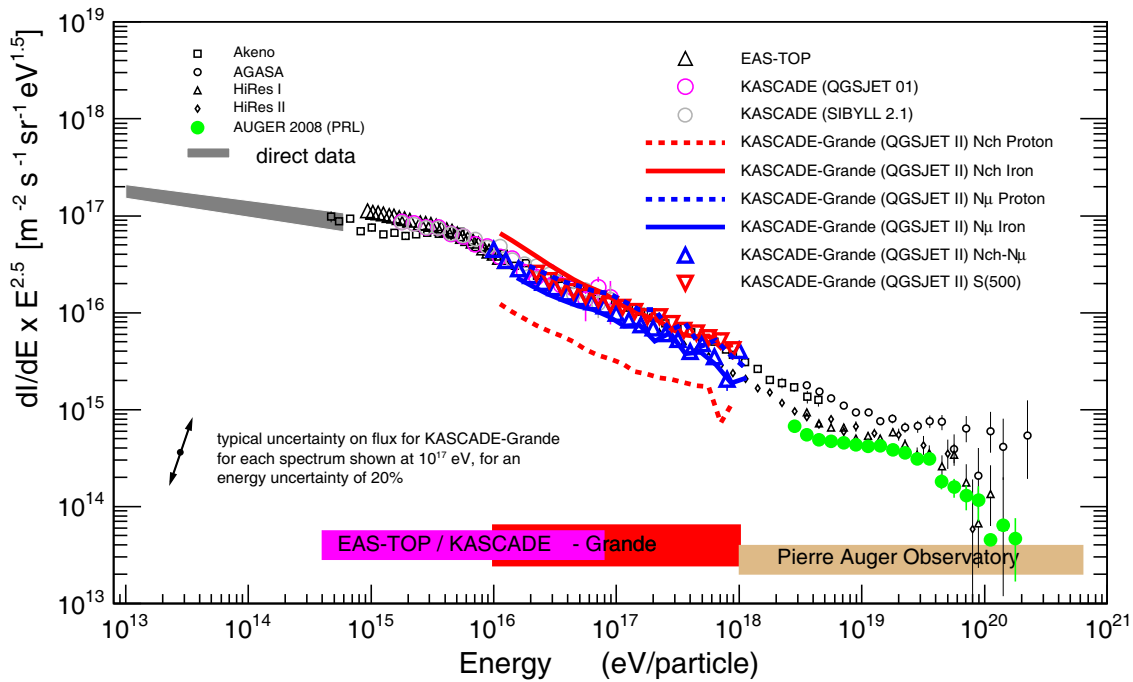

Fig. 5: Compilation of the all-particle energy spectrum obtained by four different methods applied to KASCADE-Grande data and in comparison to results of other experiments.

*C. Hadronic interaction models*

By now, for all the considerations the models QGSJet-II and FLUKA [9], [10], [11] are used, only. Other interaction models would probably change the interpretation of the data. We investigated the influence of the hadronic interaction model exemplarily by performing the $N_{ch}$-method based on simulations with the hadronic interaction model EPOS vers.1.61 [12]. As the Monte-Carlo statistics is limited in case of EPOS, both spectra were obtained by generating the calibration curve with an equally mixed composition of five primaries (H,He,C,Si,Fe). Figure 4 compares the all-particle energy spectrum obtained with the KASCADE-Grande data set for both cases. The interpretation of the KASCADE-Grande data with EPOS leads to a significantly higher flux compared to the QGSJet-II result. Though we know, that version 1.61 of the EPOS model is not consistent with air shower data (in particular, it cannot describe the correlation of hadronic observables with the muon and electron content of the EAS [13]) this example shows that by applying and comparing various reconstruction methods on the same data set will be useful for a better understanding of the interaction processes in the air shower development.

## IV. CONCLUSION

Applying various different reconstruction methods to the KASCADE-Grande data the obtained all-particle energy spectra are compared for cross-checks of reconstruction, for studies of systematic uncertainties and for testing the validity of the underlying hadronic interaction model. The resulting energy spectra are consistent to each other and in the overlapping energy range in a very good agreement to the spectrum obtained by the KASCADE and EAS-TOP experiments (fig. 5).

# Cosmic ray energy spectrum based on shower size measurements of KASCADE-Grande

D. Kang$^†$, W.D. Apel$^*$, J.C. Arteaga$^{†,\text{xi}}$, F. Badea$^*$, K. Bekk$^*$, M. Bertaina$^‡$, J. Blümer$^{*,†}$,
H. Bozdog$^*$, I.M. Brancus$^§$, M. Brüggemann$^¶$, P. Buchholz$^¶$, E. Cantoni$^{‡,\|}$, A. Chiavassa$^‡$,
F. Cossavella$^†$, K. Daumiller$^*$, V. de Souza$^{†,\text{xii}}$, F. Di Pierro$^‡$, P. Doll$^*$, R. Engel$^*$, J. Engler$^*$,
M. Finger$^*$, D. Fuhrmann$^{**}$, P.L. Ghia$^\|$, H.J. Gils$^*$, R. Glasstetter$^{**}$, C. Grupen$^¶$,
A. Haungs$^*$, D. Heck$^*$, J.R. Hörandel$^{†,\text{xiii}}$, T. Huege$^*$, P.G. Isar$^*$, K.-H. Kampert$^{**}$,
D. Kickelbick$^¶$, H.O. Klages$^*$, P. Łuczak$^{††}$, H.J. Mathes$^*$, H.J. Mayer$^*$,
J. Milke$^*$, B. Mitrica$^§$, C. Morello$^\|$, G. Navarra$^‡$, S. Nehls$^*$, J. Oehlschläger$^*$,
S. Ostapchenko$^{*,\text{xiv}}$, S. Over$^¶$, M. Petcu$^§$, T. Pierog$^*$, H. Rebel$^*$, M. Roth$^*$,
H. Schieler$^*$, F. Schröder$^*$, O. Sima$^{‡‡}$, M. Stümpert$^†$, G. Toma$^§$, G.C. Trinchero$^\|$,
H. Ulrich$^*$, A. Weindl$^*$, J. Wochele$^*$, M. Wommer$^*$, J. Zabierowski$^{††}$

$^†$*Institut für Experimentelle Kernphysik, Universität Karlsruhe, 76021 Karlsruhe, Germany*
$^*$*Institut für Kernphysik, Forschungszentrum Karlsruhe, 76021 Karlsruhe, Germany*
$^‡$*Dipartimento di Fisica Generale dell'Università, 10125 Torino, Italy*
$^§$*National Institute of Physics and Nuclear Engineering, 7690 Bucharest, Romania*
$^¶$*Fachbereich Physik, Universität Siegen, 57068 Siegen, Germany*
$^\|$*Istituto di Fisica dello Spazio Interplanetario, INAF, 10133 Torino, Italy*
$^{**}$*Fachbereich Physik, Universität Wuppertal, 42097 Wuppertal, Germany*
$^{††}$*Soltan Institute for Nuclear Studies, 90950 Lodz, Poland*
$^{‡‡}$*Department of Physics, University of Bucharest, 76900 Bucharest, Romania*
$^{\text{xi}}$*now at: Universidad Michoacana, Morelia, Mexico*
$^{\text{xii}}$*now at: Universidade São Paulo, Instituto de Fisica de São Carlos, Brasil*
$^{\text{xiii}}$*now at: Dept. of Astrophysics, Radboud University Nijmegen, The Netherlands*
$^{\text{xiv}}$*now at: University of Trondheim, Norway*

*Abstract*. The KASCADE-Grande (KArlsruhe Shower Core and Array DEtector and Grande array), located on site of the Forschungszentrum Karlsruhe in Germany, is designed for observations of cosmic ray air showers in the energy range of $10^{16}$ to $10^{18}$ eV. The measurement of the all-particle energy spectrum of cosmic rays is based on the size spectra of the charged particle component, measured for different zenith angle ranges and on the "Constant Intensity Cut" method to correct for attenuation effects. The all-particle energy spectrum, calibrated by Monte-Carlo simulations, is presented and systematic uncertainties discussed.

*Keywords*: cosmic rays; KASCADE-Grande; constant intensity cut method.

## I. INTRODUCTION

The energy spectrum and composition of primary cosmic rays around $10^{17}$ eV are very important since they might be related to the existence of extragalactic cosmic ray sources, which might have a significantly different elemental composition from the one observed at lower energies [1]. The aim of KASCADE-Grande is the examination of the iron-knee in the cosmic ray energy spectrum, i.e. the end of the bulk of cosmic rays of galactic origin. It is expected at around $10^{17}$ eV following previous KASCADE observations [2]. KASCADE-Grande will allow investigations in detail about the elemental composition giving the possibility to distinguish between astrophysical models for the transition region from galactic to extragalactic origin of cosmic rays. The KASCADE-Grande array covering an area of 700×700 m$^2$ is optimized to measure extensive air showers up to primary energies of 1 EeV [3]. It comprises 37 scintillation detector stations located on a hexagonal grid with an average spacing of 137 m for the measurements of the charged shower component. Each of the detector stations is equipped with plastic scintillator sheets covering a total area of 10 m$^2$. The stations contain 16 scintillator sheets read-out by photomultipliers providing a dynamic range up to about 10000 charged particles per station for the reconstruction of particle densities and timing measurements. The timing accuracy of Grande stations allows an excellent angular resolution [4]. Grande is electronically subdivided in 18 hexagonal trigger clusters formed by six external and one central stations. A trigger signal is build when all stations in a hexagon are fired, and its total rate is about 0.5 Hz. Full efficiency for the shower size is reached at the number of charged particles of around $10^6$, which approximately corresponds to a primary energy of $10^{16}$ eV, so that a large overlap for cross-checks with measurements of the original KASCADE experiment is attained. The limit at high energies is due to the restricted area of the Grande array.

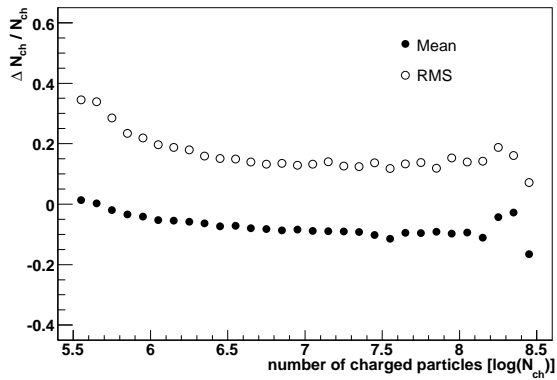

Fig. 1: Reconstruction accuracy of the number of charged particles obtained by Monte-Carlo simulations. The closed and open circles represent the values of mean and Root Mean Square, respectively.

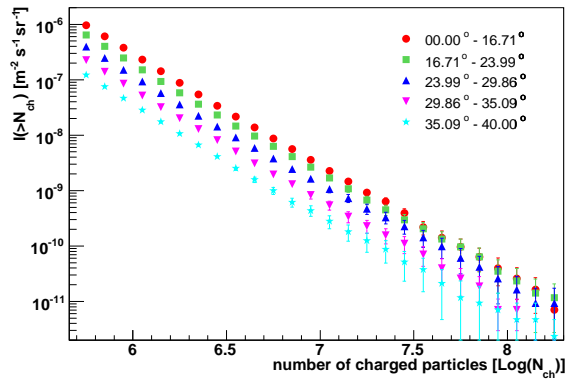

Fig. 2: Integral shower size spectra of the number of charged particles for different zenith angle ranges.

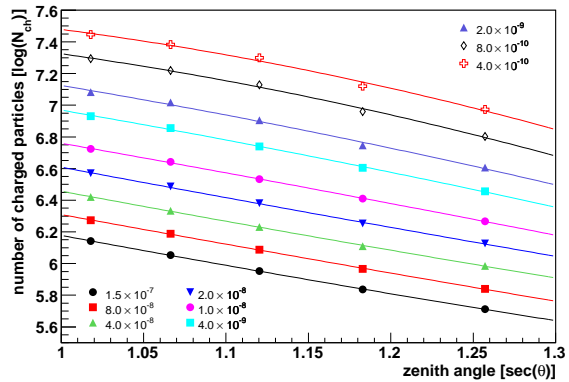

Fig. 3: The number of charged particles resulting from the constant intensity cuts as a function of the zenith angle. The unit of each flux (symbol) is $m^{-2}s^{-1}sr^{-1}$.

## II. RECONSTRUCTION ACCURACY

The primary energy of cosmic rays is reconstructed by the observed electron and muon numbers at ground. While the Grande detectors are sensitive to all charged particles, the KASCADE detectors measure separately the electromagnetic and muonic components due to the shielding above the muon counters. Therefore, the shower core position, the arrival direction, and the total number of charged particles in the shower are reconstructed from Grande array data, whereas the total number of muons is extracted from the data of the KASCADE muon detectors. Performing CORSIKA air shower simulations [5] including the detector response of the Grande array, the parameters were optimized for the lateral density distribution of KASCADE-Grande [4], and well reconstructed with sufficient accuracies for the further physics analysis. Figure 1 shows the accuracy of the reconstructed number of charged particles obtained by Monte-Carlo simulations. The accuracy could be confirmed by combining information of KASCADE with Grande reconstruction on a subsample of commonly measured events (Ref. [4]). In order to avoid misreconstruction effects of shower core positions on the border of the Grande array, a fiducial area of about 0.2 km$^2$ centered in the middle of the Grande array is chosen. The statistical uncertainty of the shower size is of the order of 20% for the total number of charged particles. Above a threshold of $10^6$ charged particles, the reconstruction accuracies of the core position and the arrival direction are better than 8 m and 0.5°, respectively, for zenith angles below 40°.

## III. DATA ANALYSIS

KASCADE-Grande has started combined data acquisition with all detector components since the end of 2003. The data presented here were taken from December 2003 to March 2009. It corresponds to the effective measuring time of 987 days, where all components of KASCADE and KASCADE-Grande were operating without failures in data acquisition. In this analysis, all showers with zenith angles smaller than 40° have been analyzed. After some quality cuts approximately 10 million events are available for the physics analysis. As the first step to determine the all-particle energy spectrum, the constant intensity cut method was introduced. This method assumes that cosmic rays arrive isotropically from all directions, i.e. the primary energy of a cosmic ray particle corresponds to a certain intensity regardless of its arrival direction. In KASCADE-Grande, an isotropic distribution is assumed in the considered energy range up to $10^{18}$ eV, so that it allows us to apply the constant intensity method to the integral shower size spectra for different zenith angular bins (Fig. 2).

Above a certain shower size, the intensity should be constant due to the assumption of the uniform intensity distribution when binned in $\cos^2\theta$. For a given intensity, the number of charged particles is calculated for each zenith angle range. The intensity cut is mostly located in between two neighboring points of the distributions, and thus the exact values of the corresponding shower size are estimated by interpolation between these two points. From the values of $N_{ch}$ we obtain the attenuation curves (Fig. 3). It represents how the number of charged particles for a given intensity attenuates through the atmosphere with increasing zenith angle, i.e. with increasing atmospheric depth. Each curve is individually fitted by a second-degree polynomial function, where the reference angle $\theta_{ref}$ of 20° is chosen by the mean value from a Gaussian fit to the measured zenith angle distribution. Due to a negligible variation of the fit parameters only one value is used to correct the number of charged particles event by event for the attenuation in the atmosphere. The zenith angle dependence of the number of charged particles was therefore eliminated by using informations from the measurements only.

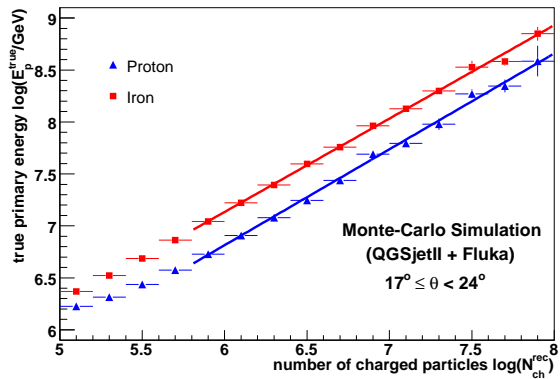

Fig. 4: The true primary energy as a function of the number of charged particles for proton and iron components. The lines show the applied fits to the points.

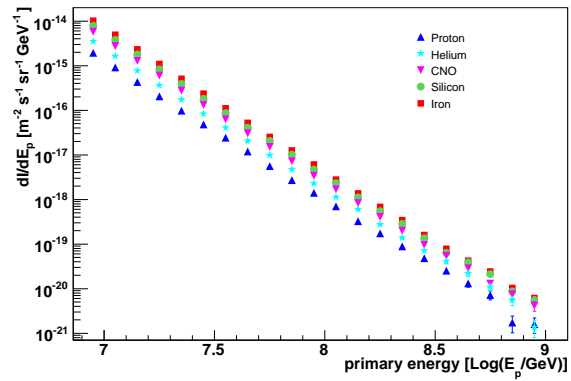

Fig. 5: All-particle energy spectrum reconstructed from KASCADE-Grande shower size for five different primary particle type assumptions together with its statistical uncertainties.

In order to determine the energy conversion relation between the number of charged particles $N_{ch}$ and the primary energy $E_p$, Monte-Carlo simulations were used. Extensive air showers were simulated using the program CORSIKA with QGSjetII [6] and FLUKA as hadronic interaction models, including full simulations of the detector response. The simulated data sets contain air shower events for five different primary mass groups: proton, helium, carbon, silicon and iron. For the simulation, events for the zenith angle ranges of $17° \leq \theta < 24°$, i.e. around the reference angle, are selected to reduce systematic effects. The relation of the primary energy as a function of the number of charged particles is shown in Fig. 4. Assuming a linear dependence $\log E_p = a + b \cdot \log N_{ch}$, the correlation between the primary energy and the number of charged particles is obtained, where the fit is applied in the range of full trigger and reconstruction efficiencies. The fit yields $a = 1.28 \pm 0.08$ and $b = 0.92 \pm 0.01$ with a reduced $\chi^2$ of 1.42 for proton, and $a = 1.74 \pm 0.07$ and $b = 0.90 \pm 0.01$ with a reduced $\chi^2$ of 0.98 for iron. The same procedure is also performed for helium, carbon and silicon to examine the dependence of the calibration on the assumed primary particle type, where the fit parameters are also in between above values. Using these correlations the all-particle energy spectrum is obtained. The reconstructed energy spectra for the assumption of five different primary particle types are shown in Fig. 5.

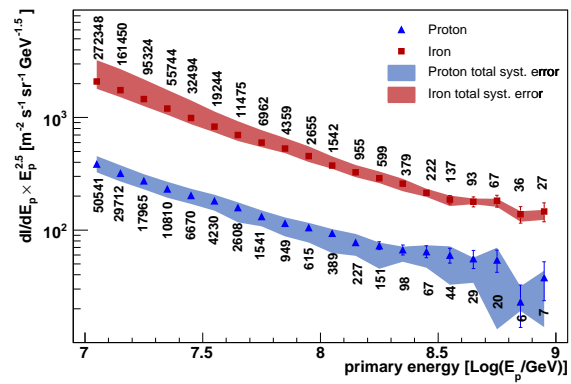

Fig. 6: The reconstructed energy spectrum for proton and iron, together with statistical error bars (vertical lines) and uncertainties (energy resolution and systematics) (bands). The numbers indicate the events per bin for the points.

## IV. SYSTEMATIC UNCERTAINTY

The energy resolution is estimated from the difference between simulated energy and derived energy, where the derived energy is obtained by applying the measured attenuation correction to the Monte-Carlo simulation. The energy resolutions for proton and iron are about 31% and 15% over the whole energy ranges, where the uncertainties of the reconstructed number of charged particles give the largest contribution. In addition, the systematic uncertainties on the reconstructed energy spectrum are investigated considering various possible contributions. Firstly, the fit of the attenuation curve was performed in order to correct the zenith angle dependence of the number of charged particles. Each fit parameter has an associated error and it effects on the determination of the energy spectrum. The fit parameters are correlated with each other, so that the propagation of errors is used to calculate the systematic uncertainty induced by the attenuation fit. This uncertainty is estimated to be less than 2% for both proton and iron in the full energy range. Secondly, for the correlation of $N_{ch}$ with $E_p$ a power law fit was applied. As the same procedure above, the estimated systematic uncertainties due to the fit of the energy calibration contribute with 1% for proton and 3% for iron to the total uncertainty. The shower fluctuations are another source of systematic uncertainties, which is basically caused by the nature of the development of extensive air showers. The influence of these fluctuations on the reconstructed primary energy spectrum is estimated by using simulation data. For the calibration a spectral index of $\gamma = -3$ is used in the simulations. By varying this spectral index, i.e. varying the effect of fluctuations on the reconstructed spectrum, the systematic uncertainty is estimated. The systematic deviation due to the shower fluctuation is evaluated by this procedure to be about 15% and 4% for proton and iron, respectively. Uncertainties due to a possible misdescription of the attenuation in the Monte-Carlo simulation are estimated by using two different reference zenith angles, together with the corresponding correlation of $N_{ch}$ and $E_p$. The systematic uncertainties of 14% for proton and 11% for iron on the energy estimation are obtained for varying the reference angle from 10° to 30° for the calibration. All these individual systematic contributions were considered to be uncorrelated, and combined thus in quadrature to obtain the total systematic uncertainty (Fig. 6), where the composition dependence was not taken into account. The systematic uncertainty (i.e. sum in quadrature of all terms discussed above except the energy resolution) in the energy scale is of the order of about 20% for proton and 12% for iron at the primary energy of $10^{17}$ eV. The uncertainties on the flux for proton and iron are 32% and 21%, respectively, at energies of $10^{17}$ eV. Further checks are currently being performed to reduce the systematic uncertainties on the energy estimation.

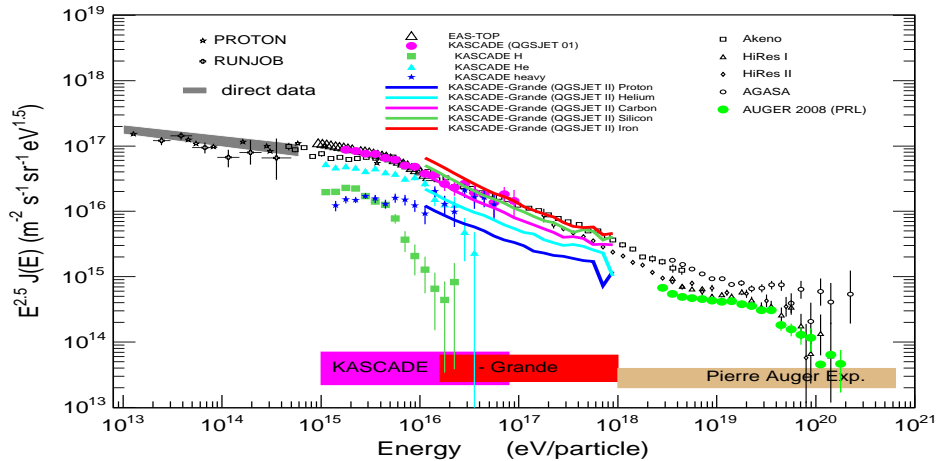

Fig. 7: All-particle energy spectrum in comparison with results of other experiments. The lines represent the KASCADE-Grande spectrum for the calibration assuming five different primary mass groups.

## V. CONCLUSION

The air shower experiment KASCADE-Grande measures cosmic rays in the energy range of $10^{16}$-$10^{18}$ eV. The multi detector setup of KASCADE-Grande allows us to reconstruct charged particles, electron and muon numbers of the showers separately with high accuracies. In the present contribution the reconstructed all-particle energy spectrum by means of the shower size measurements of the charged particle component by Grande array is presented by using the hadronic interaction model QGSjetII. The resulting spectrum is shown in Fig. 7 in comparison with results of other experiments. The obtained intensity values have been shown to depend on the nature of the primary particle as expected for an observable dominated by the electromagnetic components. Such values, inside the systematic uncertainties, are consistent with other KASCADE-Grande analysis based on different observables and methodologies. Their combination is a basic tool to provide an unbiased measurement of the primary energy spectrum and first indications on average primary composition [7].

# The Energy Spectrum of Primary Cosmic Rays Reconstructed with the KASCADE-Grande Muon Data

J.C. Arteaga-Velázquez[†,xi], W.D. Apel[*], F. Badea[*], K. Bekk[*], M. Bertaina[‡], J. Blümer[*,†],
H. Bozdog[*], I.M. Brancus[§], M. Brüggemann[¶], P. Buchholz[¶], E. Cantoni[‡,∥], A. Chiavassa[‡],
F. Cossavella[†], K. Daumiller[*], V. de Souza[†,xii], F. Di Pierro[‡], P. Doll[*], R. Engel[*], J. Engler[*],
M. Finger[*], D. Fuhrmann[**], P.L. Ghia[∥], H.J. Gils[*], R. Glasstetter[**], C. Grupen[¶],
A. Haungs[*], D. Heck[*], J.R. Hörandel[†,xiii], T. Huege[*], P.G. Isar[*], K.-H. Kampert[**],
D. Kang[†], D. Kickelbick[¶], H.O. Klages[*], P. Łuczak[††], H.J. Mathes[*], H.J. Mayer[*],
J. Milke[*], B. Mitrica[§], C. Morello[∥], G. Navarra[‡], S. Nehls[*], J. Oehlschläger[*],
S. Ostapchenko[*,xiv], S. Over[¶], M. Petcu[§], T. Pierog[*], H. Rebel[*], M. Roth[*],
H. Schieler[*], F. Schröder[*], O. Sima[‡‡], M. Stümpert[†], G. Toma[§], G.C. Trinchero[∥],
H. Ulrich[*], A. Weindl[*], J. Wochele[*], M. Wommer[*], J. Zabierowski[††]

[*]*Institut für Kernphysik, Forschungszentrum Karlsruhe, 76021 Karlsruhe, Germany*
[†]*Institut für Experimentelle Kernphysik, Universität Karlsruhe, 76021 Karlsruhe, Germany*
[‡]*Dipartimento di Fisica Generale dell'Università, 10125 Torino, Italy*
[§]*National Institute of Physics and Nuclear Engineering, 7690 Bucharest, Romania*
[¶]*Fachbereich Physik, Universität Siegen, 57068 Siegen, Germany*
[∥]*Istituto di Fisica dello Spazio Interplanetario, INAF, 10133 Torino, Italy*
[**]*Fachbereich Physik, Universität Wuppertal, 42097 Wuppertal, Germany*
[††]*Soltan Institute for Nuclear Studies, 90950 Lodz, Poland*
[‡‡]*Department of Physics, University of Bucharest, 76900 Bucharest, Romania*
[xi]*now at: Instituto de Física y Matemáticas, Universidad Michoacana, Morelia, Mexico*
[xii]*now at: Universidade de São Paulo, Instituto de Fîsica de São Carlos, Brasil*
[xiii]*now at: Dept. of Astrophysics, Radboud University Nijmegen, The Netherlands*
[xiv]*now at: University of Trondheim, Norway*

*Abstract*. A detailed analysis based on the Constant Intensity Cut method was applied to the KASCADE-Grande muon data in order to reconstruct an all-particle energy spectrum of primary cosmic rays in the interval $2.5 \times 10^{16} - 10^{18}$ **eV. To interpret the experimental data, Monte Carlo simulations carried out for five different primary nuclei (H, He, C, Si and Fe) using the high-energy hadronic interaction model QGSJET II were employed. For each case, the derived all-particle energy spectrum is presented. First estimations of the main systematic uncertainties are also shown.**

*Keywords*: Ground arrays, cosmic ray energy spectrum, muons

## I. INTRODUCTION

One of the main goals of the cosmic ray research is the measurement of the primary energy spectrum, which encloses important keys about the origin, acceleration and propagation of cosmic rays. This task can be done directly or indirectly, depending on the energy of the primary particle. At high energies, above $10^{15}$ eV, where direct detection is not feasible, the energy spectrum must be determined indirectly from the measured properties of the extensive air showers (EAS) that cosmic rays induce in the Earth's atmosphere. Depending on the experimental apparatus and the detection technique, different sets of EAS observables are available to estimate the energy of the primary cosmic ray [1]. In ground arrays the total number of charged particles in the shower and the corresponding density at observation level are more commonly employed [1], [2]. However, the muon content is also at disposal for this enterprise [3]. One reason in favor of this observable is that, in an air shower, muons undergo less atmospheric interactions than the charged component (dominated by electromagnetic particles for vertical EAS) and present in consequence less fluctuations. Another reason is that, according to MC simulations, the muon shower size ($N_\mu$) grows with the energy of the primary particle following a simple power law. Although these advantages, the muon number as an energy estimator is expected to be limited by the hadronic-interaction model, the experimental error and the uncertainty in the primary composition, among other things. In this work, the KASCADE-Grande muon data is used as a tool to derive the energy spectrum for cosmic rays in the range from $2.5 \times 10^{16}$ to $10^{18}$ under different composition scenarios. The method is explained and the main systematic uncertainties behind the calculations are presented.

## II. DESCRIPTION OF THE DETECTOR AND THE DATA

The KASCADE-Grande experiment was conceived as a ground-based air shower detector devoted to the search of the *iron knee* in the cosmic ray spectrum [4]. KASCADE-Grande, with an effective area of $0.5\,\text{km}^2$, is composed by several types of particle detectors dedicated to study different components of the EAS. Important for this analysis is the $200 \times 200 \, \text{m}^2$ muon detector array integrated by $192 \times 3.2 \, \text{m}^2$ shielded scintillator detectors, which are sensitive to muons with energy threshold above $230 \, \text{MeV}$ for vertical incidence [4], [5], [6]. The muon array was implemented to measure the lateral distribution of muons in the shower front and to extract the muon shower size. The latter is performed, event by event, from a fit to the observed lateral muon densities [6].

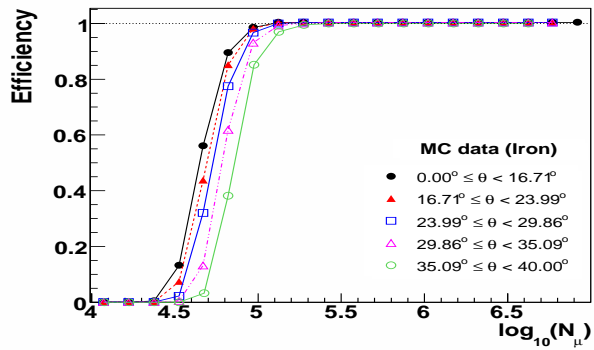

Fig. 1: KASCADE-Grande triggering and reconstruction efficiency shown as a function of the muon number for different zenith angle intervals. The efficiency was estimated from MC simulations assuming a pure iron composition.

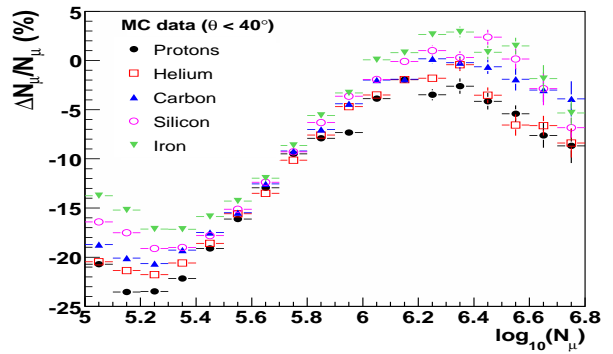

Fig. 2: Average values of the muon correction functions plotted versus the total muon number. Results for different primaries are displayed.

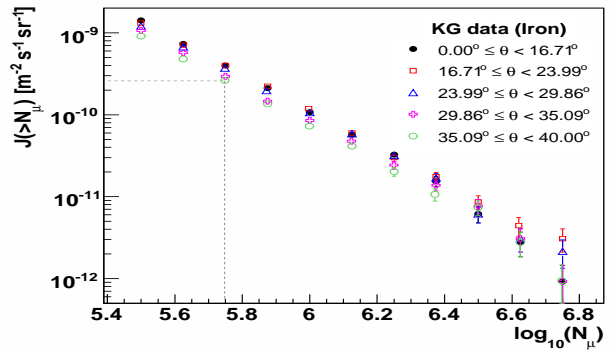

Fig. 3: Integral muon spectra obtained from the KASCADE-Grande data using the muon correction function for iron nuclei. The vertical error bars represent statistical uncertainties. In this figure, a CIC cut is represented by the horizontal line. The intersection of the CIC cut with a given integral flux defines a corresponding $N_\mu$ value, here indicated by the vertical line.

The present analysis was based on a muon data set collected with the KASCADE-Grande array during the period December 2003 - February 2009 for zenith angles, $\theta$, below $40°$. In order to reduce the influence of systematic uncertainties in this data, a fiducial area of $370 \times 520 \, \text{m}^2$ located at the center of KASCADE-Grande was employed. Moreover various experimental cuts were imposed. As a result the effective time of observation of the selected data was approximately 754.2 days. For the conditions above described, full efficiency is achieved for $\log_{10}(N_\mu) > 5.1 - 5.4$, according to MC simulations. Here the lower threshold corresponds to the case of light primaries and/or vertical air showers (see, for example, Fig. 1).

The systematic uncertainties of the instrument and the reconstruction procedures were also investigated with MC simulations. The EAS events were generated with an isotropic distribution with spectral index $\gamma = -2$ and were simulated with CORSIKA [7] and the hadronic MC generators FLUKA [8] and QGSJETII [9]. MC data sets were produced for five different representative mass groups: H, He, C, Si and Fe. In each case the simulated data was weighted with a proper function to describe a steeper energy spectrum with $\gamma = -3$, which was chosen as reference for the purpose of this study.

### III. THE PATH TO THE SPECTRUM

For the current analysis the experimental muon data was divided in five zenith angle intervals ($\Delta\theta$), each of them with the same value of acceptance. In addition, $N_\mu$ was corrected for systematic effects through a correction function, which was parameterized in terms of the zenith and azimuth angles, the core position and the muon size according to MC results. The precise magnitude of the corrections change with the primary mass, but on average they are under 25 % and tend to decrease with $N_\mu$ in the region of full efficiency (see Fig. 2). Along the paper muon correction functions were already applied to the data according to the primary composition assumed.

In order to reconstruct the all-particle energy spectrum, in a first step the CIC method was applied to the corrected muon data [2], [3], [10]. The objective was to extract a muon attenuation curve to correct the muon shower size at different atmospheric depths and convert it into an equivalent $N_\mu$ for a given zenith angle of reference to combine in this way muon data measured at different atmospheric depths. To start with, the integral muon spectra, $J(>N_\mu)$, were calculated for all $\Delta\theta$ bins. As an example, in Fig. 3 the integral fluxes derived with a $N_\mu$ correction function for iron nuclei were plotted. Unless otherwise indicated an iron primary composition will be assumed from now on to illustrate the procedure. Once the integral spectra were calculated, cuts at a fixed frequency rate or integral intensity were applied covering only the region of maximum efficiency and statistics. Then from the intersection of each cut and the integral fluxes (see Fig. 3) a muon attenuation curve was built plotting the intersected muon numbers as a function of $\sec\theta$, as displayed in Fig. 4. This can be in principle done since, according to the CIC method (where isotropy of cosmic rays is assumed) all EAS muon sizes connected through a specific cut belong to showers of identical energy. A technical point should be mentioned before continuing, that linear interpolation between two adjacent points was used in order to find the crossing between a given cut and a certain spectrum. The uncertainty introduced by the interpolation procedure in the extracted value of $N_\mu$ was properly taken into account along with the statistical errors of the integral spectra.

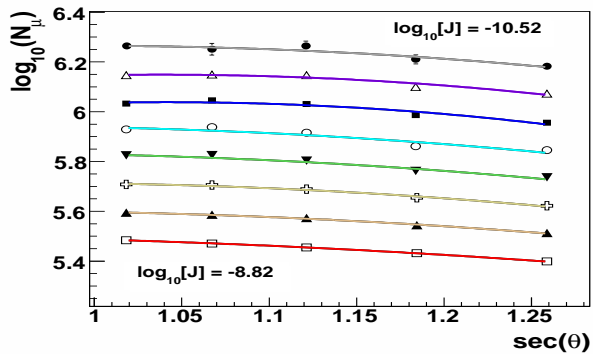

Fig. 4: Muon attenuation curves as extracted from several constant intensity cuts applied to the integral $N_\mu$ spectra of figure 3. From the bottom to the top, the CIC cuts decrease in steps of $\Delta \log_{10}[J/(m^{-2}s^{-1}sr^{-1})] \approx 0.24$.

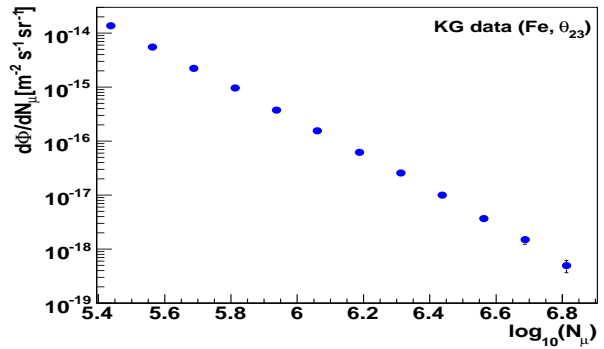

Fig. 5: The KASCADE-Grande muon spectrum obtained with the CIC method for $\theta_{ref} = 23.1°$. The muon correction function for iron nuclei was employed.

With the attenuation curves finally at disposal, one can calculate the equivalent muon number of an EAS for a zenith angle of reference, $\theta_{ref}$. This angle was chosen to be the mean of the measured zenith angle distribution, which was found around $23.1°$. Event by event, the equivalent EAS muon size for the selected atmospheric depth was estimated through the formula:

$$N_\mu(\theta_{ref}) = N_\mu(\theta)\exp\left[P(\theta_{ref}) - P(\theta)\right], \quad (1)$$

where $P(\theta)$ is a fit, with a second degree polynomial in $\sec\theta$, to the attenuation curves (see Fig. 4). In the above equation $P(\theta)$ is the closest curve to a given $N_\mu(\theta)$ data point. In Fig. 5 the equivalent muon spectrum for $\theta_{ref}$ as calculated with the CIC method is presented.

In a final step, to derive the energy spectrum from the above data a conversion relation from muon content into primary energy was invoked. The calibration formula was obtained from MC simulations by fitting the mean distribution of true energy versus $N_\mu$ for data with zenith angles around $\theta_{ref}$. The fit was done with a power law relation, $E[GeV] = \alpha \cdot N_\mu^\beta$, for the $N_\mu$ interval of full efficiency and high statistics (see Fig. 6). To test the reconstruction method the same analysis was applied to the MC data. Differences between the magnitude of the true and the estimated MC energy spectra were found. At $E \approx 10^{17}$ eV, they are smaller than 30%. Deviations are due to fluctuations in the reconstructed energy, which are bigger for light primary masses. They vary in the range of 18 − 29 % at $10^{17}$ eV. That defines our energy resolution at this energy scale.

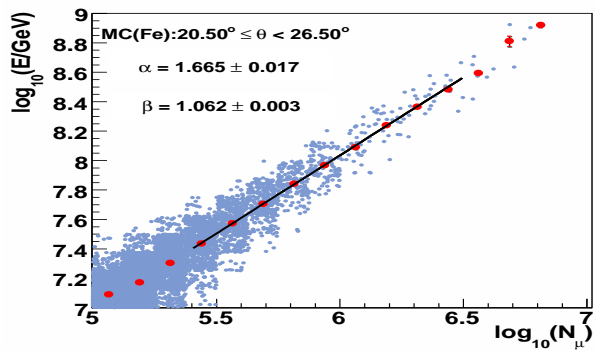

Fig. 6: Mean distribution of true energy vs muon number for iron nuclei and $\theta = 20.5° - 26.5°$ calculated with MC simulations. The fit with formula $E[GeV] = \alpha \cdot N_\mu^\beta$ is shown.

## IV. RESULTS AND CONCLUSIONS

The all-particle energy spectrum reconstructed from the KASCADE-Grande muon data is displayed in Fig. 7 for different primary composition assumptions. Measurements of the original KASCADE experiment [5] are also shown for comparison. It can be seen that the KASCADE data points are found inside the region that covers the KASCADE-Grande results, showing agreement between both experiments. Total uncertainties are also presented in the same figure. They take into account the following sources: 1) the influence of the energy resolution distribution, 2) uncertainties in the $N_\mu$ correction functions and 3) the energy conversion relation, both arising from the fits to MC data, 4) uncertainties in the estimation of the equivalent muon number, 5) a small shift observed in the estimated energy, which is introduced by the analysis, 6) uncertainties in the primary spectral index ($\gamma = -3 \pm 0.5$) and 7) the effect of selecting another reference angle $\theta_{ref}$, using for example 10 and 30°. Each of these contributions introduces a modification in the estimated energy of the events, which is propagated to the flux. The differences between the reference spectrum and the modified ones were interpreted as the corresponding uncertainties. They were added in quadrature to get the total error. In cases (2), (3) and (4) usual error propagation formulas were employed to find the energy uncertainty of the events. (1) and (5) were estimated from MC simulations. In the case of (1), the energy of an event was assigned in a probabilistic way using the energy resolution distributions per energy bin obtained with MC simulations. In (6) MC relations employed in the whole analysis were recalculated with simulations characterized by the new spectral indexes. Finally, for (7) both the equivalent $N_\mu$ and the energy calibration formula were estimated for the new values of $\theta_{ref}$. The resulting uncertainties (energy resolution (1) and systematics (2)-(7)) for the energy and the spectrum around $10^{17}$ eV are presented in table I. In general, for the interval $\log_{10}[E/GeV] = 7.4 - 8.3$ from all the estimated contributions to the total uncertainty of the flux the biggest one is related to composition ($\lesssim 50\%$). The second most important contribution ($\lesssim 35\%$) comes from (1). The uncertainties associated to (6) and (7) together occupy the third place ($\lesssim 31\%$), but sometimes they can be as important as (1). For light primaries (5) can become the next influent source ($\lesssim 12\%$). The rest contributions, (2)-(4), are always the smaller ones (added they are $\lesssim 12\%$). Both the energy and flux total uncertainties change with the value of $E$.

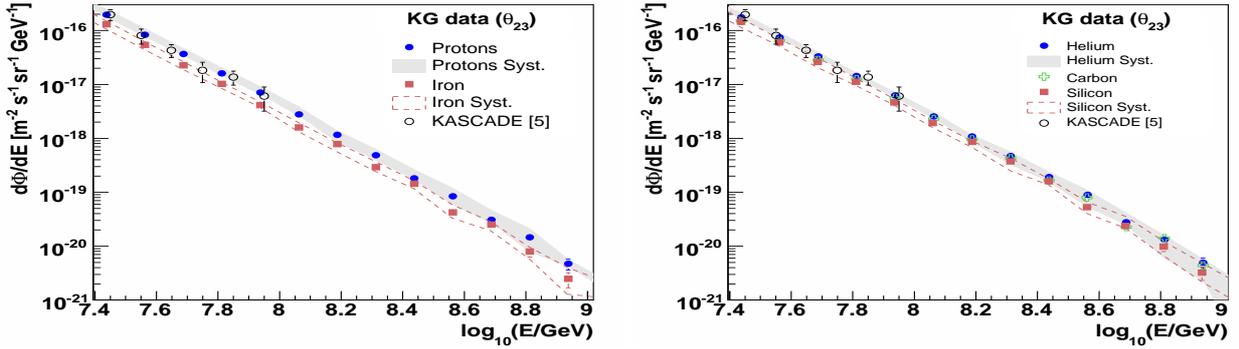

Fig. 7: The all-particle energy spectrum derived from the KASCADE-Grande muon data assuming different primary compositions. The bands for the total uncertainties (energy resolution plus systematic errors) are displayed (except for Carbon). Vertical error bars represent statistical uncertainties.

TABLE I: Percentual contributions to the total uncertainty (energy resolution and systematics) of the energy spectrum around $10^{17}$ eV. Sources are enumerated as described in the text. At the same energy scale, the average energy systematic uncertainty and energy resolution are also shown. The muon attenuation length is presented in the last column.

| Composition | $\Phi \pm$ tot. $\pm$ stat. $10^{-18}$[m$^{-2}$ s$^{-1}$ sr$^{-1}$ GeV$^{-1}$] | (1) [%] | (2) [%] | (3) [%] | (4) [%] | (5) [%] | (6) [%] | (7) [%] | $\Delta$E/E [%] $\pm$syst. $\pm$ res. | $\Lambda_\mu$ [g/cm$^2$] |
|---|---|---|---|---|---|---|---|---|---|---|
| H | $2.75^{+1.09}_{-0.21} \pm 0.07$ | +35 | $^{+2}_{-1}$ | $\pm 1$ | $^{+3}_{-1}$ | +11 | $^{+13}_{-7}$ | $^{+8}_{-0}$ | $^{+9}_{-8} \pm 29$ | $1136 \pm 115$ |
| He | $2.52^{+0.53}_{-0.20} \pm 0.07$ | +19 | $^{+1}_{-0.4}$ | $^{+1}_{-0.4}$ | $^{+2}_{-1}$ | $-3$ | $^{+6}_{-5}$ | $\pm 5$ | $^{+6}_{-12} \pm 26$ | $1111 \pm 112$ |
| C | $2.29^{+0.23}_{-0.229} \pm 0.06$ | +9 | $^{+0.2}_{-2}$ | $^{+0.2}_{-1}$ | $^{+1}_{-2}$ | $-2$ | $^{+2}_{-6}$ | $^{+3}_{-10}$ | $^{+6}_{-13} \pm 19$ | $1137 \pm 121$ |
| Si | $1.88^{+0.20}_{-0.35} \pm 0.06$ | +6 | $^{+2}_{-3}$ | $^{+0.4}_{-0.1}$ | $\pm 2$ | $-3$ | $^{+4}_{-1}$ | $^{+7}_{-18}$ | $^{+5}_{-15} \pm 20$ | $1056 \pm 146$ |
| Fe | $1.58^{+0.36}_{-0.26} \pm 0.05$ | +22 | $^{+0.5}_{-0.1}$ | $^{+0}_{-0.1}$ | $^{+1}_{-4}$ | +2 | $^{+2}_{-0.1}$ | $^{+1}_{-16}$ | $^{+4}_{-13} \pm 18$ | $1123 \pm 182$ |

They tend to increase near the energy threshold and in the high-energy region, where statistics decreases. These first estimations are very encouraging. They show that muons can be used in KASCADE-Grande as a tool to reconstruct the primary all-particle energy spectrum. More work is to come in order to improve the reconstruction method. Plans are also underway to investigate the muon attenuation length, $\Lambda_\mu$. Some values extracted from the $N_\mu$ attenuation curves are shown in Table I for $E \approx 10^{17}$ eV. A good agreement is seen among the experimental attenuation lengths under the assumption of different primary masses at this energy.

## V. ACKNOWLEDGMENTS

J.C. Arteaga acknowledges the partial support of CONACyT and the Coordinación de la Investigación Científica de la Universidad Michoacana.

# The all particle energy spectrum of KASCADE-Grande in the energy region $10^{16}$ - $10^{18}$ eV by means of the $N_{ch}$ - $N_\mu$ technique

M. Bertaina‡, W.D. Apel*, J.C. Arteaga†,xi, F. Badea*, K. Bekk*, J. Blümer*,†, H. Bozdog*, I.M. Brancus§, M. Brüggemann¶, P. Buchholz¶, E. Cantoni‡,‖, A. Chiavassa‡, F. Cossavella†, K. Daumiller*, V. de Souza†,xii, F. Di Pierro‡, P. Doll*, R. Engel*, J. Engler*, M. Finger*, D. Fuhrmann**, P.L. Ghia‖, H.J. Gils*, R. Glasstetter**, C. Grupen¶, A. Haungs*, D. Heck*, J.R. Hörandel†,xiii, T. Huege*, P.G. Isar*, K.-H. Kampert**, D. Kang†, D. Kickelbick¶, H.O. Klages*, P. Łuczak††, H.J. Mathes*, H.J. Mayer*, J. Milke*, B. Mitrica§, C. Morello‖, G. Navarra‡, S. Nehls*, J. Oehlschläger*, S. Ostapchenko*,xiv, S. Over¶, M. Petcu§, T. Pierog*, H. Rebel*, M. Roth*, H. Schieler*, F. Schröder*, O. Sima‡‡, M. Stümpert†, G. Toma§, G.C. Trinchero‖, H. Ulrich*, A. Weindl*, J. Wochele*, M. Wommer*, J. Zabierowski††

*Institut für Kernphysik, Forschungszentrum Karlsruhe, 76021 Karlsruhe, Germany
†Institut für Experimentelle Kernphysik, Universität Karlsruhe, 76021 Karlsruhe, Germany
‡Dipartimento di Fisica Generale dell'Università, 10125 Torino, Italy
§National Institute of Physics and Nuclear Engineering, 7690 Bucharest, Romania
¶Fachbereich Physik, Universität Siegen, 57068 Siegen, Germany
‖Istituto di Fisica dello Spazio Interplanetario, INAF, 10133 Torino, Italy
**Fachbereich Physik, Universität Wuppertal, 42097 Wuppertal, Germany
††Soltan Institute for Nuclear Studies, 90950 Lodz, Poland
‡‡Department of Physics, University of Bucharest, 76900 Bucharest, Romania
xi now at: Universidad Michoacana, Morelia, Mexico
xii now at: Universidade de São Paulo, Instituto de Fîsica de São Carlos, Brasil
xiii now at: Dept. of Astrophysics, Radboud University Nijmegen, The Netherlands
xiv now at: University of Trondheim, Norway

*Abstract*. The KASCADE-Grande experiment, located at Forschungszentrum Karlsruhe (Germany) is a multi-component extensive air-shower experiment devoted to the study of cosmic rays and their interactions at primary energies $10^{14}$ - $10^{18}$ eV. One of the main goals of the experiment is the measurement of the all particle energy spectrum in the $10^{16}$ - $10^{18}$ eV region. For this analysis the Grande detector samples the charged component of the air shower while the KASCADE array provides a measurement of the muon component. An independent fit of the lateral distributions of charged particle and muon densities allows to extract the charged particle and muon sizes of the shower. The size of the charged particles, combined with the ratio between charged particle and muon sizes, which is used to take into account shower-to-shower fluctuations, is used to assign the energy on an event-by-event basis, in the framework of the CORSIKA-QGSjetII model. The method itself, and the energy spectrum derived with this technique are presented.

*Keywords*: Energy spectrum, KASCADE-Grande, $10^{16}$ - $10^{18}$ eV

## I. INTRODUCTION

The KASCADE-Grande experiment [1] is a multi-component air-shower experiment with the aim of measuring the all particle energy spectrum in the $10^{16}$ - $10^{18}$ eV region by sampling the charged particle and muon densities. A fit to the lateral distribution of the charged particle densities allows to reconstruct the shower parameters (core position, angular direction) and the size of the charged component (see [2] for details). An independent fit of the lateral distribution of the muon densities (see [3],[4]), gives the size of the muon component of the shower. The performance of the KASCADE-Grande array, and, therefore, its high accuracy up to energies $10^{17}$ -$10^{18}$ eV, essential to derive an accurate energy spectrum, is summarized in [2].

The conversion between the observed quantities (charged particle and muon sizes) of the Extensive Air Shower (EAS) to the energy of the primary particle requires the assumption of a specific hadronic interaction model, whose suitability has to be verified beforehand. In this work, the energy estimations are based on the CORSIKA-QGSjetII model [5], [6], motivated by the fact that such model reproduces fairly well the distributions of the ratio of the muon and electron sizes measured by KASCADE-Grande as a function of both the electron size and the atmospheric depth [7].

The method described in this paper uses the combined information of the charged particle and muon sizes on an event-by-event basis, with the aim of reducing the systematics on the primary composition in the energy assignment, systematics which are the main sources of uncertainty on methods based on a single component information ([4], [8]).

The analysis presented here is based on ∼981 days of data collected on the central area of KASCADE-Grande array (∼0.2 km$^2$) at zenith angles $\theta < 40°$ corresponding to a total acceptance $A = 2.50 \cdot 10^9$ cm$^2$· sr (exposure $E = 2.12 \cdot 10^{17}$ cm$^2$· s·sr).

## II. TECHNIQUE

The technique has been defined on simulated data assuming a power law with index $\gamma$=-3 for the energy spectrum and then applied to the experimental ones. Proton and iron nuclei have been selected as primaries, to represent the two extreme cases. The simulation includes the full air shower development in atmosphere, the response of the detector and its electronics, as well as their uncertainties. Therefore, the reconstructed parameters from simulated showers are obtained exactly in the same way as for real data. Data have been subdivided in 5 angular bins of same acceptance ($\theta <$ 16.7, 16.7 $\leq \theta <$ 24.0, 24.0 $\leq \theta <$ 29.9, 29.9 $\leq \theta <$ 35.1, 35.1 $\leq \theta <$ 40.0) and the analysis is conducted independently in each angular bin. The difference in the results obtained among the angular bins will be considered as one of the sources in the final systematic uncertainty on the energy spectrum. In this way, possible differences in the air shower attenuation in atmosphere (i.e. the zenith angle) between real and simulated data, will be included directly into the systematic uncertainties of the measurement, without applying any correction. The energy assignment is defined as $E = f(N_{ch}, k)$ (see eq. 1), where $N_{ch}$ is the size of the charged particle component and the parameter $k$ is defined through the ratio of the sizes of the $N_{ch}$ and muon ($N_\mu$) components: $k = g(N_{ch}, N_\mu)$ (see eq. 2). The main aim of the $k$ variable is to take into account the average differences in the $N_{ch}/N_\mu$ ratio among different primaries with same $N_{ch}$, and the shower to shower fluctuations for events of the same primary mass:

$$log_{10}(E[GeV]) = [a_p + (a_{Fe} - a_p) \cdot k] \cdot log_{10}(N_{ch}) + \\ + b_p + (b_{Fe} - b_p) \cdot k \quad (1)$$

$$k = \frac{log_{10}(N_{ch}/N_\mu) - log_{10}(N_{ch}/N_\mu)_p}{log_{10}(N_{ch}/N_\mu)_{Fe} - log_{10}(N_{ch}/N_\mu)_p} \quad (2)$$

where,

$$log_{10}(N_{ch}/N_\mu)_{p,Fe} = c_{p,Fe} \cdot log_{10}(N_{ch}) + d_{p,Fe}. \quad (3)$$

The coefficients $a, b, c, d$ are obtained through the fits to the scatter plots $(N_{ch}, N_{ch}/N_\mu)$ and $(N_{ch}, E)$ in the region $6 < log_{10}(N_{ch}) < 8$, which means above the ∼100% trigger efficiency, and up to the energy for which the simulated statistics is sufficiently high. The $k$ parameter is, by definition of eq. 2, a number centered around 0 for a typical proton shower and 1 for a typical iron shower. As an example, figs. 1 and 2 show such scatter plots for the iron component in the 1$^{st}$ angular bin. Similar plots are obtained in the other 4 angular

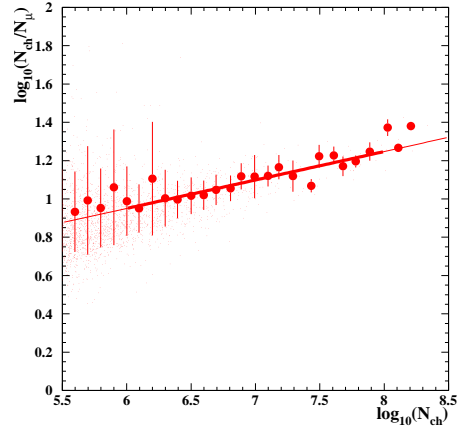

Fig. 1: Scatter plot of $N_{ch}/N_\mu$ vs $N_{ch}$ for primary iron nuclei. The small dots indicate single events, while full ones refer to the average values in each $N_{ch}$ interval ($\Delta N_{ch}$ = 0.1). The error bar of the full dots indicates the RMS of the distribution of the small dots in each $N_{ch}$ interval. The linear fit is performed on the full dots and their uncertainties in the region $6 < log_{10}(N_{ch}) < 8$ (thick line). Such fit is used to obtain the parameters $c$ and $d$ of expression 3.

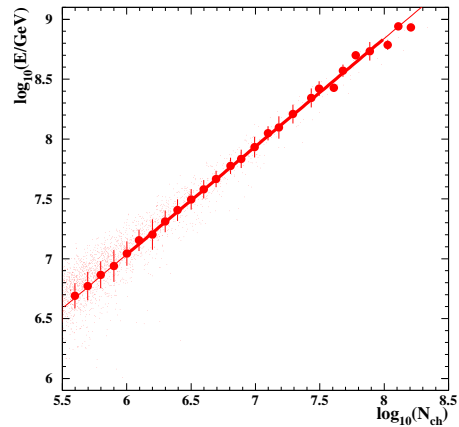

Fig. 2: Scatter plot of $E$ vs $N_{ch}$ for iron primary nuclei. See fig. 1 for detailed explanation of the meaning of dots and error bars. The fit is used to obtain the parameters $a$ and $b$ of expression 1.

bins as well as for proton primaries.

In order to check the capability of this technique of correctly reproducing the original energy spectrum, the expressions 1 and 2 have been applied to: a) the simulated energy spectra they have been derived from (H and Fe); b) to other three mass groups (He, C, Si) simulated using the same criteria; c) to the mixture of the five mass groups with 20% abundance each. Fig. 3 shows a comparison between the reconstructed and true energy spectra obtained for the 1$^{st}$ angular bin in case of iron primary nuclei. Similar plots are obtained for the other mass groups and for all angular bins. Fig. 4 summarizes the results on the ratio between reconstructed and true spectra for protons, iron and all mixed primaries. The original energy spectra are fairly well reproduced. The systematic uncertainties are discussed in section III.

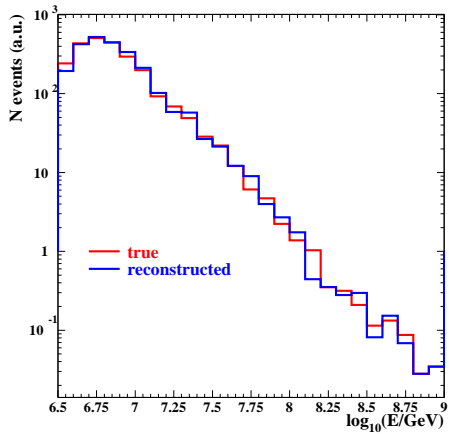

Fig. 3: True (red line), and reconstructed (blue line) energy spectrum in the 1$^{st}$ angular bin for iron primary nuclei according to expressions 1 and 2.

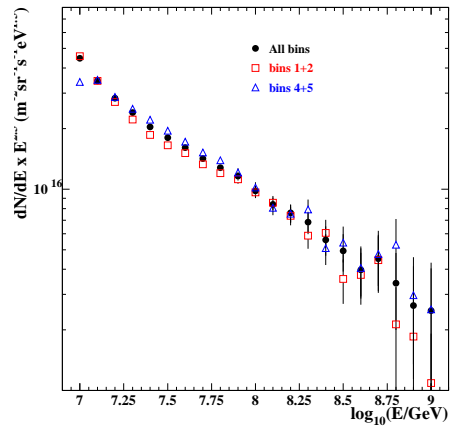

Fig. 5: The experimental energy spectrum (differential intensity multiplied by E$^{2.5}$) as a function of $\log_{10}(E/GeV)$ for vertical (bins 1+2), more inclined (bins 4+5) and all events (only statistical uncertainties).

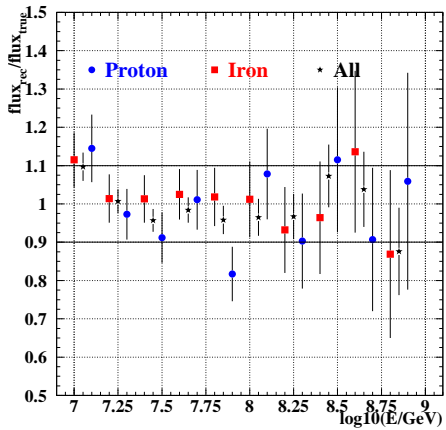

Fig. 4: Ratio between the reconstructed and true spectra (as shown in fig. 3) for protons, iron and all mixed primaries (all angular bins together).

## III. THE RECONSTRUCTED ENERGY SPECTRUM AND ITS UNCERTAINTIES

Expressions 1 and 2 have been applied to the experimental data obtaining the intensities shown in fig. 5. A detailed analysis of the systematic uncertainties on the intensities has been conducted taking into account the following effects:

a) Systematic uncertainty from the comparison of the intensity in different angular bins.
b) Systematic uncertainty on the $E(N_{ch})$ relation.
c) Systematic uncertainty related to the capability of reproducing an, *a priori* assumed, single primary spectrum with slope $\gamma$ = -3 (i.e. difference between true and reconstructed spectra of fig. 3).
d) Systematic uncertainty on the muon lateral distribution function (l.d.f.).

Possible systematic uncertainties on the reconstructed $N_{ch}$ and $N_\mu$ values compared to the true ones, are already taken into account by the technique itself as the same reconstruction procedure is applied to simulated and real data.

Concerning a), in fig. 5 data points of the 1$^{st}$ and 2$^{nd}$ angular bins have been summed together, and the same for data of the 4$^{th}$ and 5$^{th}$ bins. The semi-difference of the intensity in each energy interval (subtracted from the statistical uncertainty) between vertical and more inclined angular bins provides an estimation of the uncertainty on the relative energy calibration among the angular bins, together with the systematic uncertainty related to possible differences in the air shower attenuation in the atmosphere between real and simulated data. At E $\sim 10^{17}$ ($\sim 3 \cdot 10^{16}$) eV (at higher energies the results are dominated by the statistical uncertainty) the systematic uncertainty is $\sim$5% ($\sim$15%). This result confirms the fact that the technique is self-consistent in the entire angular range used in this analysis and that the QGSjetII model reproduces quite consistently the shower development at least up to zenith angles $\theta$ < 40 degrees. The uncertainty on the intensity provides only an indication on the relative uncertainty among expressions 1 and 2 for different angular bins, but doesn't take into account a common systematic effect of all $E = f(N_{ch}, k)$. For this reason, $E = f(N_{ch}, k)$ in simulated data have been artificially modified, at a level in which the systematic effect is clearly visible between the true and reconstructed simulated energy spectra as in fig. 3 and an upper limit has been set and used as systematic effect on the $E(N_{ch})$ conversion relation: at E $\sim 10^{17}$ eV such uncertainty is < 10%.

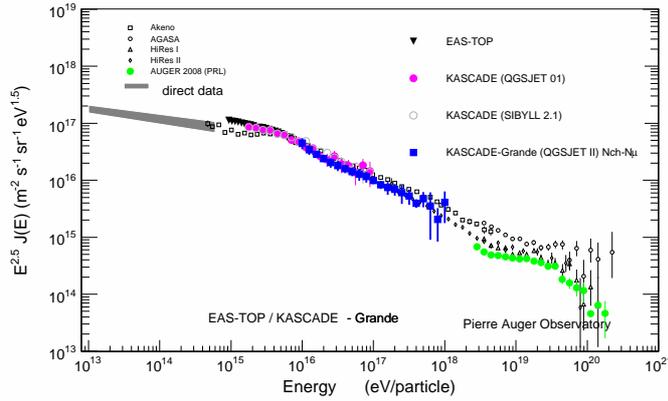

Fig. 6: All primaries energy spectrum (intensity multiplied by $E^{2.5}$) obtained with KASCADE-Grande data applying the $N_{ch}$ - $N_\mu$ technique. A comparison with other experimental results is also presented.

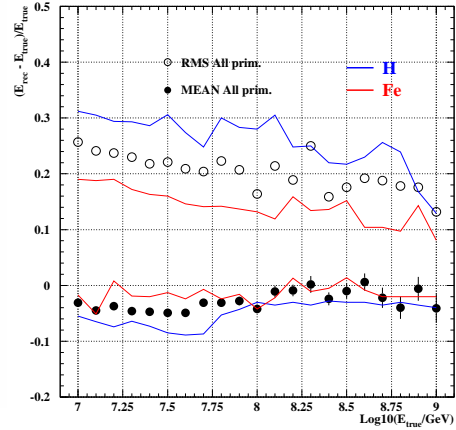

Fig. 7: Resolution in the energy assignment for a mixture of primaries of the 5 simulated mass groups (relative abundance of each group 20%), for H and Fe. The full dots show the offset of the reconstructed energy $E_{rec}$ in bins of true energy $E_{true}$. The open dots show the RMS of such distributions (see text for details).

A further systematic uncertainty comes from the capability of reproducing the original energy spectrum assuming a single mass composition as shown in fig. 4. In general the ratio between the reconstructed and true fluxes obtained in each energy bin are compatible, inside the statistical uncertainty, with unity. Only at the threshold a systematic effect of ∼12% is visible. The typical relative differences in flux at energies E<$10^{17}$ eV are <5%.

Regarding d), the energy spectrum has been obtained for different ranges of distance of the shower core from the muon detector. The relative difference in intensity as a function of energy is used to compute a systematic uncertainty due to the assumed l.d.f., and it amounts to ∼3% at E ∼ $10^{17}$, slightly increasing with energy.

Finally, it is interesting to look at the relative uncertainty in the energy assignment on an event-by-event basis. Simulated data using the mixture of all primaries have been divided in bins of true energy ($E_{true}$) and the distributions of the relative differences between reconstructed ($E_{rec}$) and true energies have been created. As shown in fig. 7 the RMS of such distributions (energy resolution) is ∼26% at the energy threshold and decreases with energy, due to the lower fluctuations of the shower development, becoming < 20% at the highest energies. The small offset in the mean values of the distributions at low energies is necessary to take into account the effect of shower fluctuations on a steep spectrum. Such offset does not appear in fig. 4, which indicates that the correct energy spectrum is well reproduced. Results for pure H and Fe primaries are also indicated by lines.

The statistical uncertainty on the intensity is <10% up to E∼3 · $10^{17}$ eV. The total uncertainty (statistical and systematic squared together) on the intensity is <20% at energies E < $10^{17}$ eV in the frame of the CORSIKA-QGSjet model.

## IV. RESULTS

The all particle energy spectrum of KASCADE-Grande in the $10^{16}$ - $10^{18}$ eV energy region using the $N_{ch}$ - $N_\mu$ technique is shown in fig. 6. The uncertainty in each intensity point is obtained as squared sum of all systematic and statistical uncertainties. At the threshold, the spectrum overlaps well with KASCADE and EAS-TOP spectra. Moreover, it is in agreement with the energy spectra of KASCADE-Grande obtained using other techniques ([4], [8], [9]) which have partially different systematic uncertainties (see [10]).

The mean of the average $\overline{k}$ obtained in different bins of $N_{ch}$ in the range $6 < log_{10}(N_{ch}) < 8$ is $<\overline{k}> = 0.64$, with bin to bin fluctuations of $\sigma_{\overline{k}} \sim 0.06$, therefore, perfectly compatible with the limits set by eq. 2 using QGSjetII simulations.

# Primary energy reconstruction from the S(500) observable recorded with the KASCADE-Grande detector array

G. Toma§, W.D. Apel*, J.C. Arteaga†,xi, F. Badea*, K. Bekk*, M. Bertaina‡, J. Blümer*,†,
H. Bozdog*, I.M. Brancus§, M. Brüggemann¶, P. Buchholz¶, E. Cantoni‡,‖, A. Chiavassa‡,
F. Cossavella†, K. Daumiller*, V. de Souza†,xii, F. Di Pierro‡, P. Doll*, R. Engel*, J. Engler*,
M. Finger*, D. Fuhrmann**, P.L. Ghia‖, H.J. Gils*, R. Glasstetter**, C. Grupen¶,
A. Haungs*, D. Heck*, J.R. Hörandel†,xiii, T. Huege*, P.G. Isar*, K.-H. Kampert**,
D. Kang†, D. Kickelbick¶, H.O. Klages*, P. Łuczak††, H.J. Mathes*,
H.J. Mayer*, J. Milke*, B. Mitrica§, C. Morello‖, G. Navarra‡, S. Nehls*,
J. Oehlschläger*, S. Ostapchenko*,xiv, S. Over¶, M. Petcu§, T. Pierog*, H. Rebel*,
M. Roth*, H. Schieler*, F. Schröder*, O. Sima‡‡, M. Stümpert†, G.C. Trinchero‖,
H. Ulrich*, A. Weindl*, J. Wochele*, M. Wommer*, J. Zabierowski††

*Institut für Kernphysik, Forschungszentrum Karlsruhe, 76021 Karlsruhe, Germany
†Institut für Experimentelle Kernphysik, Universität Karlsruhe, 76021 Karlsruhe, Germany
‡Dipartimento di Fisica Generale dell'Università, 10125 Torino, Italy
§National Institute of Physics and Nuclear Engineering, 7690 Bucharest, Romania
¶Fachbereich Physik, Universität Siegen, 57068 Siegen, Germany
‖Istituto di Fisica dello Spazio Interplanetario, INAF, 10133 Torino, Italy
**Fachbereich Physik, Universität Wuppertal, 42097 Wuppertal, Germany
††Soltan Institute for Nuclear Studies, 90950 Lodz, Poland
‡‡Department of Physics, University of Bucharest, 76900 Bucharest, Romania
xi now at: Universidad Michoacana, Morelia, Mexico
xii now at: Universidade de São Paulo, Instituto de Fîsica de São Carlos, Brasil
xiii now at: Dept. of Astrophysics, Radboud University Nijmegen, The Netherlands
xiv now at: University of Trondheim, Norway

*Abstract*. Previous EAS investigations have shown that the charged particle density becomes independent of the primary mass at large but fixed distances from the shower core and that it can be used as an estimator for the primary energy. The particular radial distance from the shower axis where this effect takes place is dependent on the detector layout. For the KASCADE-Grande experiment it was shown to be around 500 m. A notation $S(500)$ is used for the charged particle density at this specific distance. Extensive simulation studies have shown that $S(500)$ is mapping the primary energy. We present results on the reconstruction of the primary energy spectrum of cosmic rays from the experimentally recorded $S(500)$ observable using the KASCADE-Grande array. The constant intensity cut (CIC) method is applied to evaluate the attenuation of the $S(500)$ observable with the zenith angle. A correction is subsequently applied to correct all recorded $S(500)$ values for attenuation. The all event $S(500)$ spectrum is obtained. A calibration of $S(500)$ values with the primary energy has been worked out by simulations and has been used for conversion thus providing the possibility to obtain the primary energy spectrum (in the energy range accessible to the KASCADE-Grande array, $10^{16}$-$10^{18}$ eV). An evaluation of systematic uncertainties induced by different factors is also given.

*Keywords*: KASCADE-Grande, EAS, primary energy spectrum

## I. INTRODUCTION

Hillas has shown that the EAS particle density distributions at a certain distance from the shower core (dependent on the EAS detection array) becomes independent of the primary mass and can be used as a primary energy estimator [1]. Following this feature, a method can be derived to reconstruct the primary energy spectrum from the particular value of the charged particle density, observed at such specific radial ranges. The technique has been used by different detector arrays in order to reconstruct the primary energy spectrum of the cosmic radiation [2]. In the case of the KASCADE-Grande array (at Forschungszentrum Karlsruhe, Germany, 110 m a.s.l.) [3], detailed simulations [4] have shown that the particular distance for which this effect takes place is about 500 m (see fig. 1). Therefore an observable of interest in the case of KASCADE-Grande is the charged particle density at 500 m distance from the shower core, noted as $S(500)$ in the following. The study has been performed for both simulated (fig. 1) and experimental (fig. 2) events, using identical reconstruction procedures [5]. The reconstruction begins with recording the energy deposits of particles in the KASCADE-Grande detector stations and the associated temporal information (arrival times of particles). The arrival direction of the shower is reconstructed from the particle arrival times. Using appropriate **L**ateral **E**nergy **C**orrection **F**unctions (**LECF**), the energy deposits are converted into particle densities. The LECF functions are dependent on the shower zenith angle [6] and take into account the fact that an inclined particle will deposit more energy in detectors due to its longer cross path. For every event, the obtained lateral density distribution is approximated by a Linsley [7] **L**ateral **D**ensity **F**unction (**LDF**) in order to evaluate the particle density at the radial range of interest, 500 m. To ensure good reconstruction quality, the approximation is performed over a limited range of the lateral extension, namely only in the 40 m − 1000 m radial range.

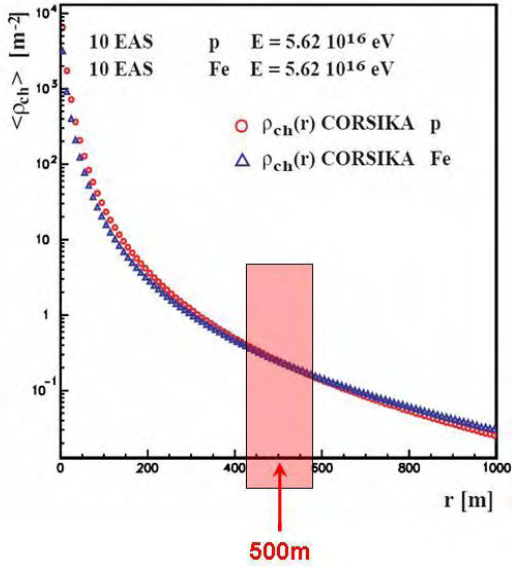

Fig. 1: Simulations show that, for the case of the KASCADE-Grande experimental layout, the particle density becomes independent of the primary mass around 500 m distance from shower core; this plot shows averaged simulated lateral distributions for different primary types with equal energy.

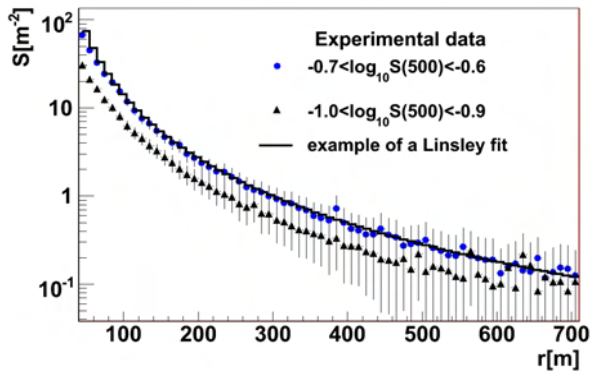

Fig. 2: Averaged lateral density distributions of experimentally recorded EAS samples for two $S(500)$ ranges.

## II. EFFICIENCY AND QUALITY CUTS

For the experimental EAS sample, the total time of acquisition was ≈902 days. Showers were detected on a 500 x 600 m² area up to 30° zenith angle. The 30° zenith angle limit was imposed due to certain systematic effects affecting the reconstruction of small showers above this threshold. In order to ensure good reconstruction quality, several quality cuts were imposed on the data. The same cuts were used for both simulated and experimental events. Only those events are accepted for which the reconstructed shower core is positioned inside the detector array and not too close to the border. A good quality of the fit to the Linsley distribution is a further important criterion. Fig. 3 shows the total reconstruction efficiency for different zenith angle intervals (the full efficiency is reached at around $\log_{10}(E_0/\text{GeV})$=7.5).

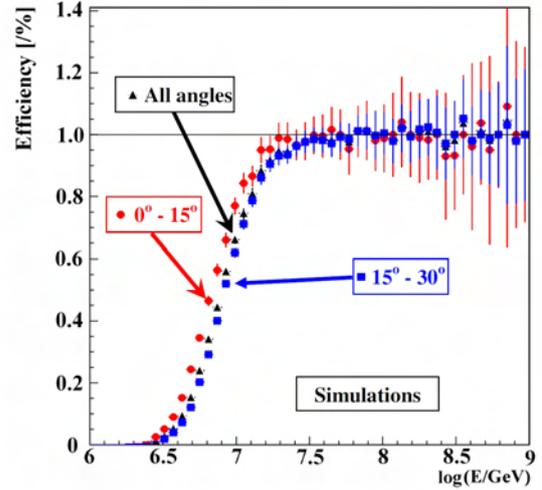

Fig. 3: Reconstruction efficiency for different zenith angle ranges and for the entire shower sample (events triggering more than 24 stations).

## III. THE CONSTANT INTENSITY CUT METHOD

Before converting the recorded $S(500)$ values into the corresponding primary energy values (via a relation derived from simulation studies), one has to take into account the atmospheric attenuation affecting the charged particle densities observed on ground. For more inclined showers, the particles have to cross a longer path through the atmosphere before reaching the detector level. In such a case, events generated by identical primaries reach the detector level at different stages of EAS development, dependent on their angles of incidence. In order to bring all recorded EAS events to the same level of consistency, one has to eliminate the influence of the zenith angle on the recorded $S(500)$ observables. This is achieved by applying the **C**onstant **I**ntensity **C**ut (**CIC**) method. The $S(500)$ attenuation is visible if $S(500)$ spectra are plotted for different EAS incident angles. For this, the recorded events are separated into several sub-samples characterized by their angle of incidence. The angular intervals are chosen in a way that they open equal solid angles: 0° - 13.2°, 13.2° - 18.8°, 18.8° - 23.1°, 23.1° - 26.7° and 26.7° - 30.0°. In fig. 4 the attenuation is visible, as $S(500)$ spectra are shifted towards lower values for increasing zenith angles. The CIC method assumes that a given intensity value in the energy spectrum corresponds to a given primary energy of particles and, since the $S(500)$ is mapping the primary energy spectrum, it is expected that this property of the intensity is true also in the case of $S(500)$ spectra. Therefore a constant intensity cut on integral $S(500)$ spectra is performed, effectively cutting them at a given primary energy. The intersection of the cut line with each spectrum will give the attenuated $S(500)$ value at the corresponding angle of incidence for a given primary energy. A linear interpolation is used between the two neighboring points in the integral spectrum in order to convert the value of the intensity into particle density for each angular bin. The observed attenuation can be corrected by parameterizing the attenuation curve and correcting all events by bringing their $S(500)$ value to their corresponding value at a given reference angle of incidence (see fig. 5; the parameterization with the lowest $\chi^2$ was chosen, namely the one corresponding to intensity 3000). For the present study this angle is considered to be 21°, since the zenith angular distribution for the recorded EAS sample peaks at this value. The CIC method implies several mathematical transformations of data before obtaining the values corrected for attenuation of the $S(500)$ observable: interpolations and analytical parameterizations (as mentioned in the above description of the CIC method). These operations introduce some systematic uncertainties on the final result of the CIC method. The CIC-induced systematic uncertainty of the corrected $S(500)$ value is evaluated by propagating the errors of fit parameters. The resulting CIC-induced error of the $S(500)$ observable will be taken into account later when evaluating the total systematic uncertainty of the reconstructed primary energy.

## IV. CONVERSION TO ENERGY

After correcting the recorded $S(500)$ values for attenuation, we can proceed to convert each of them to the corresponding primary energy value. A calibration of the primary energy $E_0$ with $S(500)$ was derived from simulations (see fig. 6). The Monte-Carlo CORSIKA EAS simulation tool was used to simulate air showers (with QGSJET II model embedded for high energy interactions). In fig. 6, two slightly different dependencies are shown for two primaries, a light primary (proton) and a heavy primary (Fe). The two dependencies are almost identical, a feature that is expected due to the mass insensitivity of the $S(500)$ observable. This calibration is used to convert all $S(500)$ values into the corresponding primary energies. The spectrum of primary energy is thus reconstructed. Fig. 7 shows the reconstructed energy spectrum compared with spectra reconstructed by other experiments. The spectrum is plotted starting from the maximum efficiency threshold (see fig. 3). For the systematic contribution to the total error, several sources of systematic uncertainties have been identified and their contributions were evaluated. Thus, the spectral index of the simulated shower sample was equal to -2 and was acting as a source of systematic uncertainty. In a similar fashion, the $S(500)$-$E_0$ calibration and the CIC method itself were also introducing systematic uncertainties. In all, these three sources were contributing with an uncertainty of $\approx 1\%$ from the total flux value. Other sources that were considered were the Monte-Carlo statistical uncertainty of the simulated shower sample and the choosing of a certain reference angle at which to perform the $S(500)$ attenuation correction (contributing with $\approx 7\%$ and $\approx 30\%$ relative uncertainty). The relative contribution of all identified sources over the full efficiency range was fairly constant for any given source and in total amounts for about 37% of the recorded flux value. The energy resolution has also been evaluated from simulations by calculating the difference between the true and the reconstructed primary energy (applying CIC to the simulated data). The energy resolution was found to be 22% for $E_0 = 10^{17}$ eV (for all primaries) and is fairly constant over the entire full efficiency range.

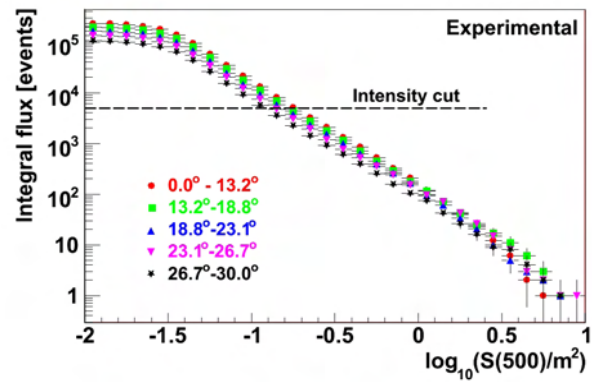

Fig. 4: Integral $S(500)$ spectra; the horizontal line is a constant intensity cut at an arbitrarily chosen intensity; attenuation length of $S(500)$ was evaluated at 347.38±21.65 g·cm$^{-2}$

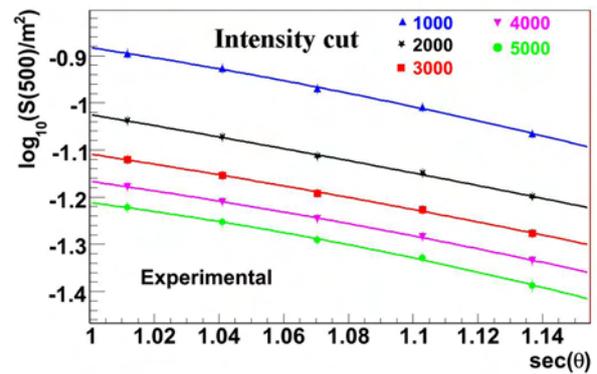

Fig. 5: Attenuation of the $S(500)$ observable with the angle of incidence; the different curves show different arbitrarily chosen intensity cuts.

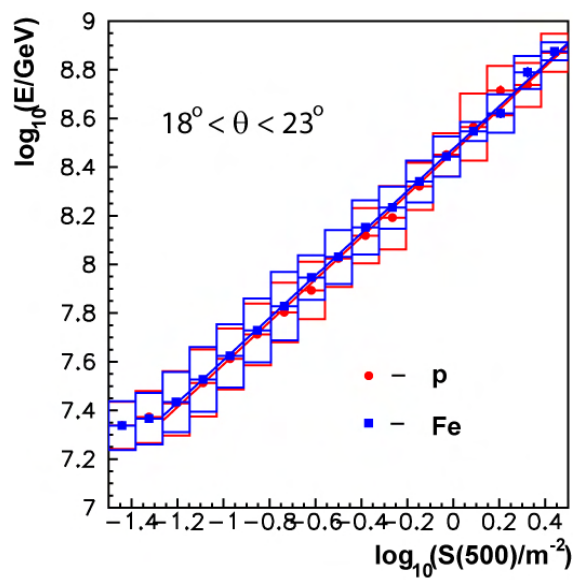

Fig. 6: $E_0$ - $S(500)$ calibration curve for two different primaries; the box-errors are the errors on the spread; the errors on the mean are represented with bars.

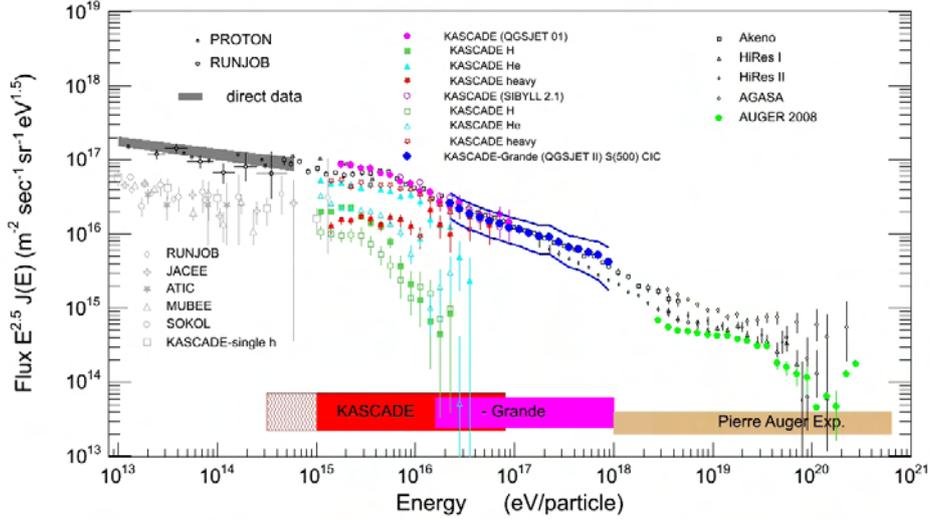

Fig. 7: Reconstructed experimental energy spectrum by KASCADE-Grande from $S(500)$/CIC, multiplied by $E^{2.5}$ compared with results of other experiments; the continuous lines above and below the spectrum are the error envelopes and show combined statistical and systematic uncertainties.

## V. CONCLUSIONS

The primary energy spectrum has been reconstructed from the particle densities recorded in the stations of the KASCADE-Grande array. In the particular case of KASCADE-Grande, the charged particle density at 500 m distance from the shower core was shown to be primary mass insensitive. The CIC method was applied on the recorded $S(500)$ spectrum in order to correct each shower for attenuation effects. Using a simulation-derived calibration between $S(500)$ and $E_0$ (based on the QGSJET II model for high energy interactions), the attenuation corrected $S(500)$ spectrum has been converted into primary energy spectrum. The $S(500)$ derived KASCADE-Grande spectrum is composition independent and comes in good agreement with the spectrum of lower energies previously reconstructed by the KASCADE array. Future investigations will concentrate also on improving the quality of the reconstruction along with gaining a better understanding of the uncertainties induced by the reconstruction technique.

ACKNOWLEDGMENT

The KASCADE-Grande experiment is supported by the BMBF of Germany, the MIUR and INAF of Italy, the the Polish Ministry of Science and Higher Education (grant 2009-2011), and the Romanian Ministry of Education and Research.

# Performance of the KASCADE-Grande array

F. Di Pierro‡, W.D. Apel∗, J.C. Arteaga†,xi, F. Badea∗, K. Bekk∗, M. Bertaina‡, J. Blümer∗,†,
H. Bozdog∗, I.M. Brancus§, M. Brüggemann¶, P. Buchholz¶, E. Cantoni‡,‖, A. Chiavassa‡,
F. Cossavella†, K. Daumiller∗, V. de Souza†,xii, P. Doll∗, R. Engel∗, J. Engler∗, M. Finger∗,
D. Fuhrmann∗∗, P.L. Ghia‖, H.J. Gils∗, R. Glasstetter∗∗, C. Grupen¶, A. Haungs∗,
D. Heck∗, J.R. Hörandel†,xiii, T. Huege∗, P.G. Isar∗, K.-H. Kampert∗∗, D. Kang†,
D. Kickelbick¶, H.O. Klages∗, P. Łuczak††, H.J. Mathes∗, H.J. Mayer∗,
J. Milke∗, B. Mitrica§, C. Morello‖, G. Navarra‡, S. Nehls∗, J. Oehlschläger∗,
S. Ostapchenko∗,xiv, S. Over¶, M. Petcu§, T. Pierog∗, H. Rebel∗, M. Roth∗,
H. Schieler∗, F. Schröder∗, O. Sima‡‡, M. Stümpert†, G. Toma§, G.C. Trinchero‖,
H. Ulrich∗, A. Weindl∗, J. Wochele∗, M. Wommer∗, J. Zabierowski††

∗*Institut für Kernphysik, Forschungszentrum Karlsruhe, 76021 Karlsruhe, Germany*
†*Institut für Experimentelle Kernphysik, Universität Karlsruhe, 76021 Karlsruhe, Germany*
‡*Dipartimento di Fisica Generale dell'Università, 10125 Torino, Italy*
§*National Institute of Physics and Nuclear Engineering, 7690 Bucharest, Romania*
¶*Fachbereich Physik, Universität Siegen, 57068 Siegen, Germany*
‖*Istituto di Fisica dello Spazio Interplanetario, INAF, 10133 Torino, Italy*
∗∗*Fachbereich Physik, Universität Wuppertal, 42097 Wuppertal, Germany*
††*Soltan Institute for Nuclear Studies, 90950 Lodz, Poland*
‡‡*Department of Physics, University of Bucharest, 76900 Bucharest, Romania*
xi *now at: Universidad Michoacana, Morelia, Mexico*
xii *now at: Universidade de São Paulo, Instituto de Fîsica de São Carlos, Brasil*
xiii *now at: Dept. of Astrophysics, Radboud University Nijmegen, The Netherlands*
xiv *now at: University of Trondheim, Norway*

*Abstract.* The KASCADE-Grande experiment consists of the basic KASCADE complex and of an extended array, Grande, made of 37x10 m$^2$ scintillation detectors spread over an area of 700 x 700 m$^2$. Grande enables triggers and reconstruction of primary cosmic rays in the energy range of $10^{16}$ to $10^{18}$ eV through the detection of the all-charged particle component of the related Extensive Air Showers. The experimental set-up allows, for a subsample of the registered showers, detailed comparisons of the data with measurements of the original KASCADE array (252 unshielded detectors, with 490 m$^2$ sensitive area and 192 shielded muon detectors with 622 m$^2$ sensitive area spread over 200 x 200 m$^2$) on an event by event basis. We discuss the Grande reconstruction procedures and accuracies. The lateral charged particle distributions are measured over a wide range of core distances, the results of the measurement are presented.

*Keywords*: KASCADE-Grande, Reconstruction, LDF

## I. Introduction

The extensive air shower experiments KASCADE and EAS-TOP have shown that the change of the slope in the energy spectrum ("the knee") for different elemental groups occurs at different energies [1], [2]. For the lightest nuclei the knee has been measured at $3 \cdot 10^{15}$ eV. These results support the general view which attributes the change of the spectral index to processes of magnetic confinement, occurring either at acceleration regions, or as diffusive leakage from the Galaxy (or both). Such processes predict that the maximum energy for a specific primary nucleus depends on its atomic number Z.

The Grande array increases the collecting area of KASCADE in order to extend the measured energy range. The aim of KASCADE-Grande [3] is to perform energy, composition and anisotropy studies up to $10^{18}$ eV, i.e. in the region where the transition from galactic to extragalactic cosmic rays is supposed to happen.

## II. Experimental setup

The KASCADE-Grande experiment is located at Forschungszentrum Karlsruhe, Germany (49.1$^o$ N, 8.4$^o$ E) at 110 m a.s.l., corresponding to an average atmospheric depth of 1023 g/cm$^2$. It consists of an extension of the KASCADE [4] experiment with Grande, an array of plastic scintillators obtained reassembling the EAS-TOP [5] electromagnetic detector, which expands the collecting area, and Piccolo a smaller array providing a fast trigger common to all components.

The KASCADE-Grande detectors and their main characteristics are listed in table I and their layout is shown in fig. 1.

The KASCADE array is composed of 252 detector stations on a square grid with 13 m spacing. It is composed of:

- e/γ-detectors which mainly consists of 2/4 liquid scintillator units (1 m diameter, 5 cm thick);

| Detector | Particle | Area m$^2$ | Threshold |
|---|---|---|---|
| Grande array (plastic scintillators) | charged | 370 | 3 MeV |
| Piccolo array (plastic scintillators) | charged | 80 | 3 MeV |
| KASCADE array (liquid scint.) | $e/\gamma$ | 490 | 5 MeV |
| KASCADE array (shielded pl. scint.) | $\mu$ | 622 | 230 MeV |
| Muon tracking det. (streamer tubes) | $\mu$ | 3×128 | 800 MeV |
| Multi wire proportional chambers | $\mu$ | 2×129 | 2.4 GeV |
| Limited streamer tubes | $\mu$ | 250 | 2.4 GeV |
| Calorimeter | h | 9×304 | 50 GeV |

TABLE I: The KASCADE-Grande detectors, their total sensitive area and threshold for vertical particle.

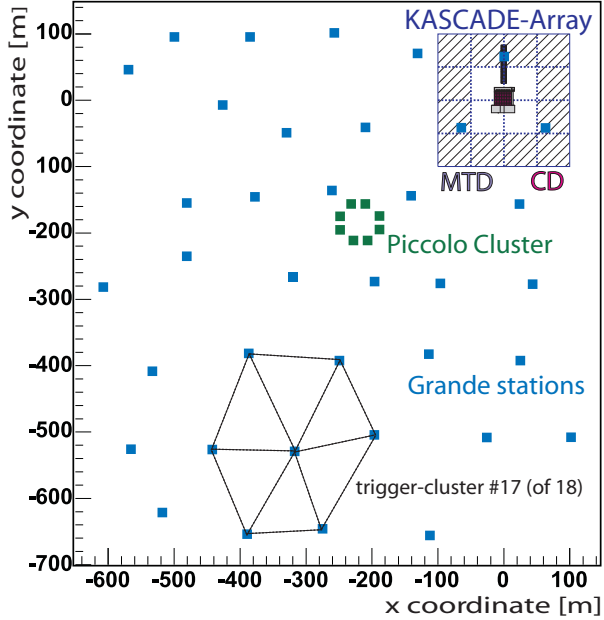

Fig. 1: Layout of the KASCADE-Grande experiment.

- $\mu$-detectors which mainly consists of 4 plastic scintillators (90 x 90 x 3 cm$^3$), placed below the e/$\gamma$-detectors and a shielding (10 cm of lead and 4 cm of iron, which entails a threshold of 230 MeV for vertical muons).

The Grande array consists of 37 stations with an average spacing of 137 m over a 700 x 700 m$^2$ area. Every detector station consists of 10 m$^2$ of plastic scintillator organized in 16 units (80 x 80 x 4 cm$^3$). Each unit is equipped with a high gain (HG) photomultiplier (pmt) and the 4 central units are additionally equipped with a low gain (LG) photomultiplier to increase the dynamic range. The signals from the pmts are added up through passive mixers, one for the HG and one for the LG pmts. The output signals are preamplified and shaped by Shaping Amplifiers into 3 analog signals, digitized by 3 Peak-ADCs, covering the dynamic ranges 0.3 ÷ 8, 2 ÷ 80, 20 ÷ 800 particles/m$^2$ respectively. The overlapping ranges between the scales are used for cross-calibration. Each detector is continuously monitored and calibrated by means of single muon spectra. The systematic uncertainty on the measured particle density by each detector is less than 15% and the statistical uncertainties are dominated by poissonian fluctuations.

The array is divided in 18 trigger clusters of 7 modules each (6 modules in an hexagon and a central one). The trigger rate is 0.5 Hz and becomes fully efficient for all primaries at $E_0 = 10^{16}$ eV.

## III. RECONSTRUCTION OF EXTENSIVE AIR SHOWERS

Core position, arrival direction and total number of charged particles are reconstructed in an iterative fit procedure of the energy deposit and timing measurements by the Grande array detectors. The energy deposit is converted to charged particle density through a function of the core distance, derived from shower and detector simulations, taking into account energy deposit of charged particles and gamma conversion [6]. The fitting functions have been derived from full shower and detector simulations. The shower front is fitted with:

$$\bar{t} = 2.43 \cdot (1 + \frac{r}{30})^{1.55} \text{ns}$$

with a time spread:

$$\sigma_{\bar{t}} = 1.43 \cdot (1 + \frac{r}{30})^{1.39} \text{ns}$$

where $r$ is the distance from shower axis in meter. The lateral distribution is fitted with a modified NKG function:

$$\rho_{ch} = C(s)N_{ch}\left(\frac{r}{30}\right)^{s-1.6}\left(1 + \frac{r}{30}\right)^{s-3.4}$$

and the normalization factor is:

$$C(s) = \frac{\Gamma(3.4-s)}{2\pi \cdot 30^2 \cdot \Gamma(s-1.6+2) \cdot \Gamma(1.6+3.4-2s-2)}$$

The total number of muons is obtained by means of a fit of the muon densities measured by the KASCADE array with a Lagutin function (core position is fixed):

$$\rho_\mu(r) = N_\mu \cdot f(r)$$

$$f(r) = \frac{0.28}{320^2}\left(\frac{r}{320}\right)^{-0.69}\left(1 + \frac{r}{320}\right)^{-2.39}\left(1 + \left(\frac{r}{10 \cdot 320}\right)^2\right)^{-1.0}$$

Reconstruction procedure and accuracy of the muonic component are described in [7].

A single event reconstruction, with the particle densities measured by Grande and by the KASCADE array muon detectors and the corresponding lateral distribution fits, is shown in figure 2.

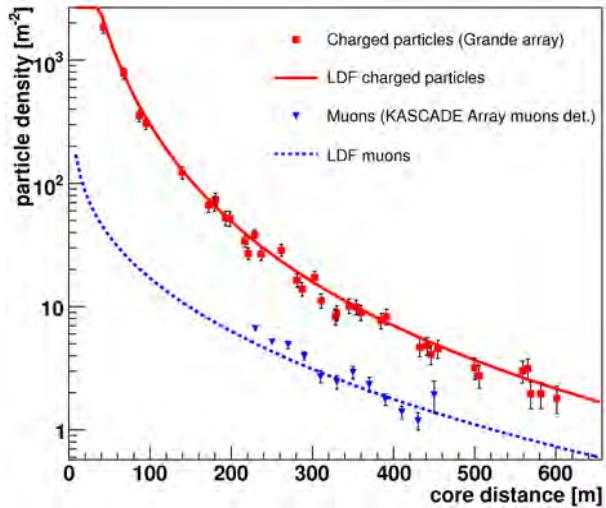

Fig. 2: Lateral distributions of a single event: muon densities measured by KASCADE array muon detector and muon ldf (black triangles represent the average densities in rings of 20 m, while the fit is performed on individual detector measurements); all-charged particles measured by Grande detectors and all-charged ldf. The parameters of the shown event are:
x$_{core}$ = -307 m, y$_{core}$ = -86 m, lg N$_{ch}$ = 7.9, lg N$_\mu$ = 6.7, $\Theta$ = 16.5$^o$, $\Phi$ = 245.5$^o$.

## IV. RECONSTRUCTION ACCURACIES

For a subsample of the events collected by the Grande array it is possible to compare on an event by event basis the two independent reconstructions of KASCADE and Grande. This provides the unique opportunity of evaluating the reconstruction accuracies of the Grande array by a direct comparison with an independent experiment instead of the usual procedure involving simulations. Since the KASCADE array is much more dense (sensitive area over fiducial area = 0.06) than Grande (sensitive area over fiducial area = 0.002), the contribution to the differences in the reconstructed observables from KASCADE is negligible and its reconstruction can be taken as reference. The subsample is obtained accordingly to the following selection criteria: maximum energy deposit in the central station of the hexagon overlapping with KASCADE, core position within a circle of 90 m radius from KASCADE center, zenith angle less than 40$^o$. The scatter plot with the shower sizes reconstructed by both arrays is shown in fig. 3. By means of such a comparison the Grande reconstruction accuracies are found to be:

- shower size (fig. 4): systematic $\leq$ 5%, statistical $\leq$ 15%;
- arrival direction (figs. 5, 6): $\sigma_\Psi \approx 0.7^o$;
- core position (fig. 7): $\sigma_{core} \approx 5$ m.

## V. MEAN LATERAL DISTRIBUTION OF CHARGED PARTICLES

In fig. 8 the experimental mean lateral distributions for vertical showers (0$^o \div$ 18$^o$) and for different shower sizes in the range 6.2 < lg N$_{ch}$ < 7.8 are shown. The lines represent the lateral distribution functions with mean N$_{ch}$ and s-parameter values of the corresponding N$_{ch}$ bin. The lateral distributions measured by the Grande array extend up to more than 700 m and the used lateral distribution function represents the data well over the whole range.

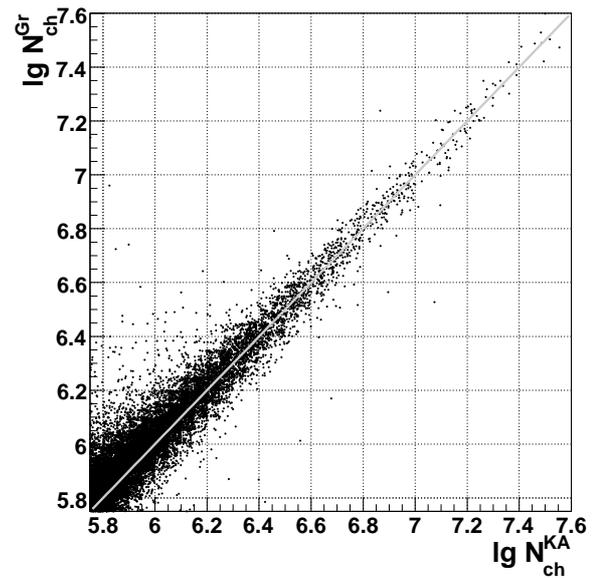

Fig. 3: Scatter plot of the shower sizes (charged particles) reconstructed by KASCADE (x-axis) and Grande (y-axis).

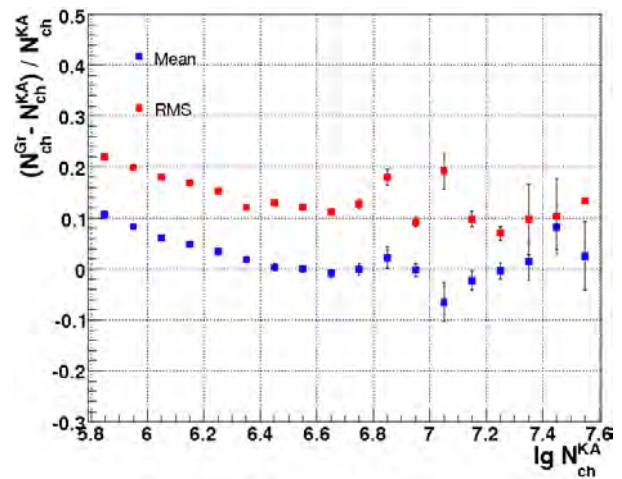

Fig. 4: Mean value and Root Mean Square of the distribution of the shower size differences over shower size reconstructed by KASCADE (N$_{ch}^{KA}$) as function of N$_{ch}^{KA}$.

In fig. 9 the lateral distributions are shown in a region closer to the core and with the core position reconstructed independently by KASCADE, showing that the good description of measured particle densities by the functions is not just a consequence of the fit procedure.

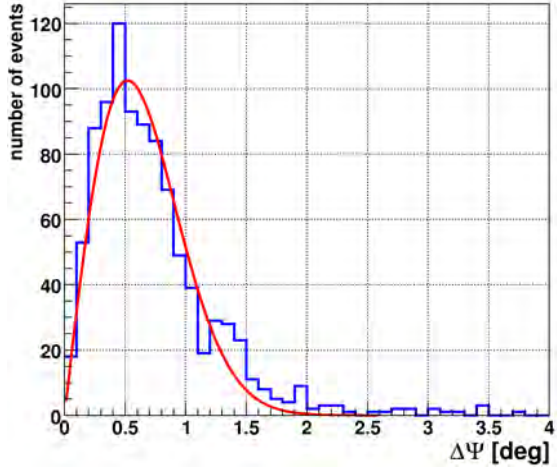

Fig. 5: Angle ($\Psi$) between the arrival directions reconstructed by KASCADE and Grande, in a bin of $N_{ch}^{KA}$ (6.2 < lg $N_{ch}^{KA}$ < 6.3).

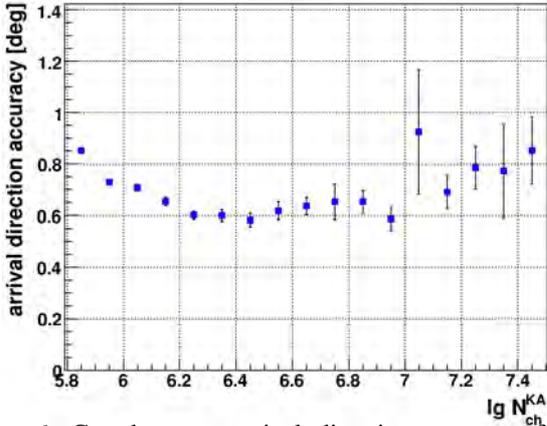

Fig. 6: Grande array arrival direction accuracy from Rayleigh distribution of angular differences ($\Psi$).

## VI. CONCLUSIONS

The reconstruction procedures and the achieved accuracies of the Grande array have been studied experimentally and shown. The result is that Grande provides a large acceptance ($2.5 \cdot 10^9$ cm$^2$·sr for $E_0 > 10^{16}$eV, zenith angle $< 40^o$) and it is accurate enough ($N_{ch}$ uncertainty: systematic $< 5\%$, statistical $< 15\%$) for the aims of present analysis [8]. The used lateral distribution function describes experimental lateral distributions over the whole 700 m range. Finally the mean lateral distributions of charged particle have been shown.

## REFERENCES

[1] T. Antoni et al. - KASCADE Coll., Astrop. Phys. 24, 1 (2005),
[2] M. Aglietta et al. EAS-TOP Coll., Astrop. Phys. 21, 583 (2004),
[3] G. Navarra et al. - KASCADE-Grande Coll., Nucl.Instr. and Meth. A 518 (2004),
[4] T. Antoni et al KASCADE Coll., Nucl. Instr. and Meth. A 513, 490 (2003),
[5] M. Aglietta et al. - EAS-TOP Coll., Nucl.Instr. and Meth. A 336 (1993),
[6] R. Glasstetter et al. - KASCADE-Grande Coll. - Proc. $28^{th}$ ICRC, Tsukuba, Japan, vol. 2 (2003) 781,
[7] D. Fuhrmann for the KASCADE-Grande Coll., these proceedings,
[8] E. Cantoni for the KASCADE-Grande Coll., these proceedings.

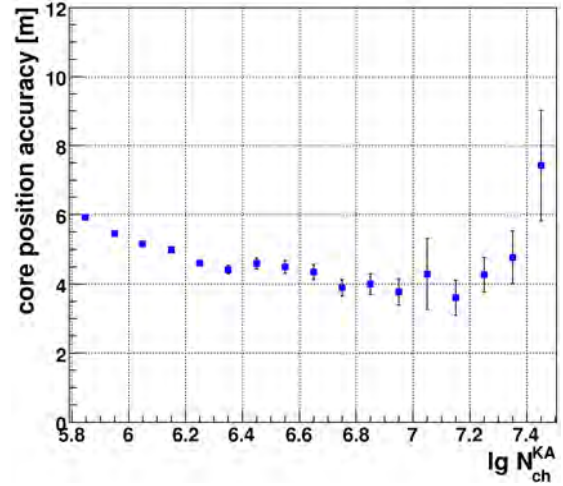

Fig. 7: Grande array core position accuracy from Rayleigh distribution of core position differences.

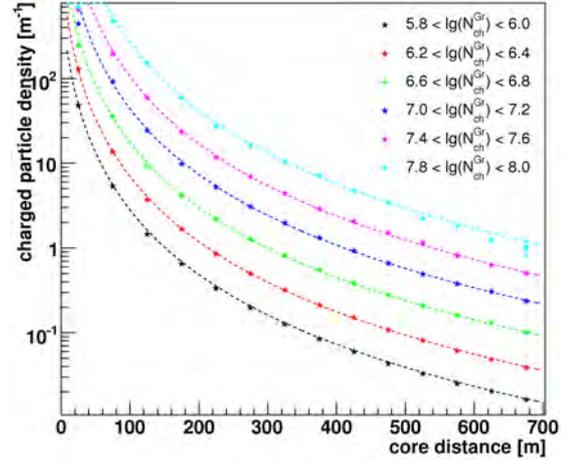

Fig. 8: Mean lateral distributions of charged particles, grouped in lg $N_{ch}^{Gr}$ bins.

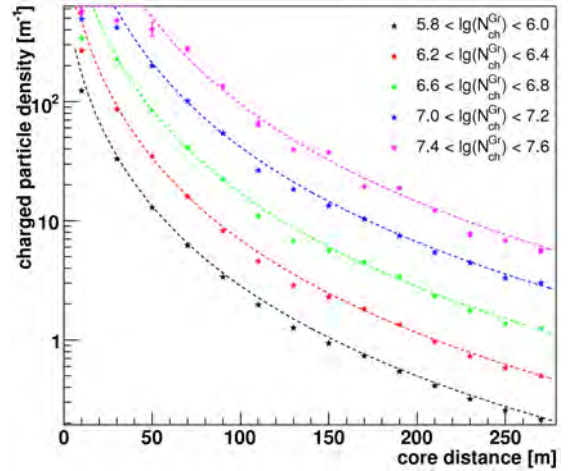

Fig. 9: Mean lateral distributions of charged particles closer to the core and with core position reconstructed independently by KASCADE.

# Muonic Component of Air Showers Measured by the KASCADE-Grande Experiment

D. Fuhrmann\*\*, W.D. Apel\*, J.C. Arteaga†,xi, F. Badea\*, K. Bekk\*, M. Bertaina‡, J. Blümer\*,†,
H. Bozdog\* I.M. Brancus§, M. Brüggemann¶, P. Buchholz¶, E. Cantoni‡,‖, A. Chiavassa‡,
F. Cossavella†, K. Daumiller\*, V. de Souza†,xii, F. Di Pierro‡, P. Doll\*, R. Engel\*, J. Engler\*,
M. Finger\*, P.L. Ghia‖, H.J. Gils\*, R. Glasstetter\*\*, C. Grupen¶, A. Haungs\*,
D. Heck\*, J.R. Hörandel†,xiii, T. Huege\*, P.G. Isar\*, K.-H. Kampert\*\*, D. Kang†,
D. Kickelbick¶, H.O. Klages\*, Y. Kolotaev¶, P. Łuczak††, H.J. Mathes\*, H.J. Mayer\*,
J. Milke\*, B. Mitrica§, C. Morello‖, G. Navarra‡, S. Nehls\*, J. Oehlschläger\*,
S. Ostapchenko\*,xiv, S. Over¶, M. Petcu§, T. Pierog\*, H. Rebel\*, M. Roth\*,
H. Schieler\*, F. Schröder\*, O. Sima‡‡, M. Stümpert†, G. Toma§, G.C. Trinchero‖,
H. Ulrich\*, W. Walkowiak¶, A. Weindl\*, J. Wochele\*, M. Wommer\*, J. Zabierowski††

\**Institut für Kernphysik, Forschungszentrum Karlsruhe, 76021 Karlsruhe, Germany*
†*Institut für Experimentelle Kernphysik, Universität Karlsruhe, 76021 Karlsruhe, Germany*
‡*Dipartimento di Fisica Generale dell'Università, 10125 Torino, Italy*
§*National Institute of Physics and Nuclear Engineering, 7690 Bucharest, Romania*
¶*Fachbereich Physik, Universität Siegen, 57068 Siegen, Germany*
‖*Istituto di Fisica dello Spazio Interplanetario, INAF, 10133 Torino, Italy*
\*\**Fachbereich Physik, Universität Wuppertal, 42097 Wuppertal, Germany*
††*Soltan Institute for Nuclear Studies, 90950 Lodz, Poland*
‡‡*Department of Physics, University of Bucharest, 76900 Bucharest, Romania*
xi *now at: Universidad Michoacana, Morelia, Mexico*
xii *now at: Universidade de São Paulo, Instituto de Física de São Carlos, Brasil*
xiii *now at: Dept. of Astrophysics, Radboud University Nijmegen, The Netherlands*
xiv *now at: University of Trondheim, Norway*

*Abstract*. The KASCADE-Grande experiment consists of a large array of scintillators for the detection of charged particles from extensive air showers in the primary energy range $10^{16}$ eV – $10^{18}$ eV. In combination with the detectors of the KASCADE array it provides the means to investigate the composition in the expected transition region of galactic to extragalactic cosmic rays and the possible existence of a second knee in the total energy spectrum at E $\sim 10^{17}$ eV caused by heavy primaries.
For the goals described it is indispensable to reconstruct the shower sizes with highest accuracy. The reconstruction of the muonic component as well as the muon lateral distribution will be discussed and the precision and systematic uncertainties in the reconstruction of the muon number will be studied based on Monte Carlo simulations.

*Keywords*: muonic component, lateral distribution, KASCADE-Grande

## I. INTRODUCTION AND EXPERIMENTAL SETUP

The combined KASCADE and KASCADE-Grande Experiment [1], located on the site of the Forschungszentrum Karlsruhe (110 m a.s.l.), consists of various detector components [2] for measuring the particles of extensive air showers in the primary energy range from $10^{16}$ eV – $10^{18}$ eV. The measurement at the upper part of that energy range is possible due to a large scintillator array, the Grande array, covering a collecting area of approximately 0.5 km$^2$. The 37 Grande stations located on a hexagonal grid with an average mutual distance of 137 m measure the total number of *charged particles* in an air shower.

With the colocated KASCADE array the *muon component* of the extensive air shower can be measured separately from the electronic one. Using an appropriate lateral distribution function, one can derive the total muon number of air showers from the muon signals measured locally with the KASCADE array. This method can be applied even in cases where the core is located in the KASCADE-Grande array, but not in the KASCADE array itself (Fig. 1, left). Subtracting the estimated number of muons from the total number of charged particles measured with KASCADE-Grande yields the total number of shower electrons [3]. The scintillators of the KASCADE detector array cover an area of 200 × 200 m$^2$ and are housed in 252 stations on a grid with 13 m spacing. While the inner stations of the KASCADE array are only equipped with liquid scintillators measuring primarily electrons and gammas, the outer stations are also containing plastic scintillators underneath a shielding[1] of 10 cm lead and 4 cm iron, which allows the measurement of muons separately from electrons and gammas (Fig. 1, right). These muon detectors consist of four plastic scintillators per station. The scintillators are of 3 cm thickness and their surface area is $90 \times 90$ cm$^2$. The light is coupled out by wavelength shifters and read out by 1.5 inch photomultipliers. The energy resolution has been determined to about 10% at 8 MeV, the mean energy deposit of a MIP[2].

---
[1] corresponding to 20 radiation lengths, muon threshold: 230 MeV.

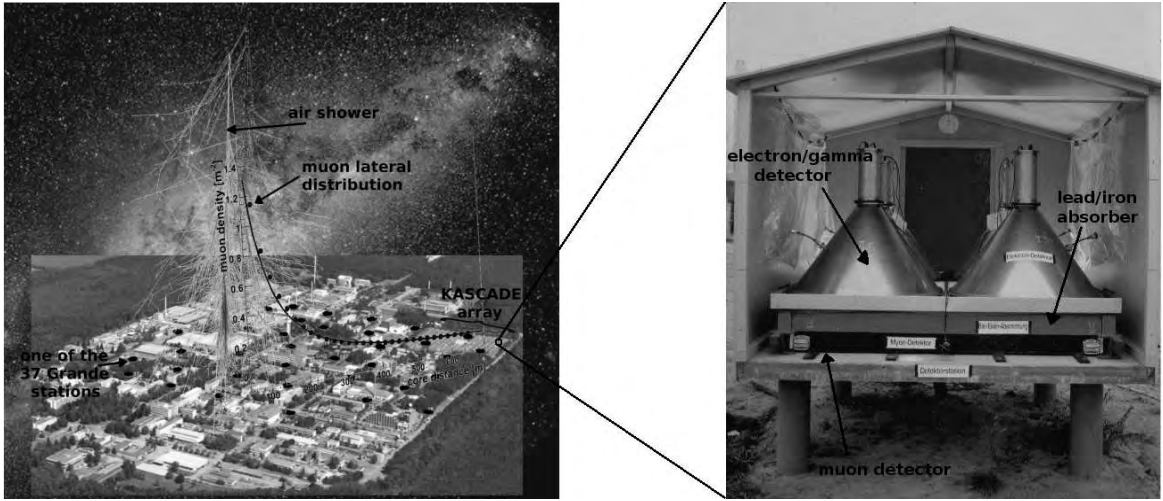

Fig. 1: *Left:* An air shower with the core located in the Grande array. Although the KASCADE detector field is far away from the core, non-zero muon densities can be measured there. *Right:* Detector station of the KASCADE-array equipped with electron/gamma and separate muon detectors.

## II. RECONSTRUCTION OF THE MUON NUMBER

As described in the previous chapter the local muon densities can be measured even in cases where the shower core is located in the KASCADE-Grande array, but not in the KASCADE array itself. For these purposes the energy deposits in the muon detectors must be converted to particle numbers by means of a conversion function, the so-called LECF[3]. The LECF is derived from simulated air showers based on CORSIKA [4] and a detailed GEANT [5] detector simulation. It has been determined based on two primaries (H and Fe) and three different simulated energies, $3 \times 10^{16}$ eV, $1 \times 10^{17}$ eV and $3 \times 10^{17}$ eV. The average energy deposit in the KASCADE muon detectors per shower muon at a distance $r$ (in meter) from the shower core is given by the following LECF:

$$\frac{E_{\text{dep}}}{muon}(r) = \left(7.461 + e^{(1.762 - 0.017 \cdot r)} + 0.0003 \cdot r\right) \text{ MeV} \quad (1)$$

For small radii up to approximately 160 m the energy deposit per muon decreases in order to correct the high energetic electromagnetic punch through close to the shower core. At larger radii the deposited energy

[2]Minimum Ionizing Particle.
[3]Lateral Energy Correction Function.

per muon reaches a constant value of approximately 7.6 MeV (Fig. 2).

For most analyses it is convenient not only to know the local muon densities given by the LECF but also the total number of muons in the shower disk. Assuming the locally detected muons fluctuate according to a poisson distribution, one can derive the total muon number $N_\mu^{\text{rec}}$ from a maximum likelihood estimation, which yields:

$$N_\mu^{\text{rec}} = \sum_{i=1}^{k} n_i \Big/ \sum_{i=1}^{k} \Big( f(r_i) \cdot A_i \cdot \cos(\theta) \Big), \quad (2)$$

where $n_i$ is the number of particles measured at a core distance $r_i$ (in meter) in one of the $k$ muon detectors within an area $A_i$ (in square meters), $\theta$ is the zenith angle (in degree) of the air shower, and $f$ is an appropriate

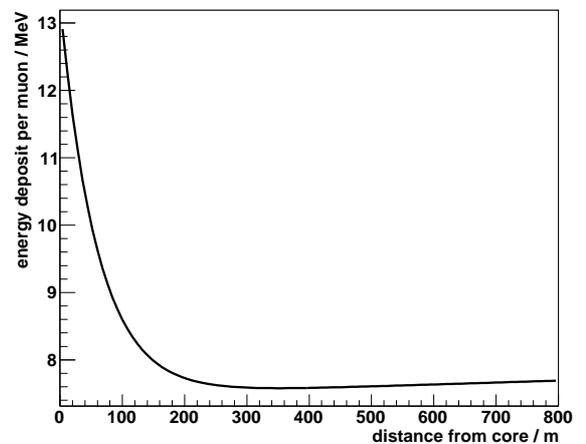

Fig. 2: Average energy deposit per muon in a KASCADE muon detector as a function of the distance of this detector from the shower core (muon LECF according to Eq. 1).

lateral distribution function.

In case of the KASCADE-Grande Experiment the lateral distribution of muon densities $\rho_\mu$ is fitted with a function based on the one proposed by Lagutin and Raikin [6] for the electron component:

$$\rho_\mu(r) = N_\mu \cdot f(r), \text{ with}$$
$$f(r) = \frac{0.28}{r_0^2} \left(\frac{r}{r_0}\right)^{p_1} \cdot \left(1 + \frac{r}{r_0}\right)^{p_2} \cdot \left(1 + \left(\frac{r}{10 \cdot r_0}\right)^2\right)^{p_3}. \quad (3)$$

The parameters $p_1 = -0.69$, $p_2 = -2.39$, $p_3 = -1.0$ and $r_0 = 320$ m are based on CORSIKA simulations using the interaction model QGSJet 01. Both proton and iron primaries were simulated at energies of $10^{16}$ eV and $10^{17}$ eV and then the average of the fit results is taken. Since the muon densities are very low, except for the highest energy showers, stable fits on the shower-by-shower basis are only obtained if the lateral distribution function is kept constant and only the muon number $N_\mu$ is taken as a fit parameter.

Substituting the lateral distribution function $f$ from Eq. 3 into Eq. 2 yields a formula for calculating the total muon number of the KASCADE-Grande event.

## III. RECONSTRUCTION ACCURACY

The muon number is reconstructed based on the local muon densities measured only on the small area of the KASCADE detector field. The measured densities are typically very small and subject to large fluctuations. The reconstruction of the total number of muons is strongly affected by these uncertainties.

The to some extent *directly measured* muon density distribution and the lateral distribution function $f$ (Eq. 3) with the muon number $N_\mu$ set to the *reconstructed* mean muon number $\overline{N_\mu^{\text{rec}}}$ (Eq. 2) in each muon size bin are shown in Fig. 3. The measured densities are in general well described by the lateral distribution function. This means a good conformity between *directly measured* sizes and *reconstructed* ones. Only in case of relatively small and large core distances one can see deviations due to the fixed shape of the lateral distribution function which does not account for the primary energy or the zenith angle of the air shower.

The reconstruction quality has been tested based on CORSIKA simulations using the interaction model QGSJet II. Different primaries (H, He, C, Si and Fe) in equal abundances, with an E$^{-3}$ power law spectrum, zenith angles up to $40°$ and cores scattered over the Grande array were considered in the simulations. The full detector response was also simulated (GEANT [5] detector simulation) and the usual reconstruction techniques were applied to the resulting data. The mean deviation of the reconstructed muon number $N_\mu^{\text{rec}}$ from the true muon number $N_\mu^{\text{tru}}$ as a function of the true muon number itself or the distance of the shower core to the centre of the KASCADE array are shown in Fig. 4. In the latter case, only events with muon numbers above $\log_{10} N_\mu^{\text{tru}} \geq 5.0$ are taken into account, that means only muon numbers above full reconstruction efficiency[4]. In case of muon numbers above a threshold of $\log_{10} N_\mu^{\text{tru}} \approx 5.6$, which corresponds to an energy of approximately $5 \times 10^{16}$ eV, the systematic deviation of the reconstructed total muon number is smaller than 5% and to some extent constant in this range (Fig. 4a). Above the mentioned threshold, the statistical error (represented by the error bars, RMS) decreases from around 20% to 7% with increasing muon number. Showers below 100% efficiency are characterized by a rather large statistical uncertainty up to 40%. In Fig. 4b the dependence of the reconstructed accuracies on the distance of the core to the centre of the KASCADE array is shown. An increase of the statistical uncertainty with increasing distances from approximately 15% at 100 m to 30% at 700 m distance is observed. The under- or overestimation of the local muon densities by the lateral distribution function (discussed above, Fig. 3) in cases of small and large core distances results in an under- or overestimation of the total muon number in these distance ranges. The deviation of the reconstructed muon number from the true one starts from $\sim -7\%$ for small distances, gets zero for $\sim 240$ m distance and increases to $\sim +12\%$ for larger core distances. Taking into account the fact that quite small particle densities are measured across a small detection area far away from the shower core, one can draw the conclusion, that the reconstruction of the total muon number works surprisingly well. Furthermore the features of the accuracies are well understood and open the possibility to correct the reconstructed muon number to the true one using appropriate correction functions and to perform analyses based on these corrected muon numbers (see [7]).

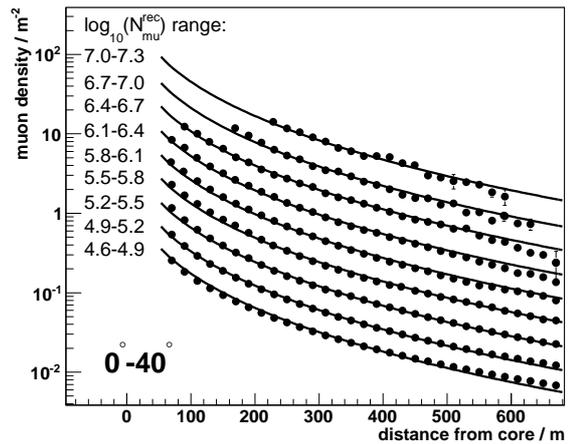

Fig. 3: Measured muon density distribution (dots) for zenith angles $0°$–$40°$ and different intervals of the reconstructed muon number. The lateral distribution function of Eq. 3 (curves) with the muon number $N_\mu$ set to the measured mean muon number $\overline{N_\mu^{\text{rec}}}$ in each interval describes the data quite well.

---

[4]100% reconstruction efficiency of muon number is obtained above $\log_{10} N_\mu^{\text{tru}} \approx 5.0$.

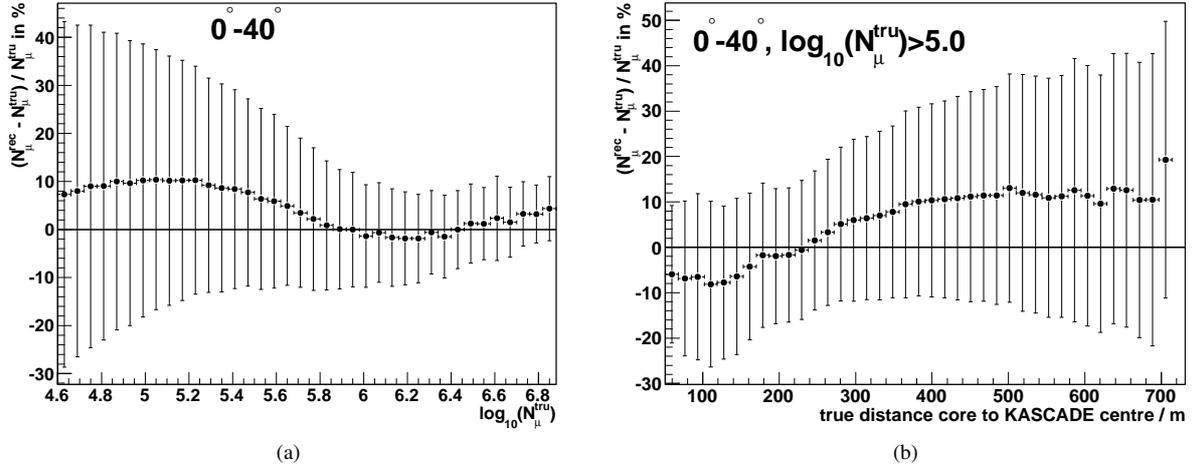

Fig. 4: Reconstruction quality tested based on Monte Carlo simulations. In case *a)* the deviation of the reconstructed to the true muon number is shown as a function of the true muon number, in case *b)* as a function of the distance of the shower core to the centre of the KASCADE array. In both cases the error bars represent the statistical uncertainty (RMS) in a single measurement.

## IV. CONCLUSION

The reconstruction of the total muon number in the shower disk has been presented using the KASCADE scintillator array as a part of the KASCADE-Grande experiment. The procedure of converting the energy deposits to particle numbers was explained as well as the calculation of the total muon number using a maximum likelihood method. The reconstruction accuracies have been discussed and reveal a good reconstruction quality, despite the fact that the total muon number is reconstructed based on just a small fraction of radial detector coverage. Deviations between reconstructed and true shower sizes are well understood such that it is possible to derive correction functions allowing to correct the reconstructed muon number to the true one. Hence, KASCADE-Grande analyses taking into account the total muon number can be performed.

# The sensitivity of KASCADE-Grande to the cosmic ray primary composition between $10^{16}$ and $10^{18}$ eV

E. Cantoni$^{\ddagger,\|}$, W.D. Apel$^{*}$, J.C. Arteaga$^{\dagger,xi}$, F. Badea$^{*}$, K. Bekk$^{*}$, M. Bertaina$^{\ddagger}$, J. Blümer$^{*,\dagger}$, H. Bozdog$^{*}$ I.M. Brancus$^{\S}$, M. Brüggemann$^{\P}$, P. Buchholz$^{\P}$, A. Chiavassa$^{\ddagger}$, F. Cossavella$^{\dagger}$, K. Daumiller$^{*}$, V. de Souza$^{\dagger,xii}$, F. Di Pierro$^{\ddagger}$, P. Doll$^{*}$, R. Engel$^{*}$, J. Engler$^{*}$, M. Finger$^{*}$, D. Fuhrmann$^{**}$, P.L. Ghia$^{\|}$, H.J. Gils$^{*}$, R. Glasstetter$^{**}$, C. Grupen$^{\P}$, A. Haungs$^{*}$, D. Heck$^{*}$, J.R. Hörandel$^{\dagger,xiii}$, T. Huege$^{*}$, P.G. Isar$^{*}$, K.-H. Kampert$^{**}$, D. Kang$^{\dagger}$, D. Kickelbick$^{\P}$, H.O. Klages$^{*}$, Y. Kolotaev$^{\P}$, P. Łuczak$^{\dagger\dagger}$, H.J. Mathes$^{*}$, H.J. Mayer$^{*}$, J. Milke$^{*}$, B. Mitrica$^{\S}$, C. Morello$^{\|}$, G. Navarra$^{\ddagger}$, S. Nehls$^{*}$, J. Oehlschläger$^{*}$, S. Ostapchenko$^{*,xiv}$, S. Over$^{\P}$, M. Petcu$^{\S}$, T. Pierog$^{*}$, H. Rebel$^{*}$, M. Roth$^{*}$, H. Schieler$^{*}$, F. Schröder$^{*}$, O. Sima$^{\ddagger\ddagger}$, M. Stümpert$^{\dagger}$, G. Toma$^{\S}$, G.C. Trinchero$^{\|}$, H. Ulrich$^{*}$, W. Walkowiak$^{\P}$, A. Weindl$^{*}$, J. Wochele$^{*}$, M. Wommer$^{*}$, J. Zabierowski$^{\dagger\dagger}$

$^{*}$*Institut für Kernphysik, Forschungszentrum Karlsruhe, 76021 Karlsruhe, Germany*
$^{\dagger}$*Institut für Experimentelle Kernphysik, Universität Karlsruhe, 76021 Karlsruhe, Germany*
$^{\ddagger}$*Dipartimento di Fisica Generale dell'Università, 10125 Torino, Italy*
$^{\S}$*National Institute of Physics and Nuclear Engineering, 7690 Bucharest, Romania*
$^{\P}$*Fachbereich Physik, Universität Siegen, 57068 Siegen, Germany*
$^{\|}$*Istituto di Fisica dello Spazio Interplanetario, INAF, 10133 Torino, Italy*
$^{**}$*Fachbereich Physik, Universität Wuppertal, 42097 Wuppertal, Germany*
$^{\dagger\dagger}$*Soltan Institute for Nuclear Studies, 90950 Lodz, Poland*
$^{\ddagger\ddagger}$*Department of Physics, University of Bucharest, 76900 Bucharest, Romania*
$^{xi}$ *now at: Universidad Michoacana, Morelia, Mexico*
$^{xii}$ *now at: Universidade de São Paulo, Instituto de Física de São Carlos, Brasil*
$^{xiii}$ *now at: Dept. of Astrophysics, Radboud University Nijmegen, The Netherlands*
$^{xiv}$ *now at: University of Trondheim, Norway*

*Abstract*. The goal of the KASCADE-Grande experiment is the study of the cosmic ray energy spectrum and chemical composition in the range $10^{16}$ - $10^{18}$ eV detecting the charged particles of the respective Extensive Air Showers. The observables here taken into account for discussion are the measured electron size ($N_e$) and muon size ($N_\mu$). It is crucial to verify: the sensitivity to different chemical components, the data reproducibility with the hadronic interaction model in use as a function of the electron size and the athmospheric depth, the consistency with the results obtained by other experiments sensitive to composition in overlapping energy regions. The analysis is presented using KASCADE as reference experiment and using QGSjetII as hadronic interaction model.

*Keywords*: KASCADE-Grande, sensitivity, interaction model.

## I. INTRODUCTION

The KASCADE-Grande experiment is located at Forschungszentrum Karlsruhe (Germany). It consists of an array of 37 scintillator modules 10 m$^2$ each (the Grande array) spread over an area of 700 x 700 m$^2$, working jointly with the co-located and formerly present KASCADE experiment [1], made of 252 scintillation detectors, 490 m$^2$ sensitive area spread over 200 x 200 m$^2$. The extension from KASCADE to KASCADE-Grande is meant to increase the experimental acceptance of a factor ∼10, the achieved accuracies showing there is no significant loss in resolution for the present analysis (see [2]). For each recorded EAS the charged particle size N$_{ch}$ is measured through Grande, the reconstruction procedure being fine tuned over the whole experimental area (see [2]), and the muon size N$_\mu$ is obtained from KASCADE, the reconstruction and accuracy of the muonic component being described in [3]. The electron size is obtained subtracting the muon from the charged particle density. As to validate the experimental results and verify the applicability of the interaction model in use for data interpretation[1] it is first of all important to achieve an accurate event reconstruction, the next step is to test the sensitivity of the extended apparatus to observables, to verify the data reproducibility with the hadronic interaction model in use and to test the consistency of this reproducibility with the former KASCADE data. For this aims, in the following analysis, the total number of electrons N$_e$ and the total number of muons N$_\mu$ of each recorded event are considered and the distribution of N$_\mu$/N$_e$ is studied in different intervals of N$_e$[2] and zenith angle (athmospheric depth).

---

[1]Both for energy measurements and composition studies.
[2]Corresponding to different energy intervals (see IV).

TABLE I: The results for the chi square minimization on the selected experimental data using just one chemical component.

| chemical element | protons (p) | Helium (He) | Carbon (CNO) | Silicium (Si) | Iron (Fe) |
|---|---|---|---|---|---|
| $\chi^2/\nu$ | 6798.09 | 404.59 | 26.44 | 17.20 | 10.44 |

## II. THE FEATURES OF THE ANALYSIS

KASCADE-Grande data are chosen, at first, in an electron size range providing full reconstruction efficiency (see [5]) and high statistics: $6.49 \leq \text{Log}(N_e) < 6.74$ in $0° \leq \theta < 23.99°$ (see figure 1). The same event selection is made on the simulated QGSjetII [4] data sets at disposal for each cosmic ray primary[3]. The experimental distribution of the observable $N_\mu/N_e$ is taken into account and fitted with a linear combination of elemental contributions from simulations, expressed as follows:

$$F_{sim}(i) = \sum_j \alpha_j f_{sim,j}(i) \quad (1)$$

where $F_{sim}(i)$ is the total theoretical fraction of simulated events falling in the channel i of the distribution considered as a histogram, $f_{sim,j}(i)$ is the fraction for the single chemical component j, $\Sigma_j$ is the sum over the different components and $\alpha_j$ is the fit parameter representing the *relative abundance* of the component j. The fit parameters fulfill the conditions

$$0. < \alpha_j < 1., \forall j \quad (2)$$

and

$$\sum_j \alpha_j = 1. \quad (3)$$

The fit is performed through the minimization of the following Chi Square function:

$$\chi^2 = \sum_i \frac{(F_{exp}(i) - F_{sim}(i))^2}{\sigma(i)^2} \quad (4)$$

where $F_{exp}(i)$ is the fraction of experimental events falling in the histogram channel i and $\sigma(i)$ is the error on the theoretical expression (1).

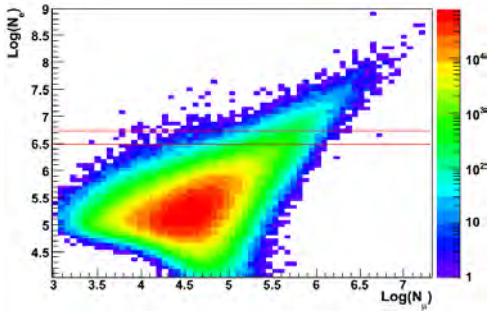

Fig. 1: The KASCADE-Grande data as they appear in the observables $N_e$ and $N_\mu$. The considered selection lies between the lines.

---

[3]p,He, C, Si, Fe. (simulated with an energy spectrum $\gamma = 3$)

### A. The fit with a single chemical component

A chi square minimization is performed at first with the use of a single chemical component. This shows to give not a good description of the data, as it can be seen in table I, summarizing the results for the chi square minimization with single elements (see also figure 2 as example).

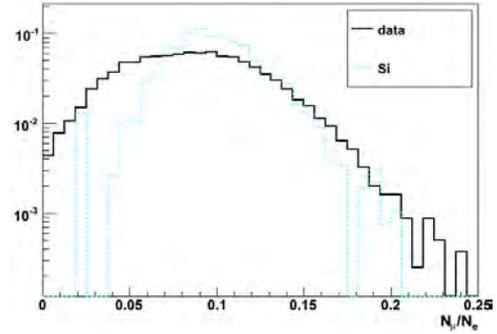

Fig. 2: The distributions (normalized to 1) of the KASCADE-Grande data in $6.49 \leq \text{Log}(N_e) < 6.74$ ($0° \leq \theta < 23.99°$) and of QGSjetII Silicium in the same selection range.

### B. The fit with two chemical components

The experimental selection is then fitted with a combination of a light and a heavy chemical component that, at a qualitative glance, seem necessary to describe well the tails of the experimental histogram. Indeed, performing a minimization with two components steps up the fit, as it can be seen in table II and figures 3 and 4. It can be observed that Iron seems necessary to describe well the right tail of the experimental distribution, while Helium seems not to fit well on the left tail. Moreover, the shapes of the fits suggest the requirement of a third element in the middle.

TABLE II: The results for the chi square minimization on the selected experimental data using two chemical components.

| chemical elements | p + Fe | He + Fe |
|---|---|---|
| $\alpha_p$ | 0.41 ± 0.02 | – |
| $\alpha_{He}$ | – | 0.49 ± 0.02 |
| $\alpha_{Fe}$ | 0.59 ± 0.02 | 0.51 ± 0.02 |
| $\chi^2/\nu$ | 3.51 | 1.48 |

### C. The fit with three chemical components

The fit is then performed with a combination of three elements: Protons, Helium and Iron are chosen, matching the light elements with the heaviest element.

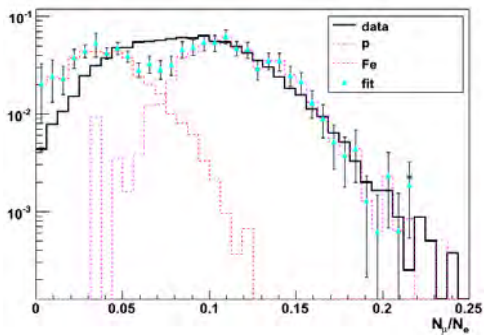

Fig. 3: The KASCADE-Grande data in $6.49 \leq \text{Log}(N_e) < 6.74$ ($0° \leq \theta < 23.99°$) described by Protons and Iron primaries. Here and in the next pictures, the experimental plot is normalized to 1 and every simulated component is normalized to its relative abundance. Each tail of the experimental distribution ($>2\cdot$RMS and $<2\cdot$RMS) is treated counting the events in a single bin: the star indicates the total experimental value the fit must be compared with in that bin.

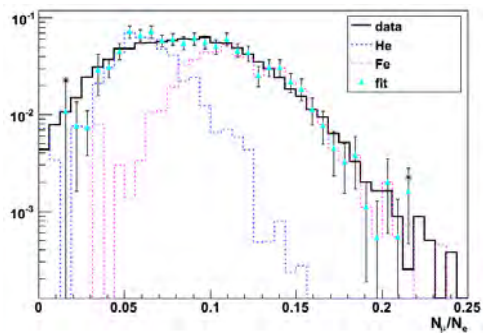

Fig. 4: The KASCADE-Grande data in $6.49 \leq \text{Log}(N_e) < 6.74$ ($0° \leq \theta < 23.99°$) described by a Helium and Iron primaries.

It can be seen that a three elements combination is well fitting the data (see table III and figure 5). It is possible to use also another combination: Protons, Carbon and Iron (see table III and figure 6). These results show that the theoretical model describes well the shape and the tails of the experimental distribution.

TABLE III: The results for the chi square minimization on the selected experimental data using three chemical components.

| chemical elements | p + He + Fe | p + CNO + Fe |
|---|---|---|
| $\alpha_p$ | 0.15 ± 0.02 | 0.23 ± 0.02 |
| $\alpha_{He}$ | 0.31 ± 0.03 | – |
| $\alpha_C$ | – | 0.34 ± 0.03 |
| $\alpha_{Fe}$ | 0.54 ± 0.02 | 0.43 ± 0.02 |
| $\chi^2/\nu$ | 0.68 | 1.14 |

### III. ANALYSIS ON INCLINED SHOWERS

To check the consistency of the result at higher zenith angles, the same fits with three chemical components are performed on a higher angular interval of equal acceptance, $29.86° \leq \theta < 40°$, in an electron size interval providing a similar number of events, $6.11 \leq \text{Log}(N_e) < 6.36$. Also at higher angles, using three elements, it is found that the model reproduces the data, as it can be seen in table IV and figure 7.

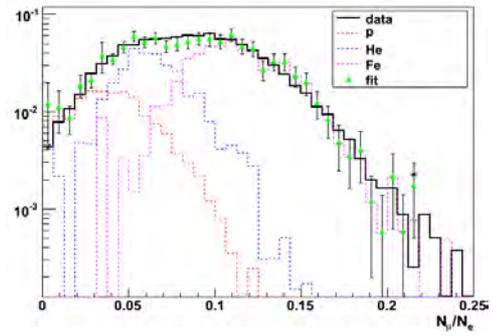

Fig. 5: The KASCADE-Grande data in $6.49 \leq \text{Log}(N_e) < 6.74$ ($0° \leq \theta < 23.99°$) described by Protons, Helium and Iron primaries.

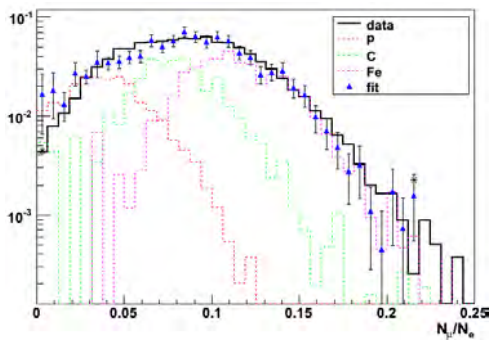

Fig. 6: The KASCADE-Grande data in $6.49 \leq \text{Log}(N_e) < 6.74$ ($0° \leq \theta < 23.99°$) described by Protons, Carbon and Iron primaries.

TABLE IV: The results for the chi square minimization on the selected experimental data using p + He + Fe. Comparison between vertical and inclined showers.

| angular bin | $0° \leq \theta < 23.99°$ | $29.86° \leq \theta < 40°$ |
|---|---|---|
| $\alpha_p$ | 0.15 ± 0.02 | 0.17 ± 0.04 |
| $\alpha_{He}$ | 0.31 ± 0.03 | 0.31 ± 0.05 |
| $\alpha_{Fe}$ | 0.54 ± 0.02 | 0.52 ± 0.03 |
| $\chi^2/\nu$ | 0.68 | 0.77 |

### IV. ANALYSIS AT HIGHER ENERGIES

Selecting the KASCADE-Grande experimental data for higher values of the electron size $N_e$ means to chose showers that were generated by higher energy events (see [5]). Also in this case, it is found that the model reproduces the data, the minimization of the $N_\mu/N_e$ distribution with three chemical components still providing a good result, as it can be seen in table V and figure 8.

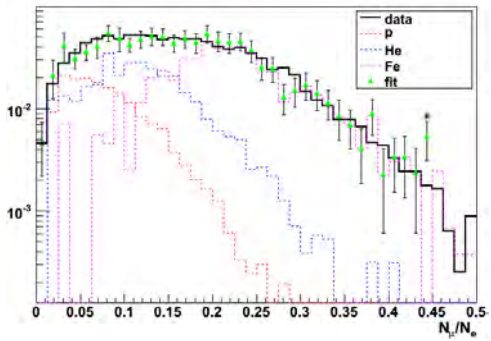

Fig. 7: The KASCADE-Grande data in $6.11 \leq \mathrm{Log}(N_e) < 6.36$ ($29.86° \leq \theta < 40°$) described by Protons, Helium and Iron primaries.

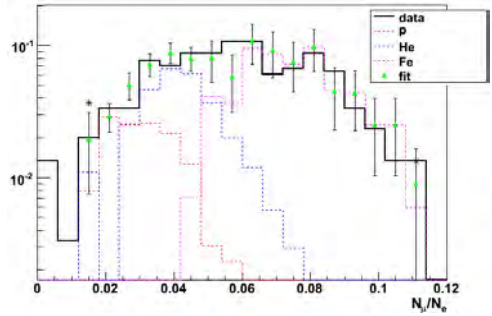

Fig. 8: The KASCADE-Grande data in $7.24 \leq \mathrm{Log}(N_e) < 7.49$ ($0° \leq \theta < 23.99°$) described by Protons, Helium and Iron primaries.

TABLE V: The results for the chi square minimization in a low and a high electron size interval, using p + He + Fe.

| $\Delta \mathrm{Log}(N_e)$ | $6.49 \leq \mathrm{Log}(N_e) < 6.74$ | $7.24 \leq \mathrm{Log}(N_e) < 7.49$ |
|---|---|---|
| $\alpha_p$ | $0.15 \pm 0.02$ | $0.13 \pm 0.02$ |
| $\alpha_{He}$ | $0.31 \pm 0.03$ | $0.29 \pm 0.04$ |
| $\alpha_{Fe}$ | $0.54 \pm 0.02$ | $0.58 \pm 0.04$ |
| $\chi^2/\nu$ | 0.68 | 0.83 |

## V. COMPARISON WITH KASCADE DATA

Being the electron size range $6.11 \leq \mathrm{Log}(N_e) < 6.36$ ($29.86° \leq \theta < 40°$) common to KASCADE and KASCADE-Grande data, the correspondent $N_\mu^{tr}/N_e$ distribution from KASCADE is taken into account. Applying the same analysis, it is found that three chemical components fit the data, as for KASCADE-Grande (see table VI and figure 9). Even with KASCADE data, the combination p + C + Fe is also fitting (see figure 10).

TABLE VI: The results for the chi square minimization on the selected data from KASCADE.

| chemical elements | p + He + Fe | p + C + Fe |
|---|---|---|
| $\alpha_p$ | $0.20 \pm 0.03$ | $0.33 \pm 0.06$ |
| $\alpha_{He}$ | $0.30 \pm 0.04$ | – |
| $\alpha_C$ | – | $0.26 \pm 0.07$ |
| $\alpha_{Fe}$ | $0.50 \pm 0.03$ | $0.41 \pm 0.04$ |
| $\chi^2/\nu$ | 1.12 | 1.25 |

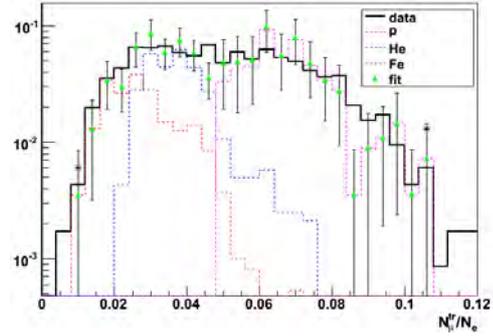

Fig. 9: The KASCADE data in $6.11 \leq \mathrm{Log}(N_e) < 6.36$ ($29.86° \leq \theta < 40°$) described by Protons, Helium and Iron primaries. Here the number of muons with distances to the shower core between 40 m and 200 m ("truncated") is considered.

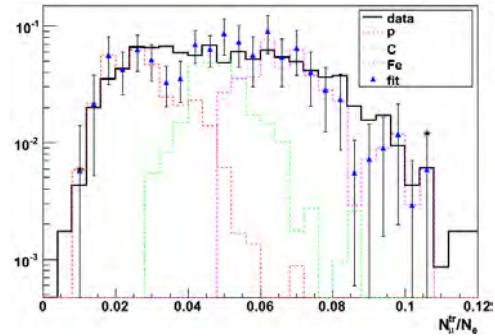

Fig. 10: The KASCADE data in $6.11 \leq \mathrm{Log}(N_e) < 6.36$ ($29.86° \leq \theta < 40°$) described by Protons, Carbon and Iron primaries.

## VI. CONCLUSIONS

In this work it has been shown that, with the use of a method exploiting a chi square minimization of a linear combination of different simulated primaries, the KASCADE-Grande $N_\mu/N_e$ distributions are fitted using at least three elements. QGSjetII, the hadronic interaction model in use, can fairly reproduce the data and, in particular, the tails of the distributions, that represent a main constraint being related to the lightest and heaviest cosmic ray primaries. Finally, this kind of analysis, performed on the correspondent $N_\mu^{tr}/N_e$ distribution from KASCADE experiment, gives a consistent result, thus showing that, in the superposition energy region, KASCADE-Grande is fairly well reproducing KASCADE data.

# A direct measurement of the muon component of air showers by the KASCADE-Grande Experiment

V. de Souza[†,xii], W.D. Apel[∗], J.C. Arteaga[†,xi], F. Badea[∗], K. Bekk[∗], M. Bertaina[‡], J. Blümer[∗,†],
H. Bozdog[∗] I.M. Brancus[§], M. Brüggemann[¶], P. Buchholz[¶], E. Cantoni[‡,∥], A. Chiavassa[‡],
F. Cossavella[†], K. Daumiller[∗], F. Di Pierro[‡], P. Doll[∗], R. Engel[∗], J. Engler[∗], M. Finger[∗],
D. Fuhrmann[∗∗], P.L. Ghia[∥], H.J. Gils[∗], R. Glasstetter[∗∗], C. Grupen[¶], A. Haungs[∗],
D. Heck[∗], J.R. Hörandel[†,xiii], T. Huege[∗], P.G. Isar[∗], K.-H. Kampert[∗∗], D. Kang[†],
D. Kickelbick[¶], H.O. Klages[∗], P. Łuczak[††], H.J. Mathes[∗], H.J. Mayer[∗],
J. Milke[∗], B. Mitrica[§], C. Morello[∥], G. Navarra[‡], S. Nehls[∗], J. Oehlschläger[∗],
S. Ostapchenko[∗,xiv], S. Over[¶], M. Petcu[§], T. Pierog[∗], H. Rebel[∗], M. Roth[∗],
H. Schieler[∗], F. Schröder[∗], O. Sima[‡‡], M. Stümpert[†], G. Toma[§], G.C. Trinchero[∥],
H. Ulrich[∗], A. Weindl[∗], J. Wochele[∗], M. Wommer[∗], J. Zabierowski[††]

[†]*Institut für Experimentelle Kernphysik, Universität Karlsruhe, 76021 Karlsruhe, Germany*
[∗]*Institut für Kernphysik, Forschungszentrum Karlsruhe, 76021 Karlsruhe, Germany*
[‡]*Dipartimento di Fisica Generale dell'Università, 10125 Torino, Italy*
[§]*National Institute of Physics and Nuclear Engineering, 7690 Bucharest, Romania*
[¶]*Fachbereich Physik, Universität Siegen, 57068 Siegen, Germany*
[∥]*Istituto di Fisica dello Spazio Interplanetario, INAF, 10133 Torino, Italy*
[∗∗]*Fachbereich Physik, Universität Wuppertal, 42097 Wuppertal, Germany*
[††]*Soltan Institute for Nuclear Studies, 90950 Lodz, Poland*
[‡‡]*Department of Physics, University of Bucharest, 76900 Bucharest, Romania*
[xi]*now at: Universidad Michoacana, Morelia, Mexico*
[xii]*now at: Universidade de São Paulo, Instituto de Física de São Carlos, Brasil*
[xiii]*now at: Dept. of Astrophysics, Radboud University Nijmegen, The Netherlands*
[xiv]*now at: University of Trondheim, Norway*

*Abstract*. The muon component of atmospheric air showers is a very relevant information in astroparticle physics due to its direct relation to the primary particle type and dependence on the hadronic interactions. In this paper, we study the muon densities measured by the KASCADE-Grande experiment and illustrate its importance in composition studies and testing of hadronic interaction models. The data analysed here was measured by the KASCADE-Grande detector and lies in the $10^{16} - 10^{18}$ eV energy range. The measured muon density is compared to predictions of EPOS 1.61 and QGSJet II hadronic interaction models.

*Keywords*: Muon Density, Composition and Simulation test.

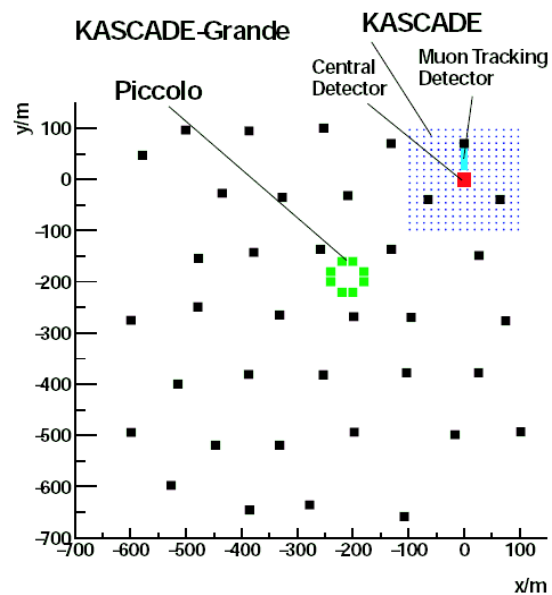

Fig. 1: Representation of the KASCADE-Grande detectors.

## I. INTRODUCTION

Cosmic rays with energy range between $10^{16}$ and $10^{18}$ eV have the potential to reveal interesting astrophysical phenomena occurring in the Universe. This might be the energy range in which a transition in the predominance of the particle flux from galactic to extragalactic sources is happening what could be followed by changes in the primary cosmic abundance. If such a transition is not occurring in this energy range, galactic sources would have an acceleration power beyond the predictions of conservative theories.

The KASCADE-Grande experiment (see figure 1) has been set up to measure primary cosmic rays in this energy range in order to help in the understanding of these questions. The experiment is located at the Forschungszentrum Karlsruhe, Germany, where, beside the existing KASCADE [1] array, two new detector set ups (Grande and Piccolo) have been installed. The experiment is able to sample different components of extensive air showers (electromagnetic, muonic and hadronic) with high accuracy and covering a surface of 0.5 km$^2$. For an overview of the actual setup of the KASCADE-Grande Experiment see ref. [2].

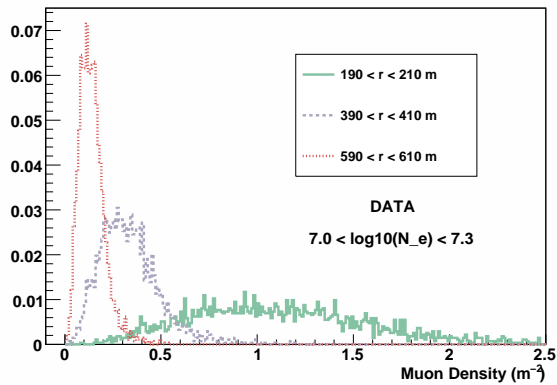

Fig. 2: Distribution of the density of muons for three distances from the shower axis.

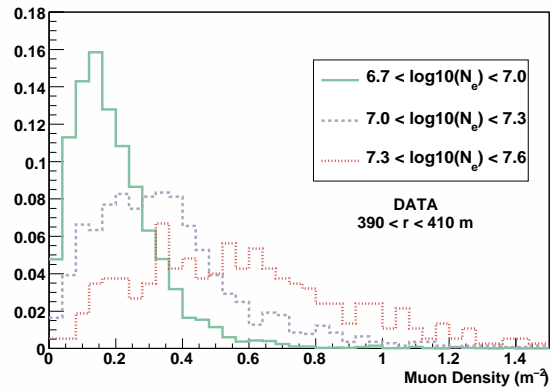

Fig. 3: Distribution of the density of muons for three cuts in the total number of electrons.

In this article we present studies of the muon component of the shower. Muons are the messengers of the hadronic interactions of the particles in the shower and therefore are a powerful tool to determine the primary particle mass and to study the hadronic interaction models.

## II. RECONSTRUCTION

The main parameters used in this study are the density of muons and the total number of electrons in the shower for which the reconstruction accuracy is going to be discussed below. For the reconstruction accuracy of the shower geometry see ref. [3].

The density of muons is directly measured by the KASCADE 622 $m^2$ scintillators. These detectors are shielded by 10 cm of lead and 4 cm of iron, corresponding to 20 radiation lengths and a threshold of 230 MeV for vertical muons. The error in the measurement of the energy deposit was experimentally determined to be smaller than 10% [1].

For each shower, the density of muons is calculated as follows. The muon stations are grouped in rings of 20 m distance from the shower axis. The sum of the signals measured by all muon stations inside each ring is divided by the effective detection area of the stations. Therefore the muon density as a function of the distance from the shower axis is measured in a very direct way. No fitting of lateral distributions is needed in these calculations.

The total number of electrons in the shower is reconstructed in a combined way using KASCADE and KASCADE-Grande stations. A lateral distribution function (LDF) of the Lagutin type can be fitted to the density of muons measured by the KASCADE detector [4]. After that, using the fitted function, the number of muons at any distance from the shower axis can be estimated. The KASCADE-Grande stations measure the number of charged particles. The number of electrons at each KASCADE-Grande stations is determined by subtracting from the measured number of charged particles the number of muons estimated with the LDF fitted to the KASCADE stations.

At this stage, the number of electrons at each KASCADE-Grande station is known. Finally, a modified NKG [5] function is fitted to this data and the total number of electrons is determined in the fit.

Quality cuts have been applied to the events in this analysis procedure. We have required more than 19 KASCADE-Grande stations with signal. The showers used in all analysis along this paper were reconstructed with zenith angle between 0 and 42 degrees. The same quality cuts were applied to the simulated events used for reconstruction studies and to the data presented in the following section. After the quality cuts, the total number of electrons can be estimated with a systematic shift smaller than 10% and a statistical uncertainty smaller than 20% along the entire range considered in this paper [3].

Figure 2 shows the measured density of muons at three distances from the shower axis for events with a total number of electrons ($N_e$) in the range $7.0 < Log10(N_e) < 7.3$ ($\approx 10^{17}$ eV). Similar plots were obtained for other $N_e$ ranges.

Figure 3 shows the density of muons at 400 m from the shower axis for events with total number of electrons ($N_e$) in the range $6.7 < Log10(N_e) < 7.0$, $7.0 < Log10(N_e) < 7.3$ and $7.3 < Log10(N_e) < 7.6$. Similar plots were obtained for other distances from the shower axis.

Figure 2 and Figure 3 show the general expected trend: a) decrease of the muon density with increasing distance from the shower axis and b) increase of the muon density with increasing total number of electrons. In the next sections we explore these relations in order to show the capabilities of the KASCADE-Grande experiment for a composition study and for tests of the hadronic interaction models.

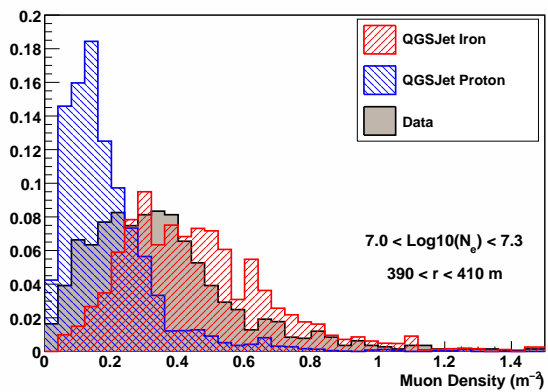

Fig. 4: Measured distribution of the density of muons at 400 m compared to the predictions of QGSJet II.

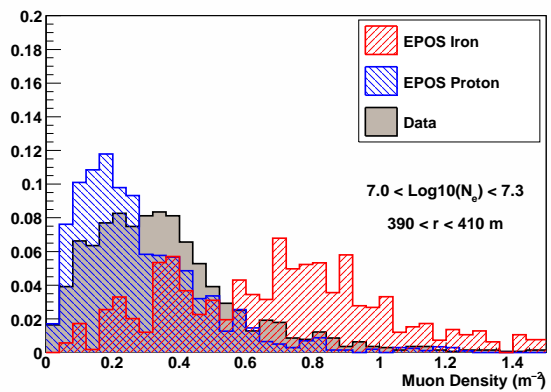

Fig. 5: Measured distribution of the density of muons at 400 m compared to the predictions of EPOS 1.61.

We present data for $7.0 < Log10(N_e) < 7.3$ and $390 < r < 410\ m$, these cuts have been chosen in order to minimize the fluctuation of the signal and the reconstruction inaccuracy and to maximize the number of showers for which we have data, however the same conclusions would be drawn for all parameter cuts.

## III. SIMULATION

For all studies in this paper we have used the COR-SIKA [6] simulation program with the FLUKA [7] option for low energy hadronic interactions. Two high energy hadronic interaction models were used EPOS 1.61 [8] and QGSJet II [9]. No thinning is used [6].

CORSIKA showers are simulated through the detectors and reconstructed in the same way as the measured data, such that a direct comparison between data and simulation is possible.

Figures 4 and 5 show the comparison of the measured density of muons to values predicted by QGSJet II and EPOS 1.61. For both hadronic interactions models we show the limiting cases of proton and iron nuclei as primary particles. It can be seen in figures 4 and 5 that the data lie well within the proton and iron limits for QGSJet II and EPOS 1.61. These graphics are going to be further discussed in the next sections.

## IV. ANALYSIS

Figure 6 shows the mean muon density as a function of the distance from the shower axis compared to the predictions of QGSJet II and EPOS 1.61. Both hadronic interaction models include the data within the proton and iron limits for the entire range of distances from 100 to 750 meters. For distances further than 750 meters the statistics is not enough for a conclusion.

Interesting to note is also the slope of the LDF. Considering an equal probability trigger for protons and iron primaries as a function of distance from the shower axis, one should expect the LDF to be parallel to pure composition primaries. Note that the LDF of simulated proton and iron shower are parallel. However the measured LDF is not parallel to the QGSJet II nor to the EPOS 1.6 curves. That shows that the slope of the LDF can not be well described by neither models.

Figure 7 shows the evolution of the mean muon density as a function of $N_e$. The calculations done with QGSJet II and EPOS 1.61 using proton and iron nuclei as primary particles bracket the data in the entire range of $5 < Log10(N_e) < 8$.

Nevertheless, both figures 6 and 7 show that EPOS 1.61 would require a very light primary composition in order to fit the data. On the other hand, QGSJet II could fit the data with an intermediate primary abundance between proton and iron nuclei.

Besides that, in figure 7 it is possible to analyse a possible transition of the primary component with increasing total number of electrons. The analysis done with both models show no abrupt change in the compositon in the entire energy range.

The change in slope seen in figure 7 for $Log10(N_e) < 6.0$ corresponds to the threshold of the experiment and the fact that both data and simulation show the same behavior illustrates the good level of understanding of our detectors.

## V. CONCLUSIONS

The Grande array is in continuous and stable data taking since December 2003. The quality of the detector can be illustrated by the smooth data curve and small fluctuations in figures 6 and 7.

In this article, we have briefly described the procedure used to measure the density of muons with the KAS-CADE array and we have studied its correlation with the distance from the shower axis and the total number of electrons in the shower.

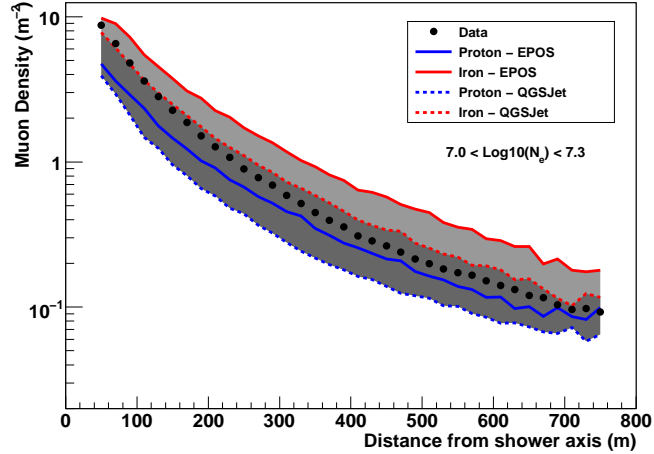

Fig. 6: Lateral distribution of muons compared to the predictions of QGSJet II and EPOS 1.61.

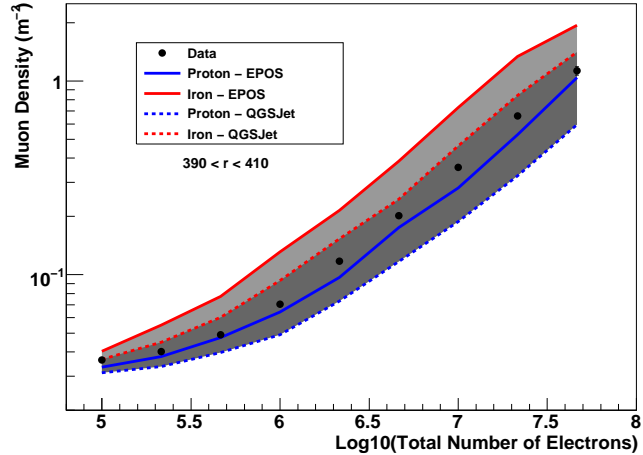

Fig. 7: Muon density as a function of the total number of electrons compared to the predictions of QGSJet II and EPOS 1.61.

The density of muons in the shower is measured directly by the KASCADE detectors. We have used this data to study the hadronic interaction models QGSJet II and EPOS 1.61. The data taken with KASCADE-Grande confirms at higher energies the recent results published by the KASCADE [10] experiment. EPOS 1.61 would require a very light abundance of primary particles in order to fit the data. QGSJet II could fit the data with an intermediate primary abundance.

Figure 7 shows no abrupt change with increasing total number of electrons up to $Log10(N_e) = 7.5 \approx 5 \times 10^{17}$ eV. The mean primary mass estimation would depend on the hadronic interaction model used.

# Study of EAS development with the Muon Tracking Detector in KASCADE-Grande

J. Zabierowski*, W.D. Apel†, J.C. Arteaga‡,xi, F. Badea†, K. Bekk†, M. Bertaina§, J. Blümer†,‡,
H. Bozdog† I.M. Brancus¶, M. Brüggemann∥, P. Buchholz∥, E. Cantoni§,**, A. Chiavassa§,
F. Cossavella‡, K. Daumiller†, V. de Souza‡,xii, F. Di Pierro§, P. Doll†, R. Engel†, J. Engler†,
M. Finger†, D. Fuhrmann††, P.L. Ghia**, H.J. Gils†, R. Glasstetter††, C. Grupen∥,
A. Haungs†, D. Heck†, J.R. Hörandel‡,xiii, T. Huege†, P.G. Isar†, K.-H. Kampert††,
D. Kang‡, D. Kickelbick∥, H.O. Klages†, Y. Kolotaev∥, P. Łuczak*, H.J. Mathes†,
H.J. Mayer†, J. Milke†, B. Mitrica¶, C. Morello**, G. Navarra§, S. Nehls†,
J. Oehlschläger†, S. Ostapchenko†,xiv, S. Over∥, M. Petcu¶, T. Pierog†, H. Rebel†,
M. Roth†, H. Schieler†, F. Schröder†, O. Sima‡‡, M. Stümpert‡, G. Toma¶,
G.C. Trinchero**, H. Ulrich†, W. Walkowiak∥, A. Weindl†, J. Wochele†, M. Wommer†,

*Soltan Institute for Nuclear Studies, 90950 Lodz, Poland
†Institut für Kernphysik, Forschungszentrum Karlsruhe, 76021 Karlsruhe, Germany
‡Institut für Experimentelle Kernphysik, Universität Karlsruhe, 76021 Karlsruhe, Germany
§Dipartimento di Fisica Generale dell'Università, 10125 Torino, Italy
¶National Institute of Physics and Nuclear Engineering, 7690 Bucharest, Romania
∥Fachbereich Physik, Universität Siegen, 57068 Siegen, Germany
**Istituto di Fisica dello Spazio Interplanetario, INAF, 10133 Torino, Italy
††Fachbereich Physik, Universität Wuppertal, 42097 Wuppertal, Germany
‡‡Department of Physics, University of Bucharest, 76900 Bucharest, Romania
xi now at: Universidad Michoacana, Morelia, Mexico
xii now at: Universidade São Paulo, Instituto de Fisica de São Carlos, Brasil
xiii now at: Dept. of Astrophysics, Radboud University Nijmegen, The Netherlands
xiv now at: University of Trondheim, Norway

*Abstract*. The Muon Tracking Detector (MTD) in KASCADE-Grande allows to measure with high accuracy muon directions in EAS up to 700 m distance from the shower center. According to the simulations this directional information allows to study longitudinal development of showers by means of such quantities like muon radial angles and, derived out of radial and tangential angle values, muon pseudorapidities. Shower development depends on the hadronic interactions taking place in the atmosphere, therefore, such study is a good tool for testing interaction models embedded in the Monte-Carlo shower simulations.

Sensitivity of the muon radial angles and their pseudorapidities to the shower development will be discussed and examples of measured distributions will be shown. Experimental results will be compared with simulation predictions showing the possibility to validate hadronic interaction models with the MTD data.

*Keywords*: KASCADE-Grande, muon tracking,

## I. INTRODUCTION

The Muon Tracking Detector (MTD) [1], registering muons above an energy threshold 800 MeV, is one of the detector components in the KASCADE-Grande EAS experiment [2] operated on site of the Research Center Karlsruhe in Germany by an international collaboration (see Fig.1). The directions of muon tracks in EAS are measured by the MTD with excellent angular resolution of $\approx 0.35°$. These directional data allow to investigate the longitudinal development of the muonic component in showers which is a signature of the development of the hadronic EAS core. Among the various EAS components there are only four, that are the penetrating ones: optical photons, muons, neutrinos and radio emission. As a result of their penetrating ability they provide practically undisturbed information about their origin. Out of these four, optical photons (of eV energy or smaller) as the most numerous particles, have been used most successfully so far (e.g. ref. [3]) in the study of the longitudinal shower development of individual showers. Muon information has usually been integrated over a large sample of showers and over the whole longitudinal profile.

However, muons have some advantage compared with optical photons and the radio emission: they reflect the development of the nuclear cascade with no mediation from the electromagnetic part of the shower. They are also "seen" the whole day long, not only on clear moonless nights. This feature they share with the EAS radio emission. Evident disadvantage of muons is that they are less numerous than photons and are therefore subject to large fluctuations. Moreover, being charged particles they are subjected to deflection in the geomagnetic field. Therefore, attempts to use them as an independent source of information on EAS development were rather rare in the past, but now, with the development of such sophisticated detectors as the MTD, they become more feasible and gain importance. On the other hand, muons have never been used up to now to reconstruct the hadron longitudinal development of EAS with sufficient accuracy, due to the difficulty of building large area ground-based muon telescopes.

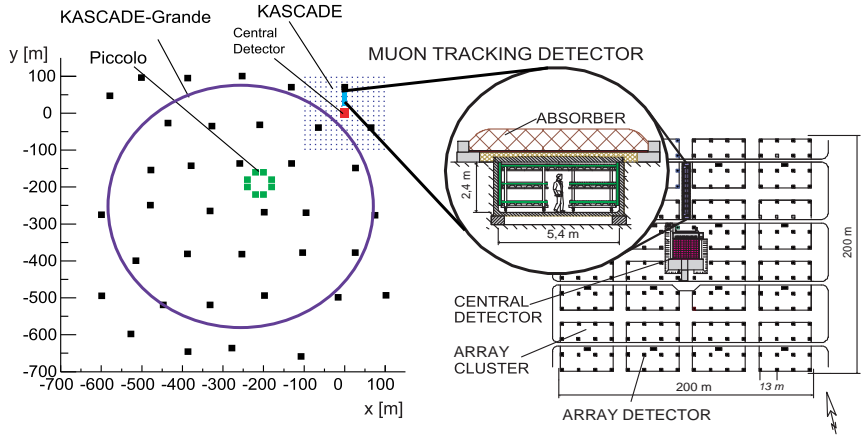

Fig. 1: Layout of the KASCADE-Grande experiment distributed over the Research Center Karlsruhe. KASCADE is situated in the North-East corner of the Center: note the position of the Muon Tracking Detector.

Muons are produced mainly in decay processes of charged pions and kaons - most numerous products of the hadronic interations driving the development of EAS. The longitudinal profile of the hadronic cascade depends on the primary mass, and thus, can be used for testing the hadronic interaction models.

The most straightforward method of investigation of the longitudinal shower development is to reconstruct the muon production heights by means of triangulation [4], [5]. Results of such a research are presented on this conference by P. Doll et al. [6]. The longitudinal development of a shower has its imprint also in the lateral distribution of muon densities, presented on this conference by P. Łuczak et al. [7]. Here we will show, that the directional data of muons in EAS obtained with the MTD can be used to reconstruct such quantities like radial ($\rho$) and tangential ($\tau$) angles and, as a next step, muon pseudorapidies [8], which are also sensitive to the longitudinal shower development. Therefore they can also serve to validate hadronic interaction models used in Monte-Carlo EAS simulations [9].

## II. RADIAL, TANGENTIAL ANGLES, AND MUON PSEUDORAPIDITY

Investigation of muons registered in the MTD is based on the two ortogonal projections of the muon angle in space with respect to the shower axis direction, namely the radial ($\rho$) and tangential ($\tau$) angles. Their definition is given in Fig. 2 and their properties are discussed in [8]. Here we remind only, that the value of muon radial angle

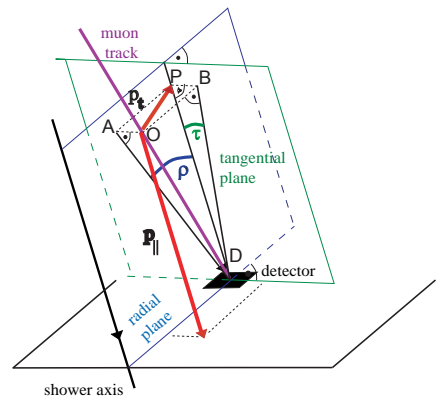

Fig. 2: Definition of radial ($\rho$) and tangential ($\tau$) angles.

is dominated by the value of the transverse momentum of its parent meson.

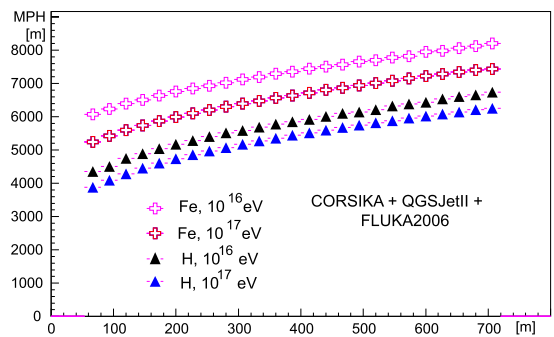

Fig. 3: Lateral distribution of the mean muon production height (MPH) in vertical showers for the two primary particle types and two energies.

In [8] it was also shown that using $\tau$ and $\rho$ one can reconstruct the pseudorapidity of muons in the shower reference system (z-axis parallel to the shower direction):

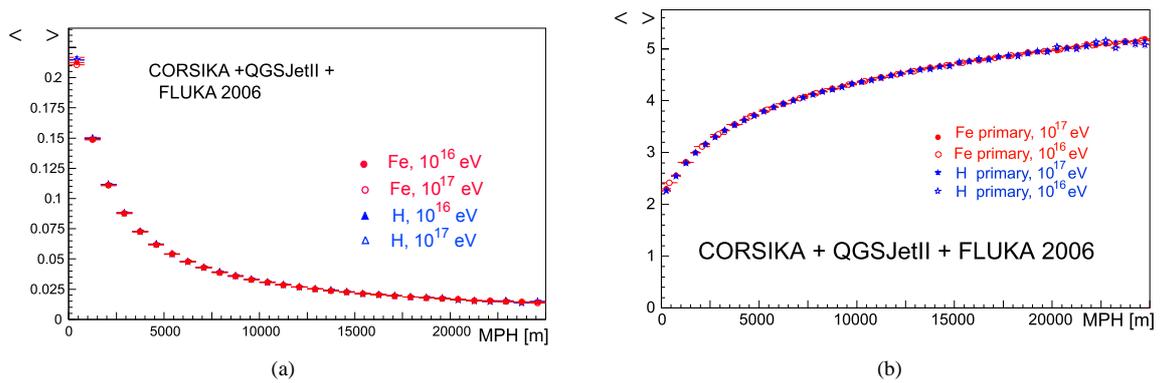

Fig. 4: Mean radial angle (a) and mean pseudorapidity (b) of muons registered in the MTD and produced at a given height above the detector is independent of the primary mass and energy.

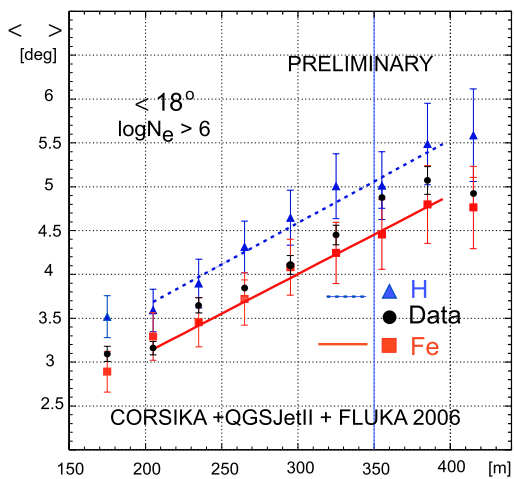

Fig. 5: Reconstructed lateral distribution of the mean radial angle compared with CORSIKA simulation results for proton and iron primaries. Lines are fits to the simulations.

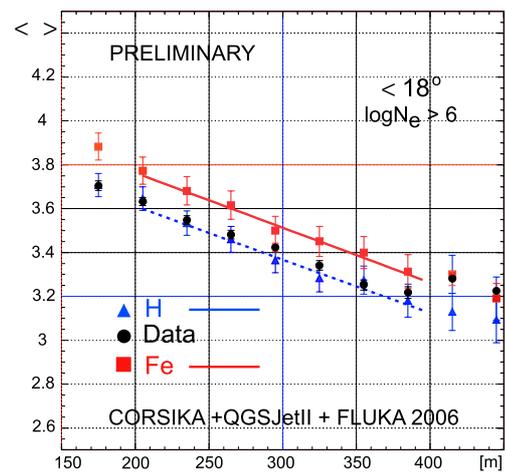

Fig. 6: Reconstructed lateral distribution of the mean muon pseudorapidity compared with CORSIKA simulation results for proton and iron primaries. Lines are fits to the simulations.

$\eta = -\ln\frac{\zeta}{2}$, where $\zeta = \sqrt{\tau^2 + \rho^2}$. This pseudorapidity is closely related with the rapidity of parent mesons [10], thus being a good tool for testing interaction models.

In Fig. 3 the sensitivity of the MTD to the longitudinal shower development is demonstrated. At each distance to the shower core registered muons have certain average production height (MPH), being dependent on the type of primary and its energy. Muons in proton induced showers are per average produced deeper than in iron showers; with increase of the primary energy mean MPH moves deeper into the atmosphere.

As it is seen from the CORSIKA [11] simulation results shown in Fig. 4 muons from a given production height carry to the observation level a certain mean value of $\rho$ and $\eta$, irrespective to the primary type and energy. Having in mind Fig. 3 we conclude, that radial angles and pseudorapidity of muons in the showers are parameters sensitive to the longitudinal shower development. Therefore, they can be used to test the hadronic interaction models.

### III. EXPERIMENTAL DATA AND SIMULATION RESULTS

The MTD data collected in the period from March 2004 to November 2008 have been used to reconstruct muon mean radial angles and mean muon pseudorapidities. Vertical showers ($\theta \leq 18°$) with the size $\lg N_e > 6$ have been selected. The fiducial area for core positions was: -550 m $\leq$ x$_{core}$ $\leq$ 50 m and -580 m $\leq$ y$_{core}$ $\leq$ 20 m.

Then the lateral distributions of those two quantities have been obtained and compared with the results reconstructed out of the simulated data for two primary species: proton and iron. The simulations were done with CORSIKA code using the QGSJETII model [12] for high energy interactions above 200 GeV and FLUKA2006 [13] below that energy.

In Fig. 5 experimental and simulated radial angle lateral distributions are compared. The comparison is done in limited ranges of muon distances to the core, where the saturation effects (seen below 150 m) and trigger inefficiencies (seen above 400 m) are not present. The lines are linear fits to the simulation results. The error bars in the simulations are still too large and the number of simulated data will be increased, thus the results are marked "Preliminary".

We can conclude here that the experimental data is compatible with the CORSIKA simulations done using QGSJETII - FLUKA model combination - data points are in-between the simulated ones. Similar concusions about intrinsic consistency of these models is found in [14]. We can also notice that experimental data tend to be positioned closer to the radial angles in iron initiated showers rather than proton ones.

In Fig. 6 lateral distributions of mean muon pseudorapidity for the same data sets, experimental and simulated, are compared. Here, one can also conclude that the experimental data points are bracketed by the simulated distributions showing compatibility of the simulations with the experiment, same as it is in the case of radial angles, discussed above.

However, there is one striking difference in this figure compared to Fig. 5. The data points here are closer to the proton simulation results rather than to the iron ones, as it is in the case of radial angle distribution. And for the shower sizes in our investigation one would really expect the result being shown by the radial angle distributions (at primary energies in the region of $10^{16}eV$ and above rather heavier than lighter composition is seen in the analyses of many other shower parameters).

This difference may be an indication of the features of the models. Mean radial angle distribution (Fig. 5) suggests that the transverse momentum of pions produced in hadronic interactions is reproduced by the models in a way close to the reality.

On the other hand, Fig. 6 shows that the rapidity of those pions is in simulations too large, by 0.05 - 0.1 in the mean values.

Mean radial angles and mean pseudorapidities of muons registered by the MTD in a given distance range from the shower core are quantities sensitive to the primary mass (what was shown also in section II).

However, for the investigation of the mass composition (e.g. in terms of the determination of the <lnA> parameter) with the model combination used in this research, one should rather wait for the increased statistics of Monte-Carlo simulations.

IV. ACKNOWLEDGEMENTS

The KASCADE-Grande experiment is supported by the BMBF of Germany, the MIUR and INAF of Italy, the Polish Ministry of Science and Higher Education (grant for 2009-2011), and the Romanian Ministry of Education, Research and Innovation. This work was supported in part by the German-Polish bilateral collaboration grant (PPP - DAAD) for the years 2009-2010.

# Lateral distribution of EAS muons measured with the KASCADE-Grande Muon Tracking Detector

P. Łuczak*, W.D. Apel†, J.C. Arteaga‡,xi, F. Badea†, K. Bekk†, M. Bertaina§, J. Blümer†,‡,
H. Bozdog† I.M. Brancus¶, M. Brüggemann‖, P. Buchholz‖, E. Cantoni§,**, A. Chiavassa§,
F. Cossavella‡, K. Daumiller†, V. de Souza‡,xii, F. Di Pierro§, P. Doll†, R. Engel†, J. Engler†,
M. Finger†, D. Fuhrmann††, P.L. Ghia**, H.J. Gils†, A. Haungs†, R. Glasstetter††,
C. Grupen‖, A. Haungs†, D. Heck†, J.R. Hörandel‡,xiii, T. Huege†, P.G. Isar†, K.-H. Kampert††,
D. Kang‡, D. Kickelbick‖, H.O. Klages†, Y. Kolotaev‖, H.J. Mathes†, H.J. Mayer†,
J. Milke†, B. Mitrica¶, C. Morello**, G. Navarra§, S. Nehls†, J. Oehlschläger†,
S. Ostapchenko†,xiv, S. Over‖, M. Petcu¶, T. Pierog†, H. Rebel†, M. Roth†,
H. Schieler†, F. Schröder†, O. Sima‡‡, M. Stümpert‡, G. Toma¶, G.C. Trinchero**,
H. Ulrich†, W. Walkowiak‖, A. Weindl†, J. Wochele†, M. Wommer†, J. Zabierowski*

*Soltan Institute for Nuclear Studies, 90950 Lodz, Poland
†Institut für Kernphysik, Forschungszentrum Karlsruhe, 76021 Karlsruhe, Germany
‡Institut für Experimentelle Kernphysik, Universität Karlsruhe, 76021 Karlsruhe, Germany
§Dipartimento di Fisica Generale dell'Università, 10125 Torino, Italy
¶National Institute of Physics and Nuclear Engineering, 7690 Bucharest, Romania
‖Fachbereich Physik, Universität Siegen, 57068 Siegen, Germany
**Istituto di Fisica dello Spazio Interplanetario, INAF, 10133 Torino, Italy
††Fachbereich Physik, Universität Wuppertal, 42097 Wuppertal, Germany
‡‡Department of Physics, University of Bucharest, 76900 Bucharest, Romania
xi now at: Universidad Michoacana, Morelia, Mexico
xii now at: Universidade de São Paulo, Instituto de Fîsica de São Carlos, Brasil
xiii now at: Dept. of Astrophysics, Radboud University Nijmegen, The Netherlands
xiv now at: University of Trondheim, Norway

*Abstract*. The KASCADE-Grande Muon Tracking Detector (MTD) allows to measure with high accuracy directions of EAS muons with energy above 0.8 GeV up to 700 m distance from the shower center. Lateral distribution of muon densities reflects the longitudinal development of the muonic shower component, thus comparison of experimental distributions from different detectors, as well as with the simulated results, allows to check the contemporary understanding of shower physics. Experimental results for EAS muons above 0.8 GeV obtained for the first time with the tracking detector in a wide range of distances from the core will be shown. They will be compared with the lateral distributions of muons above 0.23 GeV, measured with KASCADE Array muon scintillation counters. Comparison with the simulation results will also be shown.

*Keywords*: KASCADE-Grande, Muon Tracking Detector, lateral muon density distributions

## I. INTRODUCTION

Ivestigations of muonic component in Extensive Air Shower (EAS) is of a primary importance for understanding air shower physics. Muons carry to the observation level nearly undistorted information about their parent particles: pions and kaons, which are the most numerous products of hadronic interactions responsible for the development of the shower in the atmosphere.

A perfect tool for such investigations is the KASCADE-Grande EAS experiment [1], being an extension of the KASCADE experimental setup [2]. It is a multi-detector system located on site of the Research Centre (Forschungszentrum) Karlsruhe in Germany at 110 m a.s.l., measuring all three EAS components: hadrons, electrons and muons (at 4 energy thresholds) in a wide range of distances (up to 700 m) from the shower core, and primary particle energies ($5 \times 10^{14}$–$10^{18}$ eV). High precision measurements of particle densities and tracks, the latter by means of a dedicated Muon Tracking Detector (MTD) [3] - at different energy thresholds allow to investigate many features of EAS and are the basis for multiparameter analyses (e.g.: [4], [5]). These features of KASCADE-Grande make it also to a very good test field for the development of other shower detection techniques, like radio detection (LOPES [6]).

## II. KASCADE-GRANDE

### A. The KASCADE experiment

The KASCADE experiment (Fig.1) consists of several detector systems. A description of the performance of the experiment can be found elsewhere ([2]). An array of 252 detector stations 200 m × 200 m (called the Array), is organized in a square grid of 16 clusters, and equipped with scintillation counters, which measure the electromagnetic (threshold 5 MeV) and in the outer 12 clusters, below a lead iron shielding imposing the energy threshold of 230 MeV, also the muonic parts of EAS. In its centre, a 16 m × 20 m iron sampling calorimeter (a main part of the Central Detector) detects the hadrons in the shower core [7].

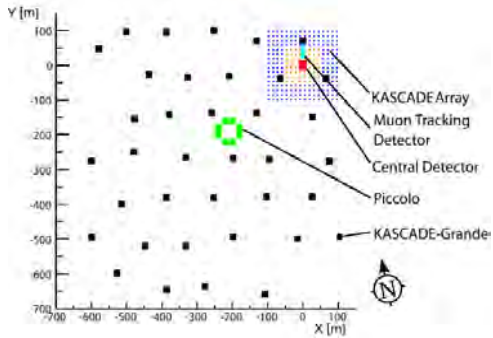

Fig. 1: The layout of the KASCADE-Grande experiment.

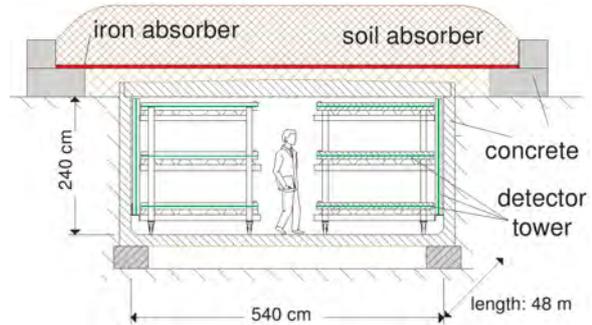

Fig. 2: Cross-section of the Muon Tracking Detector tunnel.

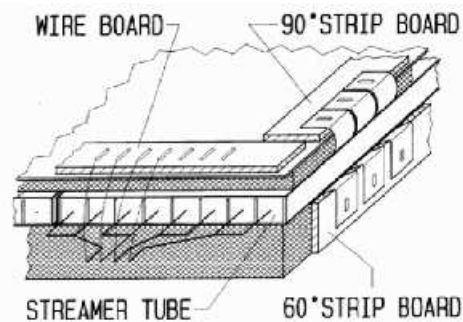

Fig. 3: The MTD module design.

Muon detectors located in the third gap of the calorimeter provide a trigger for the calorimeter and additional information about the lateral and time distribution of muons (above 490 MeV energy) near the shower core [2], [8]. Underneath the calorimeter two layers of multi-wire proportional chambers (MWPC) are used to measure tracks of muons with energy above 2.4 GeV. In the northern part of the KASCADE Array the 128 m$^2$ large Muon Tracking Detector is situated.

*B. Grande part of the experiment*

Grande is an extension of the KASCADE Array. It is an array of 37 detector stations organized in a hexagon grid of 18 clusters covering an area of 0.5 km$^2$. Each station contains 10 m$^2$ of plastic scintillators for registration of charged particles. In the centre there is a small trigger array of plastic scintillation stations, called Piccolo, build to provide additional trigger for the MTD and other KASCADE components.

## III. THE MUON TRACKING DETECTOR

The Muon Tracking Detector is installed below ground level in a concrete tunnel. Under the shielding of 18 r.l., made out of concrete, sand and iron (Fig.2), 16 muon telescopes (called detector towers) register tracks of muons which energy exceeds 800 MeV. Each tower contains limited streamer tube (ST) detector modules: three horizontal and one vertical. All towers are connected with a gas supply system, high voltage and electronic chain readout system.

Each ST chamber houses 16 anode copper-beryllium wires in two cathode comb profiles, extruded for eight parallel ST cells of 9×9 mm$^2$ cross-section and 4000 mm length.

In the MTD an efficient chain-type readout system is used. Front-end electronics boards, mounted to the detector modules are acquiring signals from wires and strips. Each of three wire and nine strip boards in a module creates digital signals being used to reconstruct the tracks. Information from all modules, under certain trigger condition, is send to the acquisition system. Detailed information about the design of the MTD may be found in [3] and [9].

When a particle is passing through the modules of the tower it ionizes the gas in the streamer tubes and a streamer is created. As a result we have a large increase of charge in a small volume of the tube. This charge is inducing a certain charge in the aluminum strips above and below the tubes (perpendicular and diagonal, 60° with respect to the wires), respectively (Fig.3). Coincidence of the signal from the wires and strips in each layer is called a hit. The tracks are reconstructed out of three or two hits, in three or two modules, respectively. The algorithm is first searching for three hit tracks and the remaining hits are used next to create two hit tracks out of them.

## IV. TRACKING MUONS IN EAS

Combined information of the muon tracks, direction of the shower axis and the shower core position allows to investigate the muonic component of the EAS more precisely than it is done with the scintillator array alone. With the MTD we count muons and, in addition, have very precise (better than 0.3°) information about their directions. This allows to investigate the longitudinal development of the muon component, and due to its close relation to EAS hadrons, the development of showers themselves. This investigation is done by studying quantities derived from the experimental data, like mean muon production height [10] and shower muon pseudorapidities [11]. The way shower develops in the atmosphere (and its muon component in particular) leaves its imprint in the lateral distributions of muons – also a subject of our investigations with the MTD data.

## V. LATERAL MUON DENSITY DISTRIBUTIONS

Lateral distribution of EAS particles is an important characteristic of the shower cascade in the atmosphere. In particular, such distributions of EAS muons, being closely related to the hadronic shower component, are a good tool to test the quality of experimental detector setup and our understanding of shower physics. Therefore, every EAS experiment, equipped with sufficiently large muon detectors, provides such distributions. Also KASCADE experiment has done so [4] and first preliminary distributions from KASCADE-Grande were reported [12], [13].

Usually results were obtained with arrays of shielded scintillator detectors, the most popular device in EAS experiments. With the MTD in KASCADE-Grande, for the first time with high angular resolution, it is possible to obtain lateral distributions of muons registered with the tracking devices, like limited streamer tube telescopes. Muon numbers (muon densities) are obtained by counting particle tracks instead of measuring energy deposits, as it is the case with shielded scintillator arrays.

### A. Selection of events

This analysis is based on the showers measured in a period from March 2004 till November 2008 fulfilling the following conditions:
1) All clusters in the KASCADE-Grande array and the MTD work properly,
2) Reconstruction of shower parameters from Grande array was succesful,
3) Zenith angle of the shower $\Theta \leq 18°$,
4) Shower core was reconstructed in fiducial area where $x_{core} \in \langle -550\text{m}; 50\text{m} \rangle$ and $y_{core} \in \langle -580\text{m}; 20\text{m} \rangle$.

### B. Calculation of the number of tracks and the area of the MTD

The detector area is divided into 30 meter radial bins around the reconstructed shower core position (see Fig.4). Muon tracks are reconstructed from hits in two or three MTD modules and the position of each hit is known. Distance from the hit in the middle module to the shower axis is the muon distance.
The area of the detector in each distance bin is calculated in the following way:
From very precise measurements the position of every wire pair and perpendicular strip is known. Point where the wire pair is crossing the perpendicular strip is a centre of a basic detection unit (cell) in the MTD. Each cell has constant area of $\sim 4 \text{ cm}^2$. Distance of each cell to the shower axis is being calculated and the number of cells is accumulated in each distance bin. This number of cells gives information about the detector area in that bin. The number of muon tracks in each distance bin is corrected for the reconstruction efficiency. The track reconstruction efficiency is calculated from the following formula (1):

$$\epsilon = \frac{1}{1 + \frac{N_{tr2}}{3 \cdot N_{tr3}}} \quad (1)$$

where $N_{tr2}$ and $N_{tr3}$ are two and three hit tracks respectively. Becasue of reconstruction procedure two and three hit tracks are not independent and it is necessary to introduce a proper correction factor $k$, given by the formula (2):

$$k = \frac{1}{3 \cdot \epsilon^3 + 2 \cdot \epsilon^2} \quad (2)$$

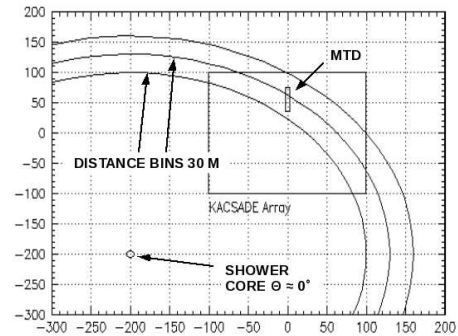

Fig. 4: 30 meter radial bins divide the MTD into parts (from one to three). In each bin the number of muons and the detector area is calculated.

Typicaly $\epsilon = 0.74$ and $k = 0.4$
The density $\rho_i$ in each distance bin is calculated as a sum of all muons from all showers corrected for reconstruction efficiency being divided by detector area corrected for zenith angle ($A_{MTD}$).

$$\rho_i = \frac{\sum_{j=1}^{N_s}(N_{tr2}^j + N_{tr3}^j) \cdot k}{\sum_{j=1}^{N_s} A_{MTD}^{j,i}} \quad (3)$$

where $i$ is distance bin number, $N_s$ is number of showers, $A_{MTD}^{j,i}$ is detector area in $i^{th}$ distance bin for $j^{th}$ shower.

In Fig.5 the preliminary results for the lateral muon density distributions are presented in four muon size bins: from $\lg(N_\mu) > 4.9$ to $\lg(N_\mu) < 6.1$. $N_\mu$ is derived from muon densities measured with KASCADE muon detectors and the above mentioned range roughly corresponds to primary energies from $10^{16}$ eV to $10^{17}$ eV. Together with the MTD results, represented by symbols, the lateral distributions based on the number of muons reconstructed out of energy deposits in shielded plastic scintillators of the KASCADE Array (represented by lines) are given. One can notice that the presented distributions can be compared in limited distance range (marked by full symbols and solid lines for the MTD and KASCADE distributions respectively). It is due to saturation effects in the MTD when the core of the shower, initiated by high energy particle, is close to the detector. At large distances the experimental setup is not able to provide an efficient trigger. The absolute values of muon densities for both muon energy thresholds (230 MeV for the KASCADE Array and 800 MeV for the MTD) are still preliminary. Systematic uncertainties (e.g shower and track reconstruction accuracies) and efficiency corrections are under investigation. However, general shape of the distributions has been already established. It can be fitted with a Lagutin-like function (4) [15],[16]. In case of the lower energy muons the function is of the form :

$$\begin{aligned} f(r) &= 0.28 \cdot r_0^{-2} \, (r/r_0)^{-0.69} \, (1 + r/r_0)^{-2.39} \\ &\quad \times \left(1 + (r/(10 \cdot r_0))^{-2}\right)^{-1} \end{aligned} \quad (4)$$

where $r_0$=320 m. For the higher energy muons registered by the MTD the distribution is steeper and can be described by similar Lagutin-like function where $r_0$ is smaller.

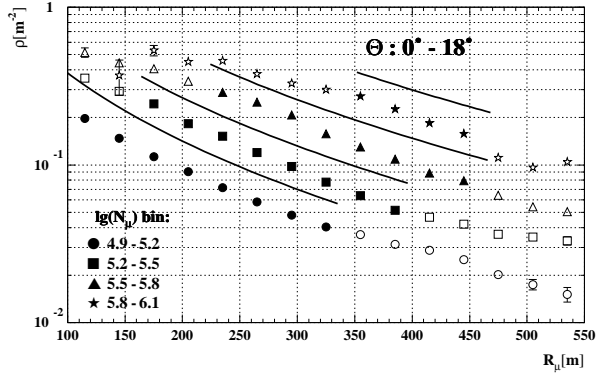

Fig. 5: Lateral muon density distributions obtained with the MTD (symbols) and with the KASCADE Array muon detectors (lines) in four muon size bins (see text).

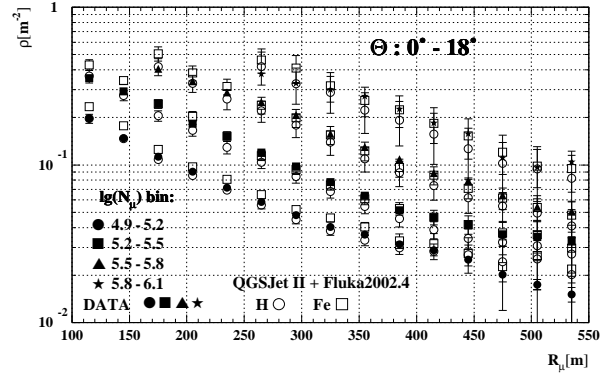

Fig. 6: Lateral muon density distributions obtained with the MTD (solid symbols) and simulations (see text).

In Fig.6 comparison of the MTD distribution with CORSIKA [14] simulations of proton and iron primaries is shown. In muon size bin $\lg(N_\mu)$ from 4.9 to 5.2 the data are between simulations. In higher bins ($\lg(N_\mu)$ from 5.2 to 6.1) the data, in the distance ranges where the MTD results can be compared with KASCADE, seem to lie on top of iron distributions. Close and far away from the shower core the data points have tendency to lie on top of proton distributions. This is due to differences in track reconstruction in data and simulations. However, within our accuracies, the experimental distributions are in a good agreement with simulations.

## Acknowledgements

The KASCADE-Grande experiment is supported by the BMBF of Germany, the MIUR and INAF of Italy, the Polish Ministry of Science and Higher Education (grant for 2009-2011), and the Romanian Ministry of Education, Research and Innovation. This work was supported in part by the German-Polish bilateral collaboration grant (PPP - DAAD) for the years 2009-2010.

# Muon Production Height and Longitudinal Shower Development in KASCADE-Grande

P. Doll[*], W.D. Apel[*], J.C. Arteaga[†,xi], F. Badea[*], K. Bekk[*], M. Bertaina[‡], J. Blümer[*,†], H. Bozdog[*]
I.M. Brancus[§], M. Brüggemann[¶], P. Buchholz[¶], E. Cantoni[‡,‖], A. Chiavassa[‡], F. Cossavella[†],
K. Daumiller[*], V. de Souza[†,xii], F. Di Pierro[‡], R. Engel[*], J. Engler[*], M. Finger[*],
D. Fuhrmann[**], P.L. Ghia[‖], H.J. Gils[*], R. Glasstetter[**], C. Grupen[¶], A. Haungs[*],
D. Heck[*], J.R. Hörandel[†,xiii], T. Huege[*], P.G. Isar[*], K.-H. Kampert[**], D. Kang[†],
D. Kickelbick[¶], H.O. Klages[*], Y. Kolotaev[¶], P. Łuczak[††], H.J. Mathes[*], H.J. Mayer[*],
J. Milke[*], B. Mitrica[§], C. Morello[¶], G. Navarra[‡], S. Nehls[*], J. Oehlschläger[*],
S. Ostapchenko[*,xiv], S. Over[¶], M. Petcu[§], T. Pierog[*], H. Rebel[*], M. Roth[*],
H. Schieler[*], F. Schröder[*], O. Sima[‡‡], M. Stümpert[†], G. Toma[§], G.C. Trinchero[§],
H. Ulrich[*], W. Walkowiak[¶], A. Weindl[*], J. Wochele[*], M. Wommer[*], J. Zabierowski[††]

[*]*Institut für Kernphysik, Forschungszentrum Karlsruhe, 76021 Karlsruhe, Germany*
[†]*Institut für Experimentelle Kernphysik, Universität Karlsruhe, 76021 Karlsruhe, Germany*
[‡]*Dipartimento di Fisica Generale dell'Università, 10125 Torino, Italy*
[§]*National Institute of Physics and Nuclear Engineering, 7690 Bucharest, Romania*
[¶]*Fachbereich Physik, Universität Siegen, 57068 Siegen, Germany*
[‖]*Istituto di Fisica dello Spazio Interplanetario, INAF, 10133 Torino, Italy*
[**]*Fachbereich Physik, Universität Wuppertal, 42097 Wuppertal, Germany*
[††]*Soltan Institute for Nuclear Studies, 90950 Lodz, Poland*
[‡‡]*Department of Physics, University of Bucharest, 76900 Bucharest, Romania*
[xi]*now at: Universidad Michoacana, Morelia, Mexico*
[xii]*now at: Universidade São Paulo, Instituto de Fisica de São Carlos, Brasil*
[xiii]*now at: Dept. of Astrophysics, Radboud University Nijmegen, The Netherlands*
[xiv]*now at: University of Trondheim, Norway*

*Abstract*. A large area ($128 m^2$) Muon Tracking Detector (MTD), located within the KASCADE experiment, has been built with the aim to identify muons ($E_\mu$ >0.8GeV) and their directions in extensive air showers by track measurements under more than 18 r.l. shielding. The orientation of the muon track with respect to the shower axis is expressed in terms of the radial- and tangential angles. By means of triangulation the muon production height $H_\mu$ is determined. By means of $H_\mu$, a transition from light to heavy cosmic ray primary particle with increasing shower energy $E_o$ from 1-10 PeV is observed.

*Keywords*: KASCADE-Grande: Muon Production Height

## I. INTRODUCTION

Muons have never been used up to now to reconstruct the hadron longitudinal development of EAS with sufficient accuracy, due to the difficulty of building large area ground-based muon telescopes [1]. Muons are produced mainly by charged pions and kaons in a wide energy range. They must not always be produced directly on the shower axis. Multiple Coulomb scattering in the atmosphere and in the detector shielding may change the muon direction. It is evident that the reconstruction of the longitudinal development of the muon component by means of triangulation [2], [3] provides a powerful tool for primary mass measurement, giving an information similar to that obtained with the Fly's Eye experiment, but in the energy range not accessible by the detection of fluorescence light. Muon tracking allows also the study of hadron interactions by means of the muon pseudorapidity [6]. Already in the past, analytical tools have been developed which describe the transformation between shower observables recorded on the observation level and observables which represent directly the longitudinal shower development [4].

Fig. 1 shows the experimental environment. Measured core position distributions for showers inside KASCADE range from 40m-140m and inside Grande from 140m-360m. These core positions stay away from the MTD more than 40m for KASCADE for shower energies $\sim 10^{14} eV - 10^{16} eV$ and more than 140m for Grande for shower energies $\sim 10^{16} eV - 10^{18} eV$. The shower core distribution for Grande covers full trigger efficiency in the Grande specific energy range as confirmed by investigations of muon lateral density distributions [8].

## II. MUON PRODUCTION HEIGHT

Usually, $X_{max}$ is the atmospheric depth at which the electrons and photons of the air shower reach their maximum numbers and is considered to be mass A sensitive [9]. Concerning muons which stem dominantly from $\pi^{\pm}$ decays, the corresponding height where most muons are created may also provide a mass A sensitive observable. For $X_{max}$, Matthews [10] in a phenomenological ansatz gives for the e.m. part the elongation rate of $\sim 60 gcm^{-2}$ per decade which is in a good agreement with simulations. For the $X_{max}$ value for nuclei ref. [10] reports: $X_{max}^A = X_{max}^p - X_o ln(A)$ ($X_o$, radiation length in air), therefore, $X_{max}$ from iron showers is $\sim 150 gcm^{-2}$ higher than $X_{max}$ from proton showers at all energies. With the integral number of muons for a proton or nucleus A induced shower:

$$N_\mu \sim E_0^\beta \quad or \quad N_\mu^A \sim A(E_A/A)^\beta \quad (1)$$

we assume that $\langle H_\mu \rangle$ exhibits a similar $lg(N_e)$ and $lg(N_\mu^{tr})$ dependence as $X_{max}$. Note however, $\langle H_\mu \rangle$, because of the long tails in the $H_\mu$ distribution towards large heights can be systematically higher than the muon production height, where most of the muons are created in a shower. Some energetic muons may stem from the first interaction and survive down to the MTD detector plane. The almost mass A independent energy assignment in equation(2) was employed.

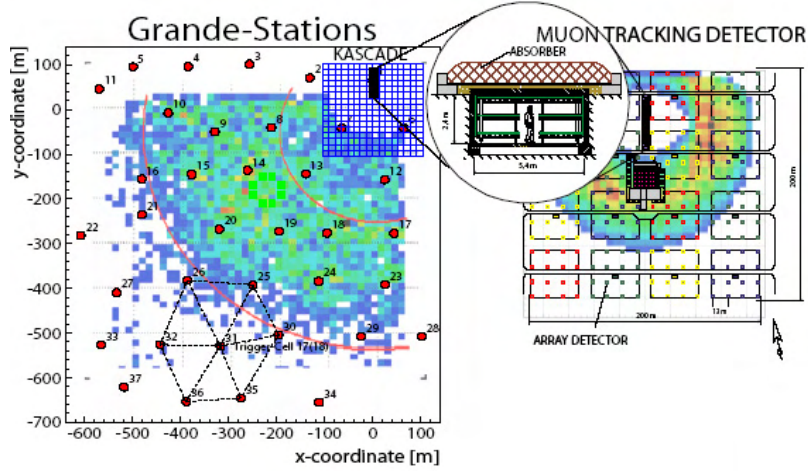

Fig. 1. Layout of the KASCADE-Grande experiment distributed over Research Center Karlsruhe. KASCADE is situated in the North-East corner of the Center: note the position of the Muon Tracking Detector.

$$lgE_0[GeV] = 0.19 lg(N_e) + 0.79 lg(N_\mu^{tr}) + 2.33 \quad (2)$$

The shower development leads also to various fluctuations in those shower parameters.

For the following analysis the elongation rate was given the value $70 gcm^{-2}$ per decade in $lg(N_\mu^{tr})$. After subtracting from each track the 'energy' dependent penetration depth

$$H_\mu^A = H_\mu - 70 gcm^{-2} lg(N_\mu^{tr}) + 20 gcm^{-2} lg(N_e) \quad (3)$$

the remaining depth $H_\mu^A$ may exhibit the mass A dependence. Note the relation $lg(N_\mu^{tr})=lg(N_\mu$-0.55.

The correction with the electron size $lg(N_e)$ in equation (3) should be of opposite sign because of fluctuations to larger size for this variable ($X_{max}$ also fluctuates to larger values).

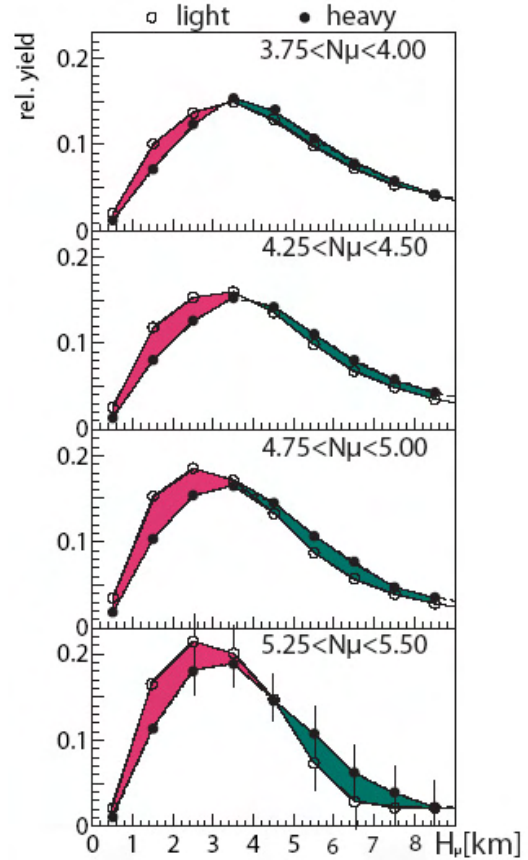

Fig. 2. Muon production height distributions for different muon size bins and different $lg(N_\mu)/lg(N_e)$ ratio above (light) and below (heavy) the solid line in Fig. 5. Colours emphasize the strong mass dependence.

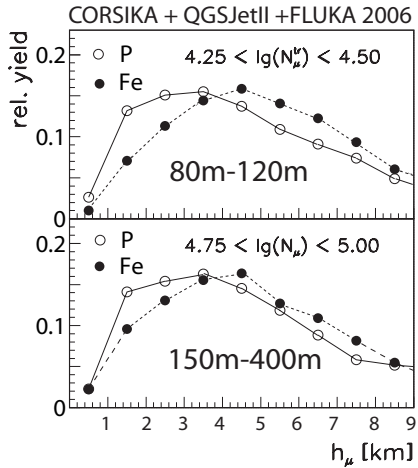

Fig. 3. Simulated muon production height distributions for different muon size bins and the KASCADE (80-120m) and Grande (150-400m) experiment components.

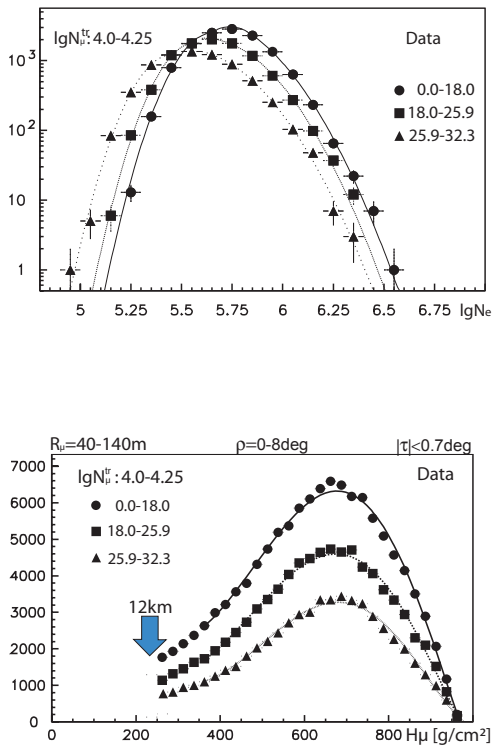

Fig. 4. (Top) Surface shower size $lg(N_e)$ distributions for 3 different angle bins. (Bottom) Muon production depth distributions for 3 different angle bins.

Investigating in a closer look the distribution of the parameters, Fig. 1 shows $H_\mu[km]$ distributions for fixed muon number bins which vary with shower energy.

In Fig. 3 simulated muon production height distributions are shown for different experimental configurations within the geometry of KASCADE (80-120m) and within the geometry of Grande (150-400m). Simulation were done with the CORSIKA code [13] using the QGSJETII model [14] for high energy interactions above 200 GeV and FLUKA2006 [15] below that energy. Because of the shift between $lg(N_\mu^{tr})$ and $lg(N_\mu)$ both distributions should be very similar. When comparing the simulated distributions to the corresponding distribution in Fig. 2, a longer tail towards larger muon production height is observed in the simulations. These tails which may stem from more abundant muon production at high altitudes and are appearing in terms of pseudorapidity [5] at large pseudorapidity.

In Fig. 2 the muon production heights $H_\mu$ are plotted for light and heavy primary mass enriched showers, employing the $lg(N_\mu)/lg(N_e)$ ratio to be larger or smaller than 0.84 as indicated by the solid line in Fig. 5. The distributions exhibit a striking dependence on the primary mass range. Further, it is known from earlier studies, that the $lg(N_e)$ parameter exhibits fluctuations to large values in agreement with simulations while the $lg(N_\mu^{tr})$ parameter exhibits little fluctuations. In contrary, the $H_\mu$ parameter in Fig. 4 is fluctuating to large heights i.e. smaller values ($gcm^{-2}$). Therefore, we may argue that the fluctuations in the corrections for $H_\mu$ for the elongation rate will cancel to some extent and, therefore, the resulting mass A dependent muon production height $H_\mu^A$ represents a stable mass A observable.

Fig. 5 shows the regions of different mass A dependent mean muon production height $\langle H_\mu^A \rangle$ in the 2-parameter $lg(N_e) - lg(N_\mu)$ space. $H_\mu^A$ in Fig. 5 is the mean $\langle H_\mu^A \rangle$ per shower and any muon track in the MTD. The picture shows regions of distinct $\langle H_\mu^A \rangle$ in a colour code with a $40 gcm^{-2}$ step size. The borders between different regions are for some cases marked with lines which exhibit a slope in the $lg(N_e) - lg(N_\mu^{tr})$ plane. While in the middle of the distribution the slope confirms the previously employed slope $lg(N_\mu) = 0.84(\pm 0.01) lg(N_e)$ for selecting light or heavy primary particles, modified slopes may be recognized for regions away from the middle of the ridge. The slope for the $600 gcm^{-2}$ line comes close to the slope of the air-shower simulations employed in [11]. Note also that the number of tracks increase with energy and exhibit a specific mass A dependent rise, which is under study.

The lines obtain their slope from the muon number-energy relation in equation (1) combined with equation (2). There, the exponent is according to ref. [10] connected to the amount of inelasticity $\kappa$ (fraction of energy used up for $\pi$ production) involved in the processes of the A-air collisions. A comparatively steeper slope $\beta =$ $(1-0.14\kappa)$ [10], corresponds to an increased inelasticity. The correction in equation (3) depending on $lg(N_e)$ and $lg(N_\mu^{tr})$ was found appropriate to get the slope of the $H_\mu^A$ profile in the 2-parameter $lg(N_e) - lg(N_\mu)$ presentation (Fig. 5). Differences between two different models in ref. [11] amount to about $20 g cm^{-2}$ on the $H_\mu^A$ scale.

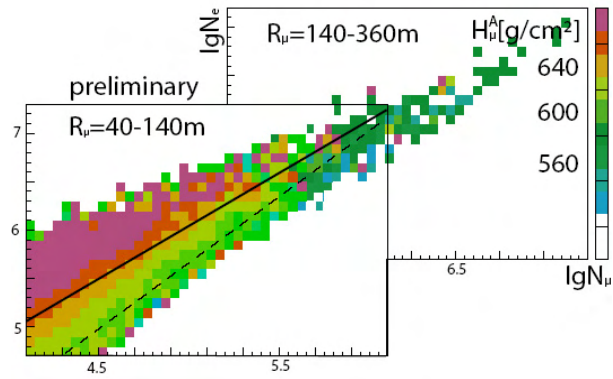

Fig. 5. Effective muon production depth $H_\mu^A$ in the 2-parameter presentation $lg(N_e) - lg(N_\mu)$ for $0^o - 18^o$. Pictures are overlayed for separate KASCADE and Grande analyses, respectively.

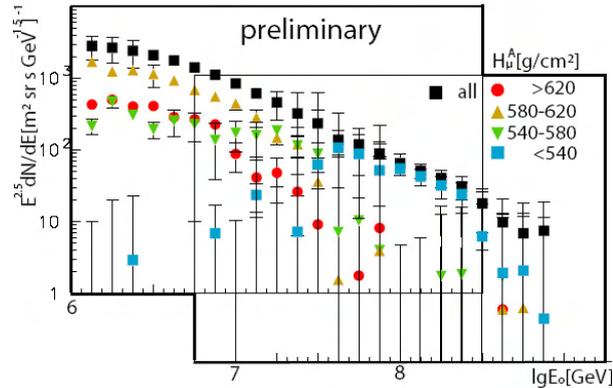

Fig. 6. Energy spectra for different effective muon production depth $H_\mu^A$ for $0^o - 18^o$. Pictures are overlayed for separate KASCADE and Grande analyses, respectively.

Sorting the $lg(N_e) - lg(N_\mu^{tr})$ events by their range in $H_\mu^A$ and employing for the same event the mass A independent equation (2) for KASCADE and a corresponding equation for Grande [7] for $lgE_o[GeV]$, energy spectra are obtained and given in Fig. 6. Sofar, no explicit mass range assignment is given as would be motivated by the equation $X_{max}^A = X_{max}^p - X_o ln(A)$. The spectra in Fig. 6 together with their preliminary error estimations are almost model independent. The preliminary spectra reveal distinct features. While low mass spectra show a rapid drop with increasing shower energy, medium mass and heavy mass spectra seem to overtake at large primary energy. Systematic errors dominate the low and high energy bins for KASCADE and Grande, respectively, and are subject of further investigations. In the KASCADE analysis the detection threshold of the MTD may be effective and a fraction of tracks may be missing leading to a light particle mass interpretation. For the large Grande geometry some flux loss for low energy muons may lead to a bias towards large primary mass.

### III. CONCLUSIONS

Triangulation allows to investigate $H_\mu$. Future analysis of other shower angle bins and a larger and improved quality data sample will provide a more detailed information on the nature of high energy shower muons. Also muon multiplicities provide valuable parameters to derive the relative contributions of different primary cosmic ray particles. A natural extension towards even larger shower energies is provided by KASCADE-Grande [12]. There is a common understanding that the high energy shower muons serve as sensitive probes to investigate [5], [6] the high energy hadronic interactions in the EAS development. Very inclined muons which can be studied with tracks recorded by the wall modules of the MTD are currently of vital interest.

### IV. ACKNOWLEDGEMENTS

The KASCADE-Grande experiment is supported by the BMBF of Germany, the MIUR and INAF of Italy, the Polish Ministry of Science and Higher Education(grant for 2009-2011),and the Romanian Ministry of Education and Research (grant CEEX05-D11-79/2005). The support by the PPP-DAAD grant for 2009-2010 is kindly acknowledged.

# Restoring Azimuthal Symmetry of Lateral Density Distributions of EAS Particles

O. Sima[‡‡], W.D. Apel[*], J.C. Arteaga[†,xi], F. Badea[*], K. Bekk[*], M. Bertaina[‡], J. Blümer[*,†],
H. Bozdog[*], I.M. Brancus[§], M. Brüggemann[¶], P. Buchholz[¶], E. Cantoni[‡,‖], A. Chiavassa[‡],
F. Cossavella[†], K. Daumiller[*], V. de Souza[†,xii], F. Di Pierro[‡], P. Doll[*], R. Engel[*], J. Engler[*],
M. Finger[*], D. Fuhrmann[**], P.L. Ghia[‖], H.J. Gils[*], R. Glasstetter[**], C. Grupen[¶],
A. Haungs[*], D. Heck[*], J.R. Hörandel[†,xiii], T. Huege[*], P.G. Isar[*], K.-H. Kampert[**],
D. Kang[†], D. Kickelbick[¶], H.O. Klages[*], P. Łuczak[††], C. Manailescu[‡‡], H.J. Mathes[*],
H.J. Mayer[*], J. Milke[*], B. Mitrica[§], C. Morariu[‡‡], C. Morello[‖], G. Navarra[‡], S. Nehls[*],
J. Oehlschläger[*], S. Ostapchenko[*,xiv], S. Over[¶], M. Petcu[§], T. Pierog[*], H. Rebel[*],
M. Roth[*], H. Schieler[*], F. Schröder[*], M. Stümpert[†], G. Toma[§], G.C. Trinchero[‖],
H. Ulrich[*], A. Weindl[*], J. Wochele[*], M. Wommer[*], J. Zabierowski[††]

[*]Institut für Kernphysik, Forschungszentrum Karlsruhe, 76021 Karlsruhe, Germany
[†]Institut für Experimentelle Kernphysik, Universität Karlsruhe, 76021 Karlsruhe, Germany
[‡]Dipartimento di Fisica Generale dell'Università, 10125 Torino, Italy
[§]National Institute of Physics and Nuclear Engineering, 7690 Bucharest, Romania
[¶]Fachbereich Physik, Universität Siegen, 57068 Siegen, Germany
[‖]Istituto di Fisica dello Spazio Interplanetario, INAF, 10133 Torino, Italy
[**]Fachbereich Physik, Universität Wuppertal, 42097 Wuppertal, Germany
[††]Soltan Institute for Nuclear Studies, 90950 Lodz, Poland
[‡‡]Department of Physics, University of Bucharest, 76900 Bucharest, Romania
[xi]now at: Universidad Michoacana, Morelia, Mexico
[xii]now at: Universidade de São Paulo, Instituto de Física de São Carlos, Brasil
[xiii]now at: Dept. of Astrophysics, Radboud University Nijmegen, The Netherlands
[xiv]now at: University of Trondheim, Norway

*Abstract*. The lateral distributions of EAS particles are affected by various kinds of azimuthal asymmetries, which arise from different effects: Geometric effects of mapping the horizontal plane observations onto the shower plane, different attenuation of particles on different sides of inclined EAS and the influence of the geomagnetic field on the particle movement. A procedure is described of minimizing the effects of azimuthal asymmetries of lateral density distributions. It is demonstrated and discussed in context of practical cases of data reconstruction by KASCADE-Grande.

*Keywords*: Extensive air showers; lateral density distribution; azimuthal asymmetry

## I. INTRODUCTION

A crucial observable for the reconstruction and analysis of Extensive Air Showers (EAS) [1] is represented by the lateral distribution of the EAS particles evaluated in the intrinsic shower plane, hereafter called normal plane. In the case of ground arrays like KASCADE-Grande [2] this observable is obtained first by converting (by the use of appropriate Lateral Energy Correction Functions, LECF) the detector signals in particle densities evaluated in the horizontal plane. In a second step the density in the horizontal plane is mapped into the normal plane by applying specific projection techniques. Typically the detectors sample only a small fraction of the EAS particles; information concerning the complete distribution is obtained by using lateral distribution functions (LDFs) fitted to the measured data. The commonly used LDFs assume that the particle density possesses axial symmetry in the normal plane. This assumption greatly simplifies the problem of fitting the LDFs, but its validity should be investigated, especially in the case of arrays which only sample a limited part of the azimuthal dependence of the particle density. The bias is more important for inclined showers and in the case of observables evaluated far from the core, e.g. the density at 500 m, which can be used as an energy estimator [3]. In this context the purpose of this work is to study the asymmetry of the reconstructed particle density in the range of the KASCADE-Grande experiment and to propose practical methods to restore the symmetry in the intrinsic shower plane.

## II. BASIC ORIGIN OF ASYMMETRY

In the absence of the Earth's magnetic field the LDF of shower particles would possess symmetry around the shower axis. Consider an inclined shower and assume for the moment that shower evolution in the vicinity of the ground is negligible. Then in the simplified description of shower particles coming on the surface of a cylinder centered on the shower axis elementary geometrical effects would distort the LDF in the horizontal plane; a simple orthogonal projection of the observed densities from the horizontal plane to the normal plane would restore the symmetry. If the shower particles would come on a surface of a cone, then the simple orthogonal projection would not completely restore the symmetry, because the particles would be projected outwards from the core with respect to their real trajectory in the region above the shower axis and towards the core in the opposite region. In the normal plane in a given radial bin the error of the reconstructed number of particles depends on the balance between the number of particles that are artificially projected into that bin and the number of particles that are artificially removed from the bin due to this imperfection of the orthogonal projection. As a result, close to the core, the reconstructed density in the region above the shower core is artificially decreased and it is artificially increased in the opposite region. Far from the core the effect is reversed.

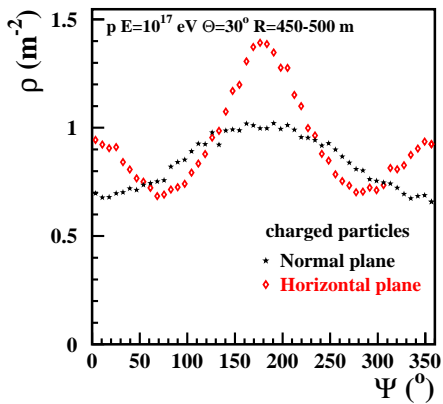

Fig. 1: Charged particle density in the horizontal plane (CORSIKA) and in the normal plane (by orthogonal projection). Coordinate system: $\Psi = 0°$ in the late region, $\Psi = 90°$ along the intersection of the horizontal plane with the normal plane.

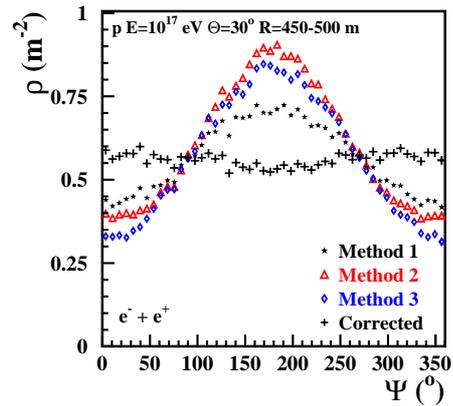

Fig. 2: Electron density reconstructed in the normal plane using the three projection methods, together with the density corrected as detailed in Section IV.

In fact shower evolution should also be taken into account. Due to shower development, the particles hitting the ground below the shower axis (the early region) represent an earlier stage of shower development with respect to the shower particles coming above the shower axis (the late region). This evolution additionally distorts the symmetry around the shower core, for e.g. the particles from the late region have traveled longer paths than the particles from the early region and are more attenuated [4], [5], [6]. The magnetic field of the Earth, producing asymmetry also for vertical showers, is especially important when the densities of particles of opposite charge are compared [7], [8].

### III. PROJECTION IN THE INTRINSIC SHOWER PLANE

In this work we analyzed proton and Fe induced showers with energy E=$10^{17}$, $1.78 \cdot 10^{17}$, $3.16 \cdot 10^{17}$ and $5.62 \cdot 10^{17}$ eV and incidence angle $\Theta$=22, 30 and 45°; proton showers with an extended range of angles ($\Theta$=22, 30, 45, 55 and 65°) were also studied for E=$10^{15}$ eV.

The showers were produced by CORSIKA version 6.01 [9] in the absence of the magnetic field of the Earth.

Clearly the simple model of shower particles coming on cylindrical surfaces with negligible shower development in the vicinity of ground is contradicted by shower simulations. Indeed, the orthogonal projection of the densities from the horizontal plane into the normal plane does not restore axial symmetry (Figure 1).

To investigate further the role played by the imperfection of the method of orthogonal projection (Method 1) and of the shower evolution we applied two other methods of mapping the particle impact point from the horizontal plane to the normal plane: projection along the particle momentum when it reaches the ground (Method 2) and a method based on triangulation using particle arrival time and assuming that the particles have been produced close to the shower axis (Method 3) [10]. Method 2 would be rigorous if the interactions in the space between the horizontal plane and the normal plane would be negligible, while Method 3 requires negligible interactions along the complete trajectory of the particle. The results demonstrate that shower evolution has an important contribution to the asymmetry of LDF, especially in the case of the electron component (Fig. 2). In the case of the muon component the three methods give almost similar results between each other and the amplitude of the early-late variation is smaller, e.g. it is 14% while for electrons it is 72% in the same conditions (Fig. 2).

### IV. CORRECTION FUNCTION

Along the intersection of the horizontal plane with the normal plane ($\Psi = 90°$ and $\Psi = 270°$) the imperfections of the projection method have minimal effects; also shower development between the two planes is negligible. The density $\rho(r, \Psi)$ in the normal plane at other azimuth angles differs from the density $\rho_{ref}(r)$ at the same radial distance and $\Psi = 90°$ or $\Psi = 270°$ due to the imperfections of the projection method and to shower evolution. The magnitude of the effects should depend on the distance $x$ between the corresponding points in the normal and horizontal planes. The dependence of the average density on $x$ is approximated by an exponential function, $\exp(-\lambda x)$. In Fig. 3 this dependence is represented for the case of muons.

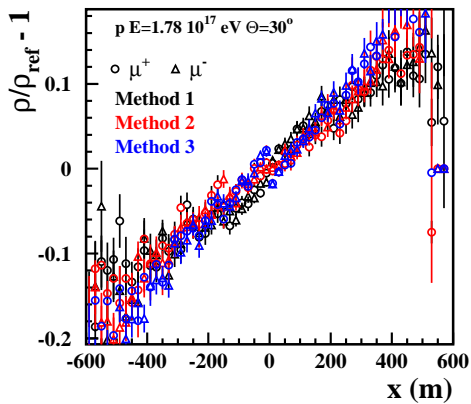

Fig. 3: The dependence of $\rho(r,\Psi)/\rho_{ref}(r) - 1$ on the distance $x$ between the corresponding points in the normal and horizontal planes ($x > 0$ in the early region).

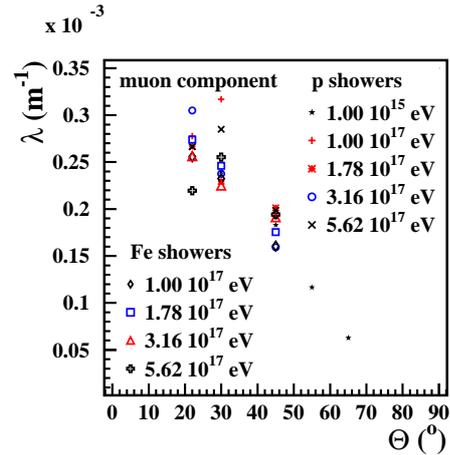

Fig. 5: Attenuation coefficient for the muon component. Note the scale difference between Fig. 4 and Fig. 5.

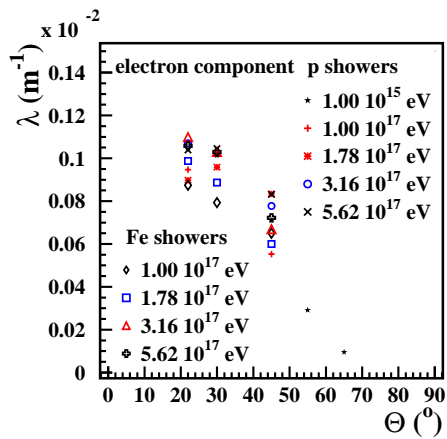

Fig. 4: Attenuation coefficient for the electron component

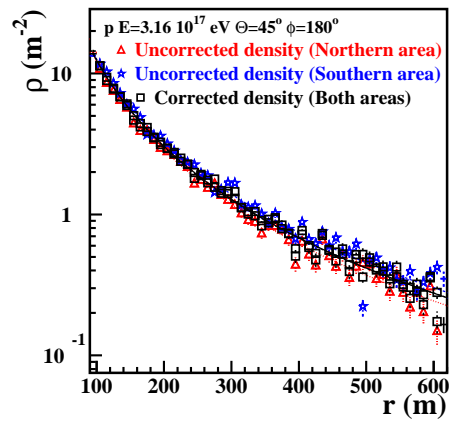

Fig. 6: Comparison of the reconstructed density in the case of showers with the core in the Northern area of Grande with the reconstructed density in the case of showers with the core in the Southern area of Grande

The values of $\lambda$ incorporate both the attenuation by shower development and the distortions due to the projection method. The results show that $\lambda$ depends mainly on the angle of the shower axis. The systematic dependence on the primary energy or composition is less obvious; certainly the sensitivity to these parameters is small. A more refined study [11], [12] shows that the imperfections of Method 1 induce a slight dependence of $\lambda$ on the radial distance from the core: it decreases when the radial coordinate increases from 0 to about 200 m and then remains practically constant. In Figs. 4 and 5 this asymptotic value of $\lambda$ in each set of simulated showers is represented for the case when Method 1 was applied for projection.

## V. RESULTS

The application of the correction procedure greatly removes the asymmetry of lateral distribution (Fig. 2). In order to test the applicability to the KASCADE-Grande experiment, a proton induced shower with E=$3.16 \cdot 10^{17}$ eV, $\Theta = 45°$ incident from North, was repeatedly positioned with the core in various points in the Northern part of KASCADE-Grande, so that most of the Grande detectors were located in the late region of the shower development. The energy deposition in the detectors was realistically simulated, then the density in the observation plane was obtained by applying an appropriate LECF. The density in the normal plane was reconstructed using the Method 1 of projection. The same procedure was applied for a second set of results, obtained in the case when the same shower was repeatedly positioned with the core in various points in the Southern part of the KASCADE-Grande, so that now

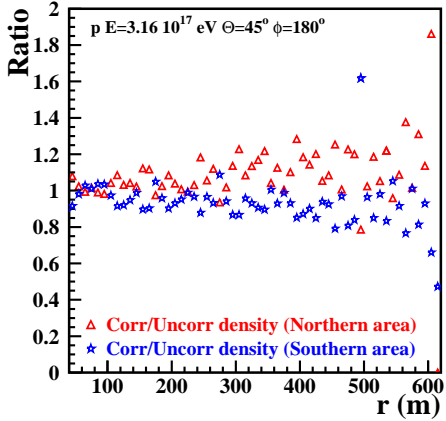

Fig. 7: The ratios of the corrected to the uncorrected densities (proton induced EAS)

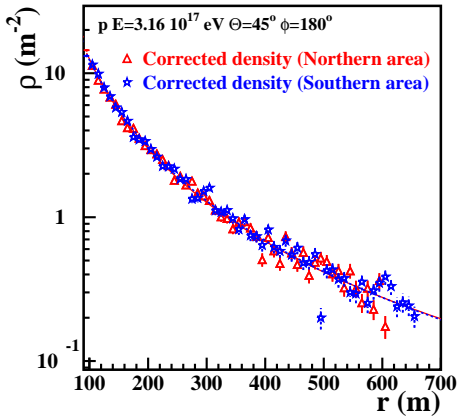

Fig. 8: Corrected lateral density distributions for p induced showers. Linsley fits of the density for showers with the core in the northern area (full line) and southern area (dashed line), respectively

most of the detectors were located in the early region. The reconstructed density in the case of the first set of showers is lower than the reconstructed density in the case of the second set of showers if no correction procedure is applied (Fig. 6).

As can be seen in Fig. 7 the corrections are important, especially at large radial distances. The difference between the mean density at 500 m in the two cases was 23% when the corrections were not applied and negligible when the corrections were applied [11]. After applying the correction method proposed the reconstructed density in the case of the set of showers with the core located in the Northern part of the Grande array does not differ significantly from the reconstructed density in the case of the set of showers with the core located in the Southern part of the array (Fig. 8).

Similar results were obtained in the case of a Fe shower with E=$5.62 \cdot 10^{17}$ eV, $\Theta = 45°$.

ACKNOWLEDGMENT

O. Sima acknowledges the support of the studies by Deutsche Forschungsgemeinschaft. C. Manailescu and C. Morariu thank for the support of the studies from the ERASMUS programme and from KIT (Karlsruhe Institute of Technology).

# Quantitative tests of hadronic interaction models with KASCADE-Grande air shower data

**J.R. Hörandel**[†,xi], **W.D. Apel**[∗], **J.C. Arteaga**[†,xii], **F. Badea**[∗], **K. Bekk**[∗], **M. Bertaina**[‡],
**J. Blümer**[∗,†], **H. Bozdog**[∗], **I.M. Brancus**[§], **M. Brüggemann**[¶], **P. Buchholz**[¶], **E. Cantoni**[‡,∥],
**A. Chiavassa**[‡], **F. Cossavella**[†], **K. Daumiller**[∗], **V. de Souza**[†,xiii], **F. Di Pierro**[‡], **P. Doll**[∗],
**R. Engel**[∗], **J. Engler**[∗], **M. Finger**[∗], **D. Fuhrmann**[∗∗], **P.L. Ghia**[∥], **H.J. Gils**[∗], **R. Glasstetter**[∗∗],
**C. Grupen**[¶], **A. Haungs**[∗], **D. Heck**[∗], **D. Hildebrand**[†,xiv], **T. Huege**[∗], **P.G. Isar**[∗],
**K.-H. Kampert**[∗∗], **D. Kang**[†], **D. Kickelbick**[¶], **H.O. Klages**[∗], **P. Łuczak**[††], **H.J. Mathes**[∗],
**H.J. Mayer**[∗], **J. Milke**[∗], **B. Mitrica**[§], **C. Morello**[∥], **G. Navarra**[‡], **S. Nehls**[∗], **J. Oehlschläger**[∗],
**S. Ostapchenko**[∗,xv], **S. Over**[¶], **M. Petcu**[§], **T. Pierog**[∗], **H. Rebel**[∗], **M. Roth**[∗], **H. Schieler**[∗],
**F. Schröder**[∗], **O. Sima**[‡‡], **M. Stümpert**[†], **G. Toma**[§], **G.C. Trinchero**[∥], **H. Ulrich**[∗],
**A. Weindl**[∗], **J. Wochele**[∗], **M. Wommer**[∗], **J. Zabierowski**[††]

[∗]*Institut für Kernphysik, Forschungszentrum Karlsruhe, 76021 Karlsruhe, Germany*
[†]*Institut für Experimentelle Kernphysik, Universität Karlsruhe, 76021 Karlsruhe, Germany*
[‡]*Dipartimento di Fisica Generale dell'Università, 10125 Torino, Italy*
[§]*National Institute of Physics and Nuclear Engineering, 7690 Bucharest, Romania*
[¶]*Fachbereich Physik, Universität Siegen, 57068 Siegen, Germany*
[∥]*Istituto di Fisica dello Spazio Interplanetario, INAF, 10133 Torino, Italy*
[∗∗]*Fachbereich Physik, Universität Wuppertal, 42097 Wuppertal, Germany*
[††]*Soltan Institute for Nuclear Studies, 90950 Lodz, Poland*
[‡‡]*Department of Physics, University of Bucharest, 76900 Bucharest, Romania*
[xi]*now at: Dept. of Astrophysics, Radboud University Nijmegen, The Netherlands*
[xii]*now at: Universidad Michoacana, Morelia, Mexico*
[xiii]*now at: Universidade de São Paulo, Instituto de Fîsica de São Carlos, Brasil*
[xiv]*now at: ETH Zürich, Switzerland*
[xv]*now at: University of Trondheim, Norway*

*Abstract*. **Quantitative tests of hadronic interaction models are described. Emphasize is given on the models EPOS 1.61 and QGSJET II-2. In addition, a new method to measure the attenuation length of hadrons in air showers is introduced. It turns out that this method is in particular sensitive to the inelastic cross sections of hadrons.**

*Keywords*: **air showers, hadronic interactions, KASCADE-Grande**

## I. INTRODUCTION

Measurements of air shower detectors are usually interpreted with an air shower model to obtain physical properties of the shower inducing primary particles. Modern detector installations, such as the KASCADE-Grande experiment comprise well calibrated particle detectors installed with high spatial density. The systematic uncertainties are dominated by uncertainties of the models used to interpret the data. For air shower interpretation the understanding of multi-particle production in hadronic interactions with a small momentum transfer is essential [1]. Due to the energy dependence of the strong coupling constant $\alpha_s$, soft interactions cannot be calculated within QCD using perturbation theory. Instead, phenomenological approaches have been introduced in different models. These models are the main source of uncertainties in simulation codes to calculate the development of extensive air showers, such as the program CORSIKA [2].

The test of interaction models necessitates detailed measurements of several shower components. The KASCADE experiment [3] with its multi-detector set-up, registering simultaneously the electromagnetic, muonic, and hadronic shower components is particularly suited for such investigations. The information derived on properties of high-energy interactions from air shower observations is complementary to measurements at accelerator experiments since different kinematical and energetic regions are probed.

In the energy range of interest, namely $10^{14}$ to $10^{17}$ eV, the composition of cosmic rays is unknown. Therefore, primary protons and iron nuclei are taken as extreme assumptions and corresponding predictions are calculated for different interaction models. The measured data should be in between the results for the extreme assumptions. If the data are outside the proton-iron range for an observable, this is an indication for an incompatibility of the particular hadronic interaction model with the observed values.

## II. EXPERIMENTAL SET-UP

KASCADE consists of several detector systems [3]. A $200 \times 200$ m² array of 252 detector stations, equipped with scintillation counters, measures the electromagnetic and, below a lead/iron shielding, the muonic parts of air showers. An iron sampling calorimeter of $16 \times 20$ m$^2$ area detects hadronic particles [4]. It has been calibrated with a test beam at the SPS at CERN up to 350 GeV particle energy [5]. For a detailed description of the reconstruction algorithms see [6].

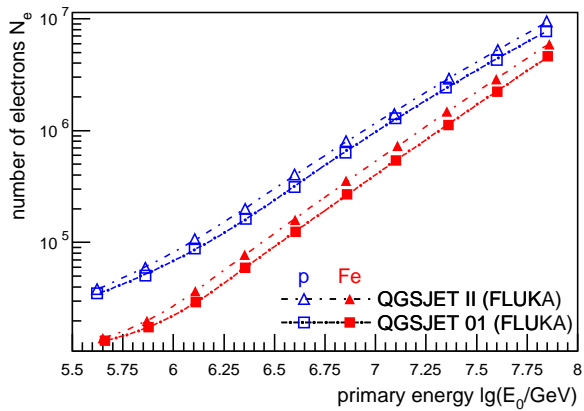

Fig. 1: Number of electrons as predicted by the hadronic interaction models QGSJET II-2 and QGSJET 01 as function of shower energy.

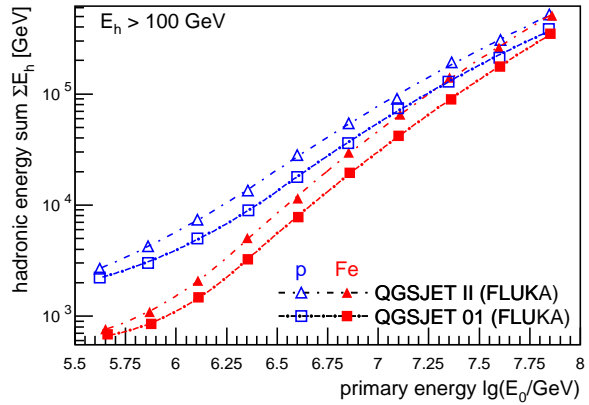

Fig. 2: Hadronic energy sum as predicted by the hadronic interaction models QGSJET II-2 and QGSJET 01 as function of shower energy.

The shower simulations were performed using COR-SIKA. Hadronic interactions at low energies ($E_h < 80$ and 200 GeV, respectively) were modeled using the GHEISHA [7] and FLUKA [8], [9] codes. Both models are found to describe the data equally well [10]. High-energy interactions were treated with different models as discussed below. In order to determine the signals in the individual detectors, all secondary particles at ground level are passed through a detector simulation program using the GEANT package [11]. For details on the event selection and reconstruction, see Ref. [10], [12], [13].

### III. EARLIER TESTS

Several hadronic interaction models as implemented in the CORSIKA program have been systematically tested over the last decade. First quantitative tests [14], [15], [16] established QGSJET 98 [17] as the most compatible code. Similar conclusions have been drawn for the successor code QGSJET 01 [10].

Predictions of SIBYLL 1.6 [18] were not compatible with air shower data, in particular there were strong inconsistencies for hadron-muon correlations. These findings stimulated the development of SIBYLL 2.1 [19]. This model proved to be very successful, the predictions of this code are fully compatible with KASCADE air shower data [20], [21], [10].

Investigations of the VENUS [22] model revealed some inconsistencies in hadron-electron correlations [16]. The predictions of NEXUS 2 [23] were found to be incompatible with the KASCADE data, in particular, when hadron-electron correlations have been investigated [10].

Analyses of the predictions of the DPMJET model yield significant problems in particular for hadron-muon correlations for the version DPMJET 2.5 [24], while the newer version DPMJET 2.55 is found to be compatible with air shower data [10].

Presently, the most compatible predictions are obtained from the models QGSJET 01 and SIBYLL 2.1.

### IV. HADRONIC MODEL EPOS

Recently, predictions of the interaction model EPOS 1.61 [25], [26], [27] have been compared to KASCADE air shower data [12]. This model is a recent development, historically emerging from the VENUS and NEXUS codes. The analysis indicates that EPOS 1.61 delivers not enough hadronic energy to the observation level and the energy per hadron seems to be too small. Most likely, the incompatibility of the EPOS predictions with the KASCADE measurements is caused by too high inelastic cross sections for hadronic interactions implemented in the EPOS code.

These findings stimulated the development of a new version EPOS 1.9 introduced at this conference [28]. Corresponding investigations with this new version are under way and results are expected to be published soon.

### V. HADRONIC MODEL QGSJET II

Also predictions of QGSJET II-2 [29], [30], [31] have been investigated. As discussed above, QGSJET 01 is found to be the most reliable interaction code. Thus, in the following, it serves as reference model and the results can easily be compared to previous publications [16], [10]. The simulations for primary protons and iron nuclei predict about equal numbers of muons as function of energy for QGSJET II and for QGSJET 01. QGSJET II predicts about 20% to 25% more electrons on observation level at a given energy for both primary species relative to QGSJET 01, see Fig. 1. Also the number of hadrons at ground level at a given energy is larger by about 30% to 35% for proton and iron induced showers. The hadronic energy sum and the maximum hadron energy registered at observation level are shown in Figs. 2 and 3, respectively. The values predicted using

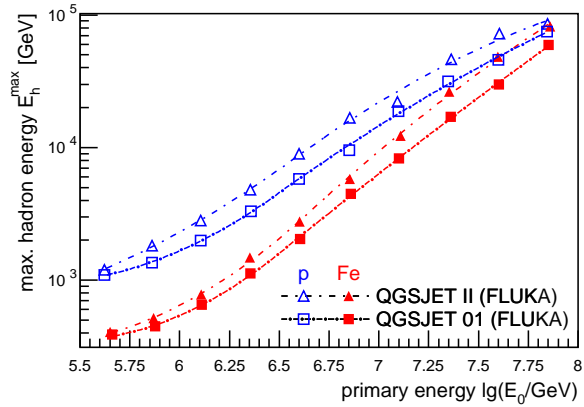

Fig. 3: Maximum hadron energy sum as predicted by the hadronic interaction models QGSJET II-2 and QGSJET 01 as function of shower energy.

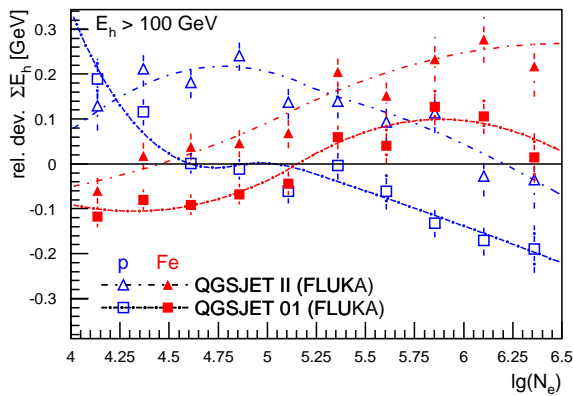

Fig. 4: Relative difference to measured values of the hadronic energy sum as predicted by the models QGSJET II-2 and QGSJET 01.

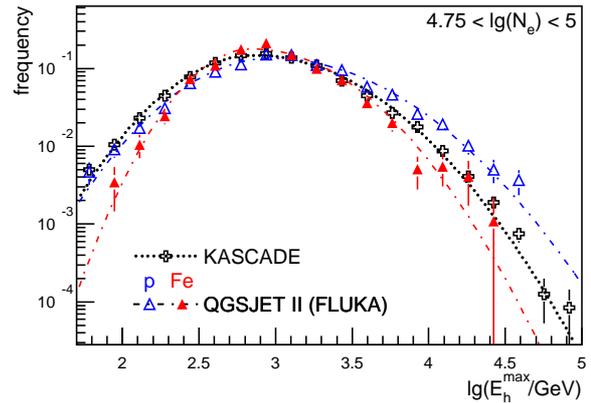

Fig. 5: Energy of the most energetic hadron reconstructed at observation level. Predictions of QGSJET II are compared to measured values.

QGSJET II exceed the ones from QGSJET 01 by a significant amount (up to ≈ 40%), as can be inferred from the figures.

The predicted values have been compared to measured data. Investigating the hadronic energy sum and the maximum hadron energy as function of the registered muon number indicates that the predictions for QGSJET II are compatible with the measurements. The measured values are in between the predictions for the extreme assumptions for proton and iron induced showers. Also the correlation between the hadronic energy sum and the number as hadrons as well as the maximum hadron energy and the number of hadrons are compatible with the measurements.

The situation is different for the correlation between the hadronic energy sum and the number of electrons, see Fig. 4. The figure displays the relative deviation of the predicted values from the measured values, i.e. the quantity $(\sum E_h^{sim} - \sum E_h^{meas})/\sum E_h^{meas}$ is shown. That means the data are at the "zero line". The predictions of QGSJET 01 are compatible with the data, since the values bracket the zero line. On the other hand, the predictions of QGSJET II are above the zero line for both primary species – an unrealistic scenario.

The energy of the most energetic hadron reconstructed at observation level is depicted in Fig. 5 for an electron number interval corresponding to a primary energy of about 1 to 2 PeV. Predictions of simulations according to QGSJET II for primary protons and iron nuclei are compared to measured values. It can be recognized that for high maximum hadron energies the measured values are in between the predictions for proton and iron-induced showers. On the other hand, QGSJET II predicts too few hadrons with low energies. A similar behavior is observed for other electron number intervals.

In summary, the investigations reveal incompatibilities in the hadron-electron correlation for the model QGSJET II-2.

## VI. ATTENUATION LENGTH

Recently, a new method to determine the attenuation length of hadrons in air has been introduced, see Ref. [13]. The energy absorbed in a material within a certain atmospheric depth $X$ is used to define an attenuation length. In this new approach we use the number of electrons $N_e$ and muons $N_\mu$ to estimate the energy of the shower inducing primary particle $E_0$. The energy reaching the observation level in form of hadrons $\sum E_H$ is measured with the hadron calorimeter. The fraction of surviving energy in form of hadrons is defined as $R = \sum E_H/E_0$. The attenuation length $\lambda_E$ is then defined as

$$\Sigma E_H = E_0 \exp\left(-\frac{X}{\lambda_E}\right) \text{ or } R = \exp\left(-\frac{X}{\lambda_E}\right). \quad (1)$$

In contrast to methods using the electromagnetic shower component, the present work focuses directly on measurements of hadrons to derive an attenuation length for this shower component. The values obtained are not a priori comparable to other attenuation lengths, given in the literature since they are based on different definitions. It should be noted that the experimentally obtained attenuation length is affected by statistical fluctuations during the development of the showers. However, in the present work we do not attempt to correct for this effect.

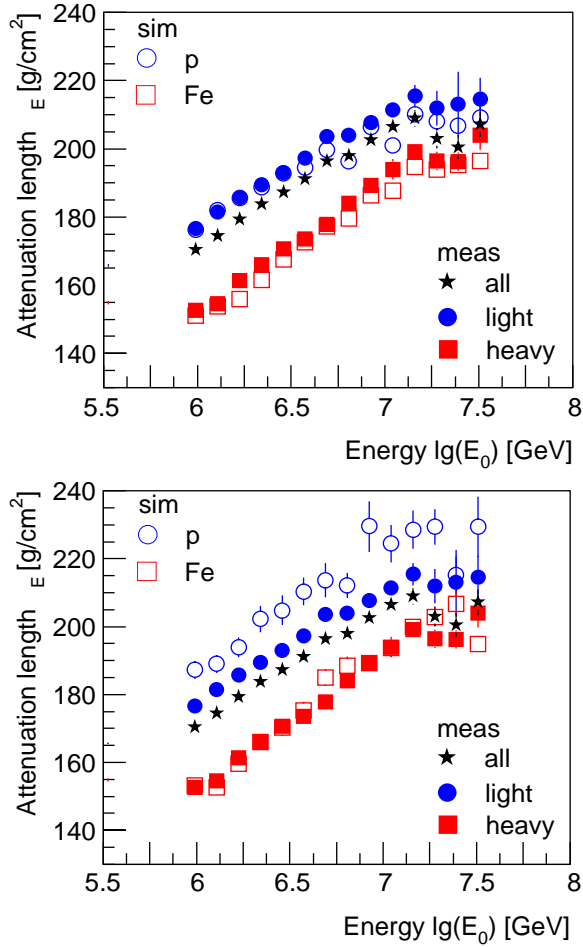

Fig. 6: Attenuation length $\lambda_E$ as function of estimated primary energy. The light and heavy groups in the measurements are compared to simulations for primary protons and iron-induced showers using CORSIKA with the hadronic interaction model QGSJET 01 (*top*) and a modified version with lower cross sections and higher elasticity (*bottom*, model 3a in Ref. [32]).

Measured values of $\lambda_E$ are shown in Fig. 6. Using a cut in the $N_e - N_\mu$ plane the data have been divided into a "light" and "heavy" sample, the corresponding values for $\lambda_E$ are depicted as well in the figure. Also predictions of air shower simulations for primary protons and iron nuclei, using the interaction code QGSJET 01 and a modified version with lower cross sections (model 3a in Ref. [32]) are shown. A closer inspection reveals that at high energies the $\lambda_E$ values of the "light" data selection are greater than the values for proton induced showers according to QGSJET 01. This is an unrealistic behavior. Lowering the inelastic hadronic cross sections by about 5% to 8% changes the situation, see lower panel. The predicted values for protons are now above the values for the "light" selection. This demonstrates the sensitivity of the observable $\lambda_E$ to hadronic cross sections applied in the simulations.

## VII. CONCLUSIONS

Quantitative tests of hadronic interaction models implemented in the CORSIKA program have been performed with KASCADE-Grande air shower data in the energy range $10^{14} - 10^{17}$ eV. They indicate that the model EPOS 1.61 is not compatible with air shower data — the new version EPOS 1.9 is presently under investigation. Predictions of the model QGSJET II-2, in particular the hadron-electron correlations are not compatible with measured values. Presently, the most consistent description of all air shower observables as obtained by the KASCADE-Grande experiment is achieved by the interaction models QGSJET 01 and SIBYLL 2.1.

The newly introduced method to measure an attenuation length of hadrons is in particular sensitive to inelastic hadronic cross sections applied in air shower simulations. A comparison of values predicted by QGSJET 01 to measured values suggests that the inelastic cross sections in QGSJET 01 are slightly too large. A version with 5% to 8% smaller cross sections is more compatible with the measurements.